\documentclass[aps,pre,twocolumn,superscriptaddress,showpacs]{revtex4-1}
\RequirePackage{mathtools}

\usepackage{mathrsfs}
\usepackage{amsfonts}
\usepackage{latexsym}
\usepackage{amsthm}
\usepackage{amsmath}
\usepackage{graphicx}
\usepackage{braket}
\usepackage{xcolor}
\usepackage[colorlinks,linkcolor=blue,citecolor=blue,urlcolor=blue]{hyperref}
\usepackage{subfigure}
\usepackage{CJK}

\newcommand{\opH}{\,{H}}

\newcommand{\opna}{\,{n}^a}
\newcommand{\opnb}{\,{n}^b}
\newcommand{\opa}{\,{a}}
\newcommand{\opb}{\,{b}}
\newcommand{\opaconj}{\,{a}^\dagger}
\newcommand{\opbconj}{\,{b}^\dagger}

\definecolor{wwblue}{RGB}{7,81,159}

\newtheorem{theorem}{Theorem}
\newtheorem{proposition}{Proposition}
\newtheorem{corollary}{Corollary}

\begin{document}
\begin{CJK*}{UTF8}{gbsn}
\title{Analysis and resolution of the ground-state degeneracy of the two-component Bose-Hubbard model}

\author{Wei Wang (王巍)}
\email{wei@ou.edu}
\affiliation{Homer L. Dodge Department of Physics and Astronomy, The University of Oklahoma, Norman, Oklahoma 73019, USA}

\author{Vittorio Penna}
\email{vittorio.penna@polito.it}
\affiliation{Dipartimento di Scienza Applicata e Tecnologia, Politecnico di Torino, Corso Duca degli Abruzzi 24, I-10129 Torino, Italy and CNISM, u.d.r., Politecnico di Torino, Corso Duca degli Abruzzi 24, I-10129 Torino, Italy}
\author{Barbara Capogrosso-Sansone}
\email{bcapogrosso@ou.edu}
\affiliation{Homer L. Dodge Department of Physics and Astronomy, The University of Oklahoma, Norman, Oklahoma 73019, USA}

\begin{abstract}
We study the degeneracy of the ground-state energy $E$ of the two-component
Bose-Hubbard model and of the perturbative correction $E_1$.
We show that the degeneracy properties of $E$ and $E_1$ are closely related to the
connectivity properties of the lattice. We determine general conditions under which $E$
is nondegenerate. This analysis is then extended to investigate the degeneracy of $E_1$. In this case, in addition to the lattice structure, the degeneracy also depends on the number of particles present in the system. After identifying the cases in which $E_1$ is degenerate and observing that the standard (degenerate) perturbation theory is not applicable, we develop a method to
determine the zeroth-order correction to the ground state by exploiting the symmetry
properties of the lattice.
This method is used to implement the perturbative
approach to the two-component Bose-Hubbard model in the case of degenerate $E_1$ and is expected to be a valid tool
to perturbatively study the asymmetric character of the Mott-insulator to superfluid transition between the particle and hole side.

\end{abstract}


\maketitle
\end{CJK*}

\section{Introduction}

Bosonic binary mixtures trapped in optical lattices have attracted considerable attention~\cite{quantummagneticphase2,bosonbosonexperiment1,bosonbosonexperiment2,bosonbosonexperiment3,Iskin1,Kuklov1,Kuklov2,Isacsson1,Pai1,Ozaki1,Nakano1,Barbara1,Barbara2} in the last decade
due to theoretic prediction of several new quantum phases originated by the interaction between the two components~\cite{Kuklov1,Kuklov2,Isacsson1,Pai1,Ozaki1,Nakano1}. The mixture can either consist of two atomic species or the same species in two different internal states, with each component being described 
within the Bose-Hubbard (BH) picture~\cite{Fisher1,Jaksch1}. The recent experimental realization of bosonic mixtures~\cite{bosonbosonexperiment1,bosonbosonexperiment2,bosonbosonexperiment3}, 
in addition to their rich phenomenology, has reinforced the interest 
for this class of systems. The mixture is described by the two-component BH model~\cite{Kuklov1}:
\begin{equation}
H= H_a + H_b +  U_{ab}\sum_{i=1}^M\opna_i \opnb_{i}
\label{model}
\end{equation}
where $U_{ab}$ represents the interspecies interaction, that is, the coupling between the two components $\cal A$ and $\cal B$. 
The local number operators 
$\opna_{i} = a^{\dagger}_i a_i$ and $\opnb_{i}= b^{\dagger}_i b_i$ are defined in terms of space-mode bosonic 
operators $a_i$ and $b_i$, relevant to species $\cal A$ and $\cal B$ respectively,
satisfying the standard commutators $[a_r, a^{\dagger}_i]=$ $[b_r, b^{\dagger}_i]= \delta_{ri}$
where $i,r \in [1,M]$ and $M$ is the number of lattice sites. 
$H_a$ and $H_b$ are defined by:
\begin{equation}
H_c = \frac{U_c}{2}\sum_{i=1}^M n^c_{i}(n^c_{i}-1) 
-T_c \sum_{(i,j)} I_{(i,j)}\,  c^{\dagger}_j c_{i} \, ,
\quad c = a,\,b\, 
\label{BHM}
\end{equation}
where $U_c$ is the onsite intraspecies interaction, 
$T_c$ the hopping amplitude describing boson tunneling,  
$I_{(i,j)}$ are the off-diagonal elements of the symmetric adjacency matrix in which bond $\{i,j\}$ runs through all pairs of sites
with $i \ne j$. 
For the common case of nearest-neighbor hopping only, one has $I_{(i,j)}=1$ on bonds $\{i,j\}$ 
connecting nearest neighboring sites and zero otherwise.
The study of mixtures by means of Monte Carlo simulations~\cite{Barbara1,Barbara2} has greatly contributed
to disclose many fundamental properties of the system and provided an accurate, unbiased study of several aspects of the global phase diagram.
On the other hand, the perturbation approach still represents a
considerably effective tool to obtain a deep insight on the structure of
the ground state and the microscopic processes governing the formation of quantum phases.
By construction, the analytic character of this method clearly shows how microscopic processes 
incorporated in the perturbation term of the Hamiltonian along with non trivial entanglement 
often characterizing mixtures influence the structure of ground states. In particular, entanglement between the two components is already present in the zeroth order correction of the ground state for certain choices 
of boson numbers $N_a$ and $N_b$, non-commensurate to $M$.
In higher dimensions, other analytic techniques such as the Gutzwiller mean-field approach~\cite{Jaksch1,Ozaki1} are able to
provide significant information for macroscopic
states characterized by no or weak entanglement. In some simpler cases, mean-field techniques can be improved by introducing `local' entanglement between the two components~\cite{Pai1,Barbara1}.

We are interested in applying the perturbation method to the two-component BH model with the ultimate goal of gaining some insight on the structure of the ground state and the role of entanglement resulting from the interspecies interaction~\cite{inprogress}. The application of the perturbation method, though,
can be challenging, certainly analytically but also numerically, owing to the remarkably-high degree of degeneracy that often characterizes
the ground state of the unperturbed Hamiltonian. In the sequel, we will
assume the hopping amplitudes $T_a$ and $T_b$ of the two species as the perturbation 
parameters for the two-component BH model.

To understand the nature of the degeneracy and the challenges of the perturbative calculation we can consider the following simple example. Let us consider the transition of bosonic component $\cal A$ from the Mott-insulator (MI) 
to the superfluid (SF) phase when component $\cal B$ is SF. This case, studied in 
Ref.~\cite{Barbara2,Barbara1} when component $\cal B$ is dilute, has revealed an 
evident asymmetric shift of the MI lobe of the (majority) component $\cal A$ between 
the particle and hole side of Mott lobe. This effect, in turn, appears to be related to 
a ground-state structure which features entanglement between $\cal A$ and $\cal B$ 
bosons which is substantially different in the particle and 
hole-excitation case~\cite{inprogress}. 
The most elementary version of this transition is found in the limit $T_a, T_b \to 0$
by considering a SF component $\cal B$ with $N_b = k_b M+1$  
together with a Mott component $\cal A$ with $N_a= k_a M$ (where $k_a, k_b$ are nonnegative integers).
The corresponding zeroth-order ground state is 
$\ket{k_a}\otimes\sum_{s}(b^{\dagger}_s/\sqrt M)\ket{k_b}$,
where $\ket{k_c}$, $c = a,b$ represents a Mott state with filling $k_c$, and
$b^{\dagger}_s\ket{k_b}$ describes the creation of a solitary boson at site $s$ causing the SF character
of species $\cal B$.
In the grand-canonical ensemble, when the energy cost for adding a boson to component $\cal A$ is zero, the transition to zeroth-order ground state of the form $\sum_{ik} F_{ik} a^{\dagger}_i b^{\dagger}_k \ket{k_a}\otimes\ket{k_b}$
corresponding to $\cal A$ and $\cal B$ both superfluid occurs. 
In order to minimize the contribution of
the interspecies-interaction term to the ground-state energy $E_0$, the diagonal elements of the 
$M \times M$ matrix $F$ must be zero. In general, the high degree
of degeneracy (represented by the arbitrariness of $F_{ik}$) is removed by imposing the 
minimization of the first-order perturbative correction $E_1$ with respect to the undetermined parameters $F_{ik}$. 
This solution scheme, however, is viable only if $E_1$ is not degenerate,
a condition whose validity can be shown to depend on the number of particles, the lattice properties, and possibly on model parameters. 

While this simple case can be solved analytically~\cite{inprogress}, for situations with 
$N_a = k_aM+A$ and $N_b = k_bM+B$, where $A$ and $B$ are arbitrary integers $A,B \in [1, M-1]$, 
the determination of the zeroth-order ground-state amplitudes (e.g matrix elements $F_{ik}$ in the example illustrated above) cannot be done analytically and can easily become numerically costly. For this reason, determining general conditions for which the ground-state energy $E$ 
of model $H$ and the first-order correction $E_1$ are nondegenerate (without resorting
to complicated either analytical or numerical calculations) represents a precious, essential
information for implementing the perturbation method.

In this paper we show that both the ground-state energy $E$ of $H$ and the lowest eigenvalue
of the perturbation term formed by the $T_a$ and $T_b$-dependent terms in $H$ are nondegenerate 
if the simple condition to have a {\it connected lattice} $\bf G$ is satisfied. 
Then, after observing that if the unperturbed ground-state energy $E_0$ is degenerate
this degeneracy can be eliminated if the first-order correction $E_1$ is nondegenerate,
we explore the conditions for which first-order correction $E_1$ is nondegenerate
for different choices of $A$ and $B$.  

This paper is organized as follows. 
Section~\ref{Sec2} is devoted to give some useful definitions and 
to recast the model interaction/hopping parameters
into a form more advantageous for our perturbation approach.
In Section~\ref{Sec3}, we define the connectedness between states of the Fock basis and 
give the sufficient and necessary condition that links the lattice connectivity to the state
connectedness. This allows us to apply the Perron-Frobenius theorem~\cite{Tasaki1}
to study the degeneracy properties of the ground-state energy. Concerning the definition of states' connectedness assumed in this paper,
it should be noted that similar definitions, followed by the application of Perron-Frobenius 
theorem, are used in Katsura and Tasaki's recent work~\cite{Katsura1} in the proof of the 
degeneracy of the spin-1 Bose-Hubbard 
model and previously in Ref.~\cite{Nagaoka1, Thouless1, Tasaki2} devoted to the study of  ferromagnetism of the 
Hubbard model. Our method differs from previous studies in the fact that we define the connectedness in a different 
way in order to give an equivalence relation. This allows us to conveniently study the degeneracy of the gorund state energy and its first order correction.

In Section~\ref{Sec4} we discuss the degeneracy of $E_1$ in two cases: 
(1) one of the two species is a MI while the other is SF; 
(2) there are $k_aM +1$ (or $k_aM -1$) 
species-$\cal A$ bosons while species $\cal B$ is SF with a generic
number of bosons. In Section~\ref{Sec5}, we extend our discussion to generic cases with the only requirement, $A+B<M-1$ or $A+B>M+1$. Our analysis shows that $E_1$ is nondegenerate if one assumes certain sufficient conditions on the connectivity of the lattice. These conditions are satisfied by most lattices.

Finally, in Section~\ref{Sec6} we show that when $A+B=M$, $E_1$ is degenerate independently on the connectivity of the lattice. We therefore discuss the determination of the unperturbed ground state in terms of symmetry properties of the lattice for $N_a$ and $N_b$ such that $A = 1$, $B= M-1$ or $A=M-1$, $B=1$.


\section{\label{Sec2} The two-component model in the strongly-interacting regime}

We intend to study the two-component Bose-Hubbard model
in the strongly-interacting regime where $T_a,T_b \ll U_a, U_b, U_{ab}$. 
We assume that $T_a\approx{T_b}$, stating that the mobility of the bosons
of the two components is essentially the same, and $0\le{I_{(i,j)}}\le 1$. 
To avoid phase separation we also assume (repulsive) onsite interactions
such that $U_{ab}<U_a,U_b$. Although in the following we will explicitly consider the case of soft-core bosons, i.e. $U_a,U_b<\infty$, the results presented are also valid for the case of hard-core bosons~\cite{inprogress1}. The application of the perturbation 
method suggests the definition of new interaction/hopping parameters
$$
T_a = T \, t_a\, ,\quad T_b = T \, t_b \, ,\quad U_a = U \, u_a\, ,\quad U_b = U \, u_b\, ,
$$
entailing that model Hamiltonian $H$ takes the form
\begin{multline}
\label{Eq0}
\opH= \, U\, \bigg[
\frac{u_a}{2}\sum_{i=1}^M\opna_{i}(\opna_{i}-1)
+\frac{u_b}{2}\sum_{i=1}^M\opnb_{i}(\opnb_{i}-1) \\
+\frac{U_{ab}}{U} \sum_{i=1}^M\opna_i\opnb_{i}\bigg] \\
+ T \bigg[-t_a\sum_{(i,j)}I_{(i,j)}\, \opaconj_i\opa_j 
-t_b\sum_{(i,j)}I_{(i,j)}\, \opbconj_i\opb_j\, \bigg],
\end{multline}
in which we call $H_0$ the $U$-dependent diagonal part of the Hamiltonian
and $T W$ represents the kinetic energy part of $H$ (fourth and fifth $T$-dependent terms 
in Eq.~\ref{Eq0}). Then ${T}W/{U}$ represents the perturbation and $\epsilon = {T}/{U}$ naturally 
identifies with the perturbation parameter.

The sites of the optical lattice and the set of all bonds $\{i,j\}$ with weight $I_{(i,j)}\ne0$ 
define an edge-weighted graph $\mathbf{G}=(\mathcal{V}(\mathbf{G}),\mathcal{E}(\mathbf{G}))$, 
where $\mathcal{V}(\mathbf{G})$ is the set of vertices, i.e. sites, $\mathcal{E}(\mathbf{G})$ 
is the set of edges, i.e. bonds (not necessarily nearest neighbors), and $I_{(i,j)}$ is the 
weight on bond $\{i,j\}$ 

In the following we will work in a finite-dimensional Hilbert space $\mathscr{H}$ 
corresponding to fixed particle numbers $N_a=k_a{M}+A$ and $N_b=k_b{M}+B$ for the two components 
respectively~\footnote{Particle number operators commute with both $H_0$ and $H$.}. 
Here, $k_a$, $k_b$, $A$ and $B$ are nonnegative integers with $0\le{A}<M$, $0\le{B}<M$. 
The space $\mathscr{H}$ is spanned by an orthonormal basis of Fock states 
$\ket{\xi}\otimes\ket{\gamma}$:
\begin{equation}
\label{Eq1}\ket{\xi}\otimes\ket{\gamma}=\frac{\prod_{l=1}^{M}{\opaconj_{l}}^{\xi_{l}}\prod_{m=1}^{M}{\opbconj_{m}}^{\gamma_m}\ket{0}}{\sqrt{\prod_{l=1}^{M}\xi_{l}!\prod_{m=1}^{M}\gamma_{m}!}},
\end{equation}
where $\ket{0}$ is the vacuum states, i.e. every site in the lattice is empty, and $\xi$, $\gamma$ are integers such that $\sum_{l=1}^{M}\xi_{l}=k_{a}M+A$, $\sum_{m=1}^{M}\gamma_{m}=k_{b}M+B$.

To simplify our notation we use $\ket{\xi,\gamma}$ to denote $\ket{\xi}\otimes\ket{\gamma}$. The ground state(s) of $H_0$ are labelled by $\ket{\sigma,\lambda}$. We call $O$ the set of states $\ket{\xi,\gamma}$s, and $O_g$ the set of the states $\ket{\sigma,\lambda}$s. The operator $W$ confined in the subspace spanned by $\ket{\sigma,\lambda}$'s is denoted by $W_g$. The matrix representation of $W$ in terms of the basis $\ket{\xi,\gamma}$'s is denoted by $\mathbf{W}$ and the matrix representation of $W_g$ in terms of $\ket{\sigma,\lambda}$'s is denoted by $\mathbf{W_g}$.

The explicit expression of matrix elements of $\mathbf{W}$ is given by:
\begin{widetext}
\begin{multline}
\label{Eq2}
\braket{\xi,\gamma|W|\xi',\gamma'}=
-t_a\, \delta_{\gamma,\gamma'} 
\sum_{(i,j)}
\Big[ 
I_{(i,j)}\sqrt{\xi_{j}+1}\sqrt{\xi'_{i}+1}\delta_{\xi_{j}+1,\xi'_j}\delta_{\xi_{i},\xi'_{i}+1}
\prod_{l\ne{i,j}}^M\delta_{\xi_l\xi'_l}
\Big]
\\
-{t_b}\, \delta_{\xi,\xi'}
\sum_{(i,j)}
\Big[
I_{(i,j)}\sqrt{\gamma_{j}+1}\sqrt{\gamma'_{i}+1} \delta_{\gamma_{j}+1,\gamma'_j}\delta_{\gamma_{i},\gamma'_{i}+1}\prod_{m\ne{i,j}}^M\delta_{\gamma_m,\gamma'_m}
\Big].
\end{multline}
\end{widetext}
It is obvious that the matrix elements are nonpositive. Moreover, a matrix element is nonzero if and only if $\xi'_{i}=\xi_{i}+1$, $\xi'_{j}+1=\xi_{j}$ on bond $\{i,j\}$ while all other $\xi'_l=\xi_l$, and $\gamma=\gamma'$; or $\gamma'_{i}=\gamma_{i}+1$, $\gamma'_{j}+1=\gamma_{j}$ on bond $\{{i,j}\}$ while all other $\gamma'_l=\gamma_l$, and $\xi=\xi'$. This property of matrix elements will be used below.

After recalling that in the strongly-interacting regime we treat the term $\epsilon W$ 
as a perturbation with $\epsilon = {T}/{U}$, the ground-state energy $E$ of $H= U( H_0+ \epsilon W)$ 
and its eigenvector(s) $\ket{\Psi_l}$ ($l=1,\cdots,f_E$, where $f_E$ is the degeneracy of $E$) 
can be expanded via the perturbative series 
$E=E_0+\sum_{n=1}^{\infty} \epsilon^{n} E_n$ and 
$\ket{\Psi_l}=\ket{\psi^0_l}+\sum_{n=1}^{\infty} \epsilon^{n}\ket{\psi^n_l}$. 
Note that if $A+B>0$, then $H_0$ has a degenerate ground-state energy $E_0$ with degeneracy $f_{E_0}$. 
If $f_{E_0}=f_E$, one can apply perturbation theory starting from any ground state of $H_0$. 
If $f_{E_0}>f_E$ and $W$ fully lifts the extra degeneracy of $E_0$, then $E_1$ and 
$\ket{\psi^0_l}$ can be uniquely determined by solving the matrix eigenvalue problem 
(degenerate perturbation theory):
\begin{equation}
\label{Eq3}\sum_{\sigma',\lambda'}\braket{\sigma,\lambda|W|\sigma',\lambda'}\braket{\psi^0_l|\sigma',\lambda'}=E_{1}\braket{\psi^0_l|\sigma,\lambda}
\end{equation}
On the other hand, if neither of the previous scenarios are true, then $\ket{\psi^0_l}$'s are not uniquely determined by solving Eq.~\ref{Eq3}. From this discussion, it becomes apparent that one needs to study the degeneracy of both $E$ and $E_1$, and, in the case Eq.~\ref{Eq3} is not applicable, find an alternative method to determine $\ket{\psi^0}$. 

As we will show in the following, the issue of degeneracy is closely related to the connectivity of the lattice. 


\section{\label{Sec3}
Degeneracy of the ground-state energy, connectivity of lattice and connectedness between states}

In this section, we discuss the degeneracy of the ground-state energy of $H$ by utilizing 
the notion of ``connectedness'' on the states of the basis and the Perron-Frobenius 
theorem (PFT).
This theorem states that if $\mathbf{X}$ is a real symmetric matrix such that (i) off-diagonal 
elements are all nonpositive, (ii) for any two different indices $p$ and $q$ there exists an $N$ 
such that $(\mathbf{X}^{N})_{p,q}\ne0$, then its lowest eigenvalue is nondegenerate and the 
corresponding eigenvector is positive~\cite{Tasaki1}. In the following we will apply PFT theorem for the case of matrix $\mathbf{W}$ and $\mathbf{H}$.

To begin with, we define the ``connectedness'' on states via a 
symmetric linear operator $X$~\footnote{A linear operator $X$ is symmetric if, for arbitrary 
states $\ket{\phi}$ and $\ket{\psi}$, $\braket{\phi|X|\psi}=\braket{\psi|X|\phi}$.} (its corresponding matrix is denoted by $\mathbf{X}$). 

Hence, we say that $\ket{\xi,\gamma}$ and $\ket{\xi',\gamma'}$ are connected by a 
symmetric linear operator $X$ if there exists a finite sequence 
$\{\ket{\alpha_1,\beta_1},\ket{\alpha_2,\beta_2},\cdots,\ket{\alpha_N,\beta_N}\}$ with $\ket{\alpha_1,\beta_1}=\ket{\xi,\gamma}$ and $\ket{\alpha_N,\beta_N}=\ket{\xi',\gamma'}$ in the set O
such that for any $1\le{i}<N$, $\braket{\alpha_i,\beta_i|X|\alpha_{i+1},\beta_{i+1}}\ne{0}$ 
or $\ket{\alpha_i,\beta_i}=\ket{\alpha_{i+1},\beta_{i+1}}$. This definition includes the 
trivial case $\ket{\xi,\gamma}=\ket{\xi',\gamma'}$. The kinetic energy operator $W$ and the Hamiltonian $H$ are indeed symmetric linear operators. The connectedness associated with 
$X$ defines an equivalence relation~\footnote{
A relation $\mathfrak{R}_X$ on 
a set $O$ is a collection of ordered pairs $(\ket{\xi,\gamma},\ket{\xi',\gamma'})$ in $O$. 
If $(\ket{\xi,\gamma},\ket{\xi',\gamma'})\in{O}$, 
we say $\ket{\xi,\gamma}\mathfrak{R}_X\ket{\xi',\gamma'}$. 
$\mathfrak{R}_X$ is an equivalence relation on a set $O$ if it satisfies 

(i) reflexivity, i.e. $\ket{\xi,\gamma}\mathfrak{R}_X\ket{\xi,\gamma}$, 
(ii) symmetry, i.e. $\ket{\xi,\gamma}\mathfrak{R}_X\ket{\xi',\gamma'}$ $\rightarrow$ 
$\ket{\xi',\gamma'}\mathfrak{R}_X\ket{\xi,\gamma}$, 
(iii) transitivity, i.e. 
$\ket{\xi,\gamma}\mathfrak{R}_X\ket{\xi',\gamma'}$, 
$\ket{\xi',\gamma'}\mathfrak{R}_X\ket{\xi'',\gamma''}$ 
$\rightarrow$ $\ket{\xi,\gamma}\mathfrak{R}_X\ket{\xi'',\gamma''}$. 

For a given relation $\mathfrak{R}_X$, $\ket{\xi,\gamma}/\mathfrak{R}_{X}$ denotes 
the set of all $\ket{\xi',\gamma'}$ related to $\ket{\xi,\gamma}$, and $O/\mathfrak{R}_{X}$ 
denotes the collection of all $\ket{\xi,\gamma}/\mathfrak{R}_{X}$'s. 
An important property of equivalence relations is that $O/\mathfrak{R}_{X}$ 
is a partition of $O$~\cite{settheory}. It's easy to check that $\mathfrak{R}_X$ is well-defined here.}
$\mathfrak{R}_X$ on $O$ such that 
$\ket{\xi,\gamma}\mathfrak{R}_X\ket{\xi',\gamma'}$ if and only if the two 
states are connected by $X$. Given this equivalence relation, we can prove that:
\medskip

\noindent
\begin{proposition}
\label{Prps1}
The following three conditions are equivalent:
(a) $\mathbf{X}$ is irreducible~\footnote{A symmetric matrix is irreducible if and only if it cannot be block-diagonalized by permuting the indices.},
(b) any $\ket{\xi,\gamma}$ and $\ket{\xi',\gamma'}$ are connected by $X$, 
(c) property-(ii) in PFT is satisfied.
\end{proposition}
\medskip

\noindent
The proof is given in the Appendix \ref{App1}.

\subsection{Connectivity of $\mathbf{G}$ and the nondegeneracy of $E$}

We want to prove that the ground state of H is nondegenerate by making use of 
PFT. Hence, we need to show that $\mathbf{H}$ satisfies 
the hypothesis of the theorem. We first notice that both $H$ and $W$ are symmetric linear operators. Moreover, 
$\braket{\xi,\gamma|( H_0 +\epsilon \, W) |\xi',\gamma'}=
\braket{\xi,\gamma|H_0|\xi',\gamma'}+\epsilon \braket{\xi,\gamma| W|\xi',\gamma'}
=\epsilon \braket{\xi,\gamma|W|\xi',\gamma'}\le{0}$ for any 
$\ket{\xi,\gamma}\ne\ket{\xi',\gamma'}$, and hence $\mathbf{H}$ and $\mathbf{W}$ 
are both real matrices with nonpositive off-diagonal elements, as requested 
by condition (i) of the PFT. Moreover, $\ket{\xi,\gamma}$ and $\ket{\xi',\gamma'}$ are connected by $H$ if and only if they are connected by $W$.
Next step is to show that condition (ii) is also satisfied. In view of Proposition~\ref{Prps1}, it is sufficient to show that 
any $\ket{\xi,\gamma}$ and $\ket{\xi',\gamma'}$ are connected by $W$. From here on, we will 
assume that $\mathbf{G}$ is connected~
\footnote{
$\mathbf{G}$ is connected if any two sites 
can be linked by a path. Two sites $i$ and $j$ are linked if there exists a path 
$\{i,\cdots,k_l,\cdots,j\}$ in which every  neighboring pair in the sequence forms a bond.}
%
(note that, the connectivity of $\mathbf{G}$ is different from the connectedness 
on the basis). We can show that:
\medskip

\noindent
\begin{proposition}
\label{Prps2}
Any $\ket{\xi,\gamma}$ and $\ket{\xi',\gamma'}$ are connected by $W$ 
if and only if $\mathbf{G}$ is connected.
\end{proposition}
\medskip

\noindent

We first prove the sufficient condition. The general idea of the proof is to connect both states $\ket{\xi,\gamma}$ and $\ket{\xi',\gamma'}$ to a state such that all particles are sitting on the same lattice site $k$, and then apply the transitivity property. Let us fix a site $k$. Then, any other site is linked to $k$ by a path. Since $\ket{\xi'',\gamma''}=c_{i}c^{\dagger}_{j}\ket{\xi,\gamma}/\|c_{i}c^{\dagger}_{j}\ket{\xi,\gamma}\|$, $c=a,b$, is connected to $\ket{\xi,\gamma}$ if $\{i,j\}$ is a bond (see Eq.~\ref{Eq2}), we can apply $a_{i}a^{\dagger}_{j}$ or $b_{i}b^{\dagger}_{j}$ subsequently on the appropriate bonds $\{i,j\}$ in order to construct the special state $\ket{\xi''',\gamma'''}$ connected to $\ket{\xi,\gamma}$. $\ket{\xi''',\gamma'''}$ is such that all bosons are sitting on site $k$. Next, we perform a similar operation on $\ket{\xi',\gamma'}$ to connect it to $\ket{\xi''',\gamma'''}$. By transitivity of connectedness we have that $\ket{\xi,\gamma}$ and $\ket{\xi',\gamma'}$ are connected to each other. A similar construction of intermediate states is also used in~\cite{Katsura1}.

Here, we prove the necessary condition by contradiction. Let us assume that the lattice is not connected. Then, there exists at least two sites $k$ and $l$ not linked by any path. Let $K$ be the set of sites linked to $k$. Then the complement of $K$ in $\mathcal{V}(\mathbf{G})$ is nonempty and is not linked to $K$~\footnote{Every site in $K$ is not linked to any site in its complement.}. Since there always exists states $\ket{\xi,\gamma}$ and $\ket{\xi',\gamma'}$ with different total number of particles on the sites belonging to $K$, then, these two states are obviously not connected. We get contradiction. 
\medskip

\noindent
We are now ready to apply PFT to conclude:

\begin{theorem}
\label{The1}
If $\mathbf{G}$ is connected, then the ground-state energy of $H$ is nondegenerate and 
it has a positive~\footnote{A vector is positive (in terms of the basis) if its expansion 
coefficients are all positive.}
ground state.
\end{theorem}

\begin{corollary}
\label{Cor1}
If $\mathbf{G}$ is connected, then $\epsilon \, W$ has a negative nondegenerate 
ground-state energy with a positive ground state.
\end{corollary}

Another interesting result can be derived by setting the hopping amplitude of one of 
the components to zero, e. g. $T_b=0$. Then, from Eq.~\ref{Eq2}, it is obvious that 
for any $\ket{\xi,\gamma}$, $\ket{\xi',\gamma'}$ with $\ket{\gamma}\ne\ket{\gamma'}$, 
the two states are not  connected by $W$. One can show the following result 
(the proof is given in Appendix \ref{App2}):

\begin{corollary}
\label{Cor2}
If $\mathbf{G}$ is connected and $T_a=0,T_b\ne{0}$ (or $T_b=0,T_a\ne{0}$), 
the degeneracy of the ground-state energy of $W$ is $M^{N_b}$ ($M^{N_a}$).
\end{corollary}

It is worth noting that the results presented in this Section are quite general. 
They hold for the case of nearest-neighbor or longer-ranged hopping. Moreover, 
they are valid  in both the weakly and strongly correlated regime and for both 
repulsive or attractive interspecies interaction. The only requirement is for 
the lattice to be connected and $T_{a,b}$ , $I_{(i,j)}$ to be nonnegative. Finally, we would like to mention that the results presented in this Section are also valid for the case of hard-core bosons although the specifics of the proofs and the degeneracy in Corollary~\ref{Cor2} are different~\cite{inprogress1}. Likewise, the results presented in the following hold for hard-core bosons as well since they are based on the results of Section.~\ref{Sec3}.

\section{\label{Sec4}Degenerate perturbation theory}

Theorem~\ref{The1} states that the ground-state energy $E$ of model~\ref{Eq0} is nondegenerate. In the general case where at least one of the two component is doped away from integer filling, the ground state corresponding to $E_0$ is degenerate. Then, the first order correction $E_1$ can either be nondegenerate (i.e. $W$ completely lifts the degeneracy of $E_0$), in which case $\ket{\psi^0}$ is uniquely determined, or degenerate, in which case $\ket{\psi^0}$ is not uniquely determined. In this section, we discuss degeneracy properties of $E_1$ in terms of graph theoretical properties of $\mathbf{G}$. In the case when $E_1$ is degenerate, we provide a method to determine $\ket{\psi^0}$ according to symmetry properties of $\mathbf{G}$, hence providing a rigorous solution to the degenerate perturbation theory Eq. \ref{Eq3}.
This case is discussed in Section \ref{Sec6}.

\subsection{\label{subsection4_1}Representing $\ket{\sigma,\lambda}$'s pictorially}

At commensurate filling, i.e. $N_a= k_a M$ and $N_b=k_b M$, the potential energy is 
minimized when, on each site, there are $k_a$ $\cal A$ bosons and $k_b$ $\cal B$ bosons 
(recall we are considering $U_{ab}<U_a,U_b$). When one or both components are doped away 
from integer filling factor, the extra particles arrange themselves in order to minimize 
the interspecies-interaction term in $H$. 
In particular, a given site will accommodate at most $k_a+1$ species-$\cal A$ 
bosons and $k_b+1$ species-$\cal B$ bosons. Hence, we can specify an arbitrary ground-state 
$\ket{\sigma,\lambda}$ of $H_0$ in terms of the sites which accommodate extra particles.

More specifically, when $A+B\le{M}$, there are no sites with both an extra $\cal A$ 
and an extra $\cal B$ boson. Thus, the set of sites with an extra $\cal A$ boson has $A$ 
elements, and the set of sites with an extra $\cal B$ boson has $B$ elements. 
Such sets have an empty intersection. On the other hand, when $A+B>M$, there 
are $A+B-M$ sites with both an extra $\cal A$ and an extra $\cal B$ boson. 
In this case, the relevant sets have a nonempty intersection containing $A+B-M$ 
sites.

\begin{figure}
\centering
\subfigure[\label{Fig001}]{\includegraphics[width=0.10\textwidth]{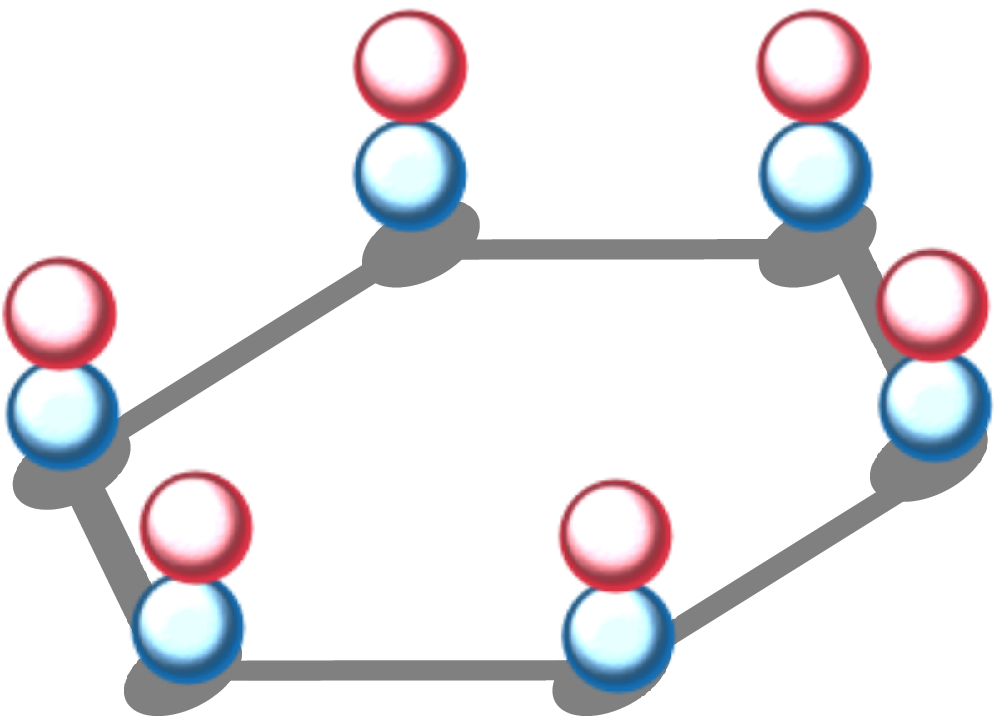}}\quad\quad
\subfigure[\label{Fig002}]{\includegraphics[width=0.10\textwidth]{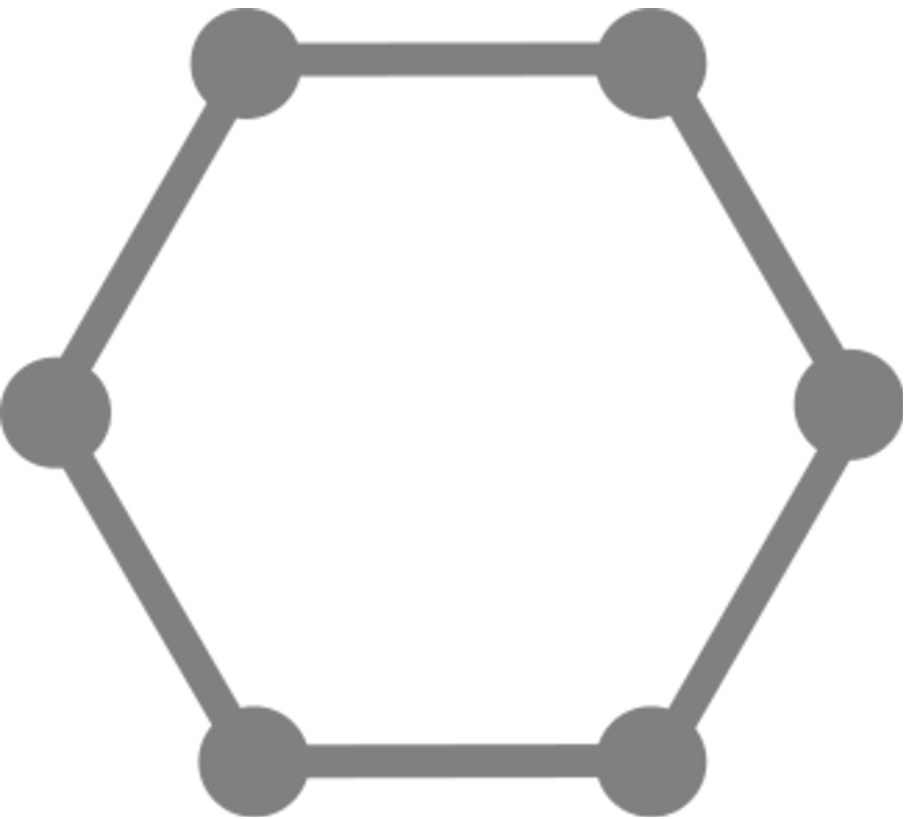}}\quad\quad
\subfigure[\label{Fig01}]{\includegraphics[width=0.10\textwidth]{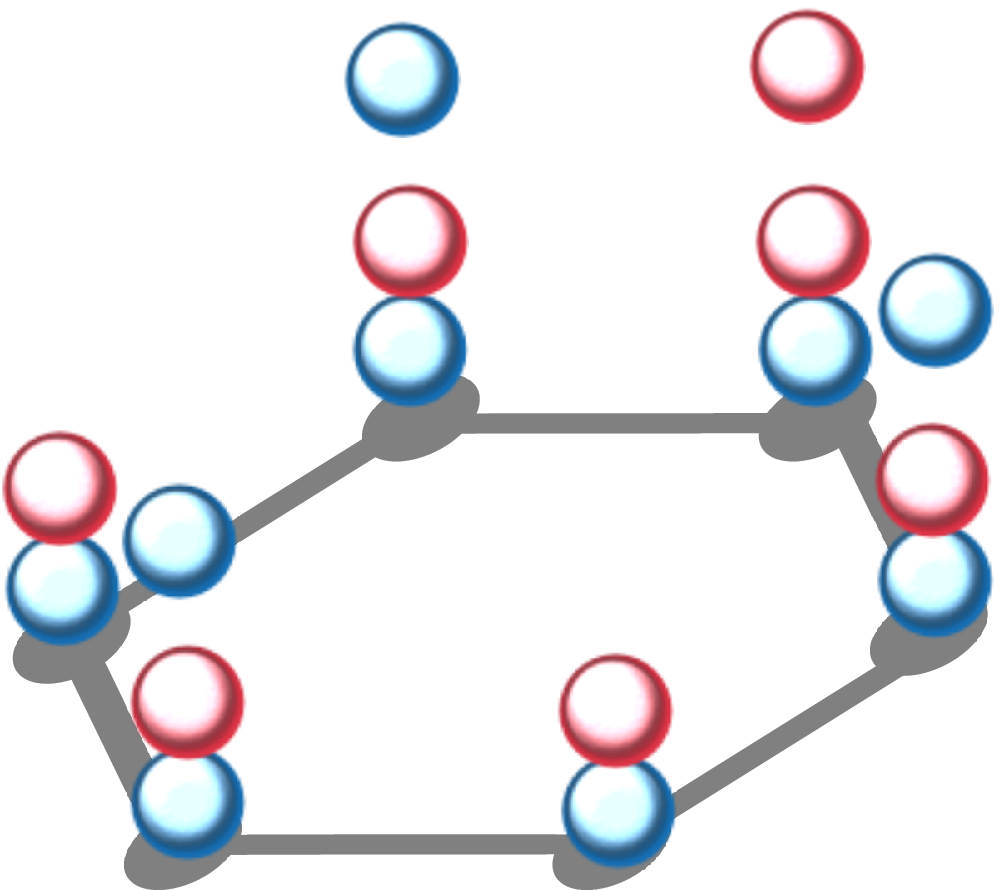}}\quad\quad
\subfigure[\label{Fig1}]{\includegraphics[width=0.10\textwidth]{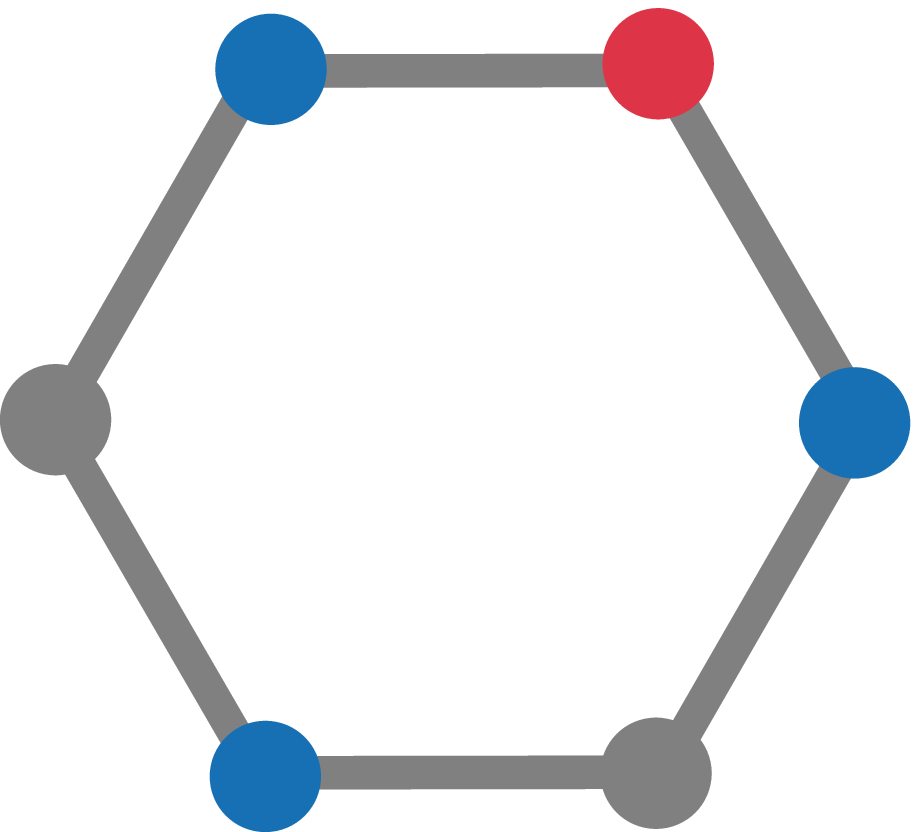}}\quad\quad
\subfigure[\label{Fig02}]{\includegraphics[width=0.10\textwidth]{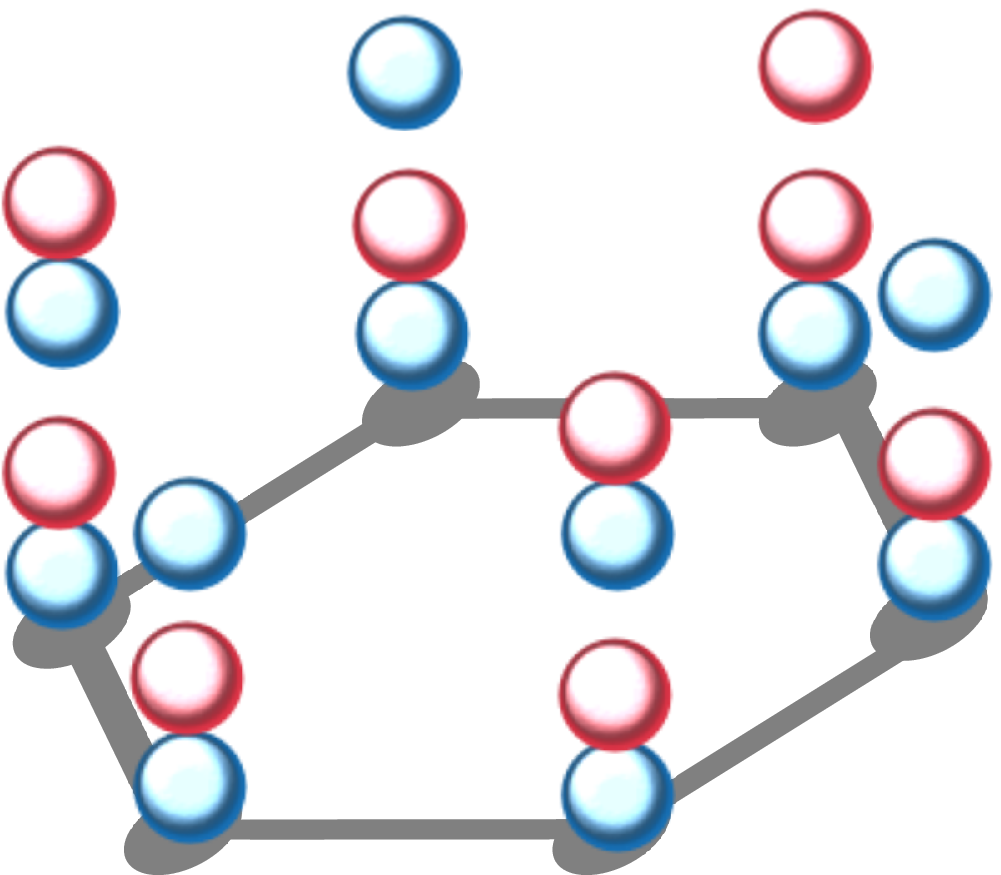}}\quad\quad
\subfigure[\label{Fig2}]{\includegraphics[width=0.10\textwidth]{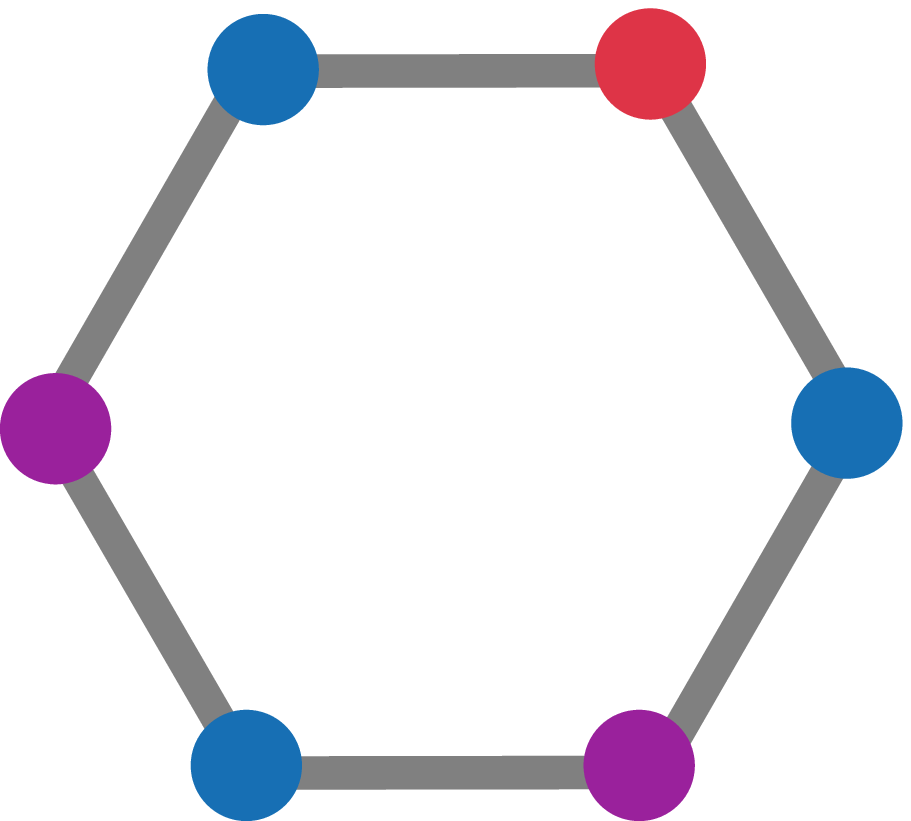}}
\caption{\label{FigOne}\ref{Fig001}, \ref{Fig01} and \ref{Fig02} are an example of states in $\ket{\sigma,\lambda}$s for the case of a one-dimensional lattice with periodic boundary condition and M=6. These states correspond to $(k_a=1,k_b=1,A=0,B=0)$, $(k_a=1,k_b=1,A=3,B=1)$, and $(k_a=1,k_b=1,A=5,B=3)$ respectively. The color blue refers to $\cal A$ bosons and red to $\cal B$ bosons. They are represented pictorially by \ref{Fig002}, \ref{Fig1} and \ref{Fig2}, where blue sites form the set $\sigma$, red sites form the set $\lambda$, and purple sites form the intersection of $\sigma$ and $\lambda$. }
\end{figure}

We can therefore identify an arbitrary ground-state $\ket{\sigma,\lambda}$ in terms of 
the sets $\sigma$ and $\lambda$ corresponding to A sites with an extra $\cal A$ boson and $B$ 
sites with an extra $\cal B$ boson, respectively. The set of ground-states $O_g$ is therefore
represented by a collection of pairs of sets $(\sigma,\lambda)$s. In the following, for the 
sake of simplicity but without loss of rigor, we represent states $\ket{\sigma,\lambda}$ 
pictorially by coloring sites belonging to $\sigma$ in blue, sites belonging to $\lambda$ in red, and sites 
belonging to the intersection between $\sigma$ and $\lambda$ in purple. 
Site with neither extra $\cal A$ nor extra $\cal B$ bosons are colored in grey. 
Examples of the mapping from $\ket{\sigma,\lambda}$ to $(\sigma,\lambda)$ are shown 
in Fig.~\ref{FigOne}, where states \ref{Fig001}, \ref{Fig01} and \ref{Fig02} are 
represented by \ref{Fig002}, \ref{Fig1} and \ref{Fig2} respectively. 

$W_g$ is a symmetric linear operator in the subspace spanned by $\ket{\sigma,\lambda}$s, 
therefore we can define the connectedness between states $\ket{\sigma,\lambda}$s by $W_g$. 
From here on, we will describe connectedness by using the representation in terms of color 
of sites. For example, according to Eq.~\ref{Eq2}, when 
$A+B<M$, $\braket{\sigma,\lambda|W_g|\sigma',\lambda'}$ 
is nonzero if and only if one grey site exchanges color with a blue or red site on a bond while all other colors remain unchanged. 
When $A=M-1,B>1$, $\braket{\sigma,\lambda|W_g|\sigma',\lambda'}$ is nonzero if and only if 
one purple site exchanges color with a blue or red site on a bond while all other colors remain 
unchanged~\footnote{These rules can also be stated formally: 
$\braket{\sigma,\lambda|W_g|\sigma',\lambda'}$ is nonzero if and only if 
either set $\lambda=\lambda'$ while sets $\sigma$, $\sigma'$ 
only differ by sites $i$ and $j$ belonging to the bond 
$\{i,j\}$, or set $\sigma=\sigma'$ while sets $\lambda$, $\lambda'$ 
only differ by sites $k$ and $l$ belonging to the bond $\{{k,l}\}$.}. 

It should now be apparent that 
purple sites behave in the same way as grey sites. Using this language, the rules to generate 
a connected state are as follows: (i) change the color of a grey (or purple) site with 
a red or blue site on a bond; (ii) ``exchange'' the color of two sites with the same color 
on a bond (this operation is trivial and results from the definition of connectedness 
where two neighboring states in the connecting sequence can be identical). This second 
rule is introduced just for the sake of convenience in the remainder of our discussion.

\begin{figure}
\subfigure[\label{Fig3}]{\includegraphics[width=0.07\textwidth]{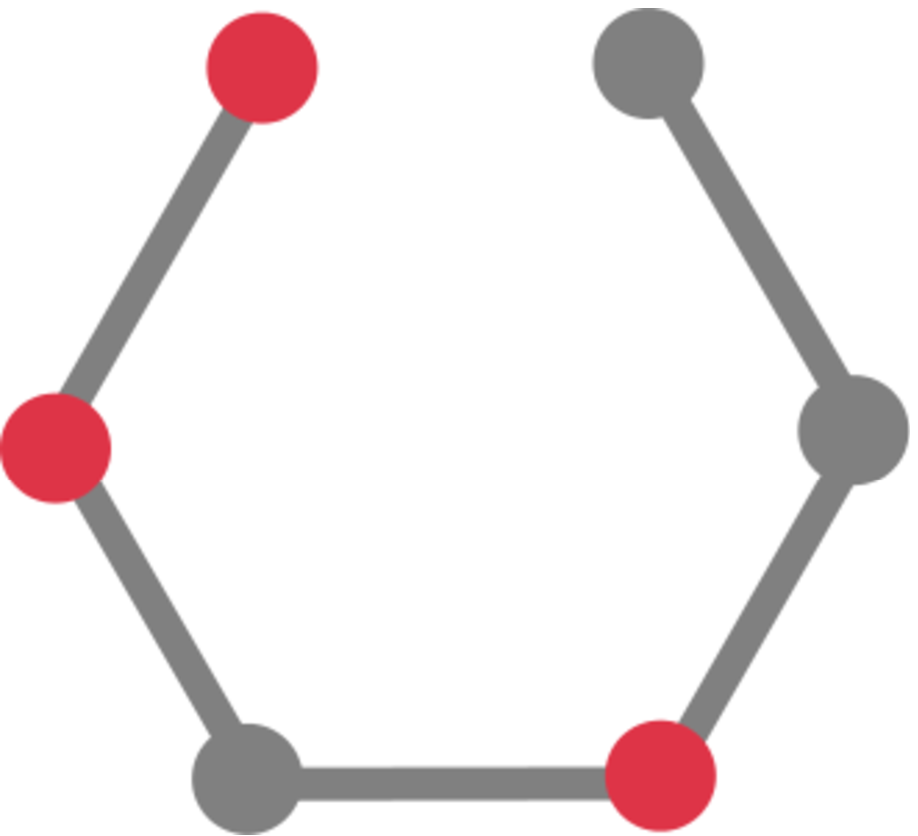}}\quad\quad
\subfigure[\label{Fig3_1}]{\includegraphics[width=0.07\textwidth]{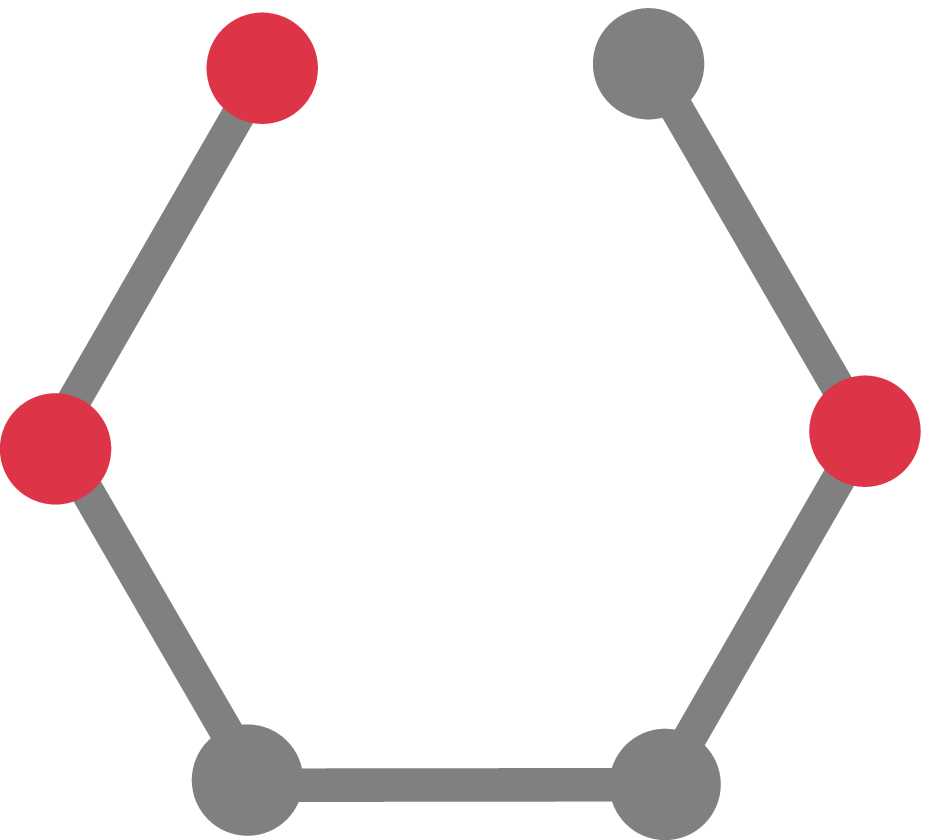}}\quad\quad
\subfigure[\label{Fig4}]{\includegraphics[width=0.07\textwidth]{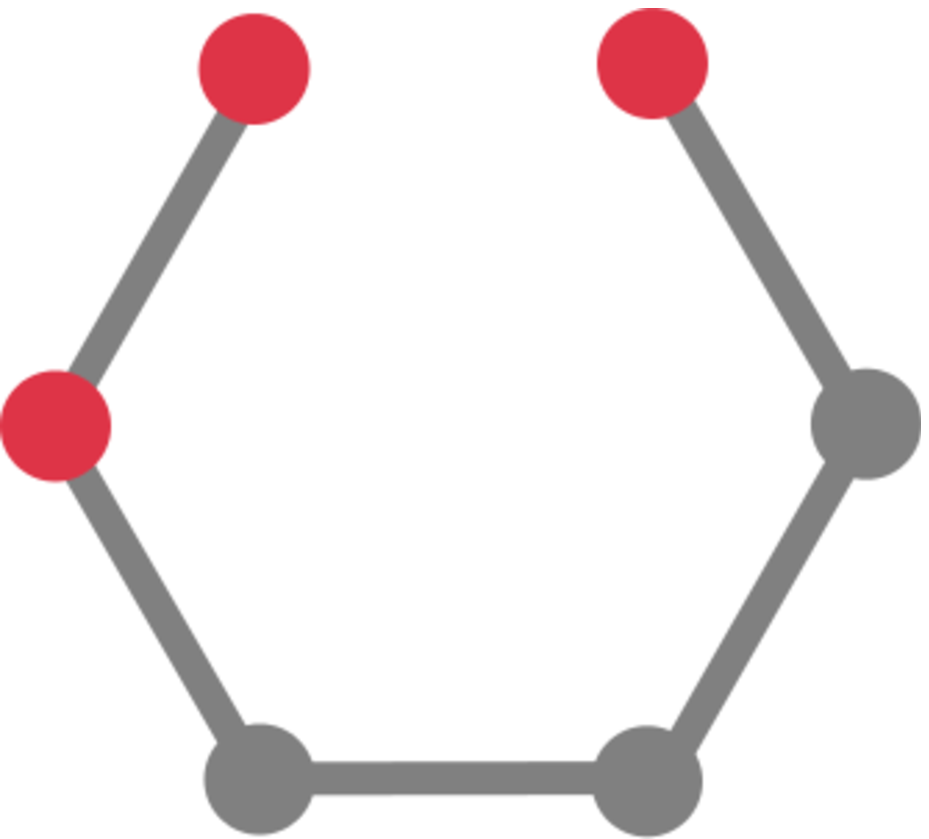}}\quad\quad
\subfigure[\label{Fig4_1}]{\includegraphics[width=0.07\textwidth]{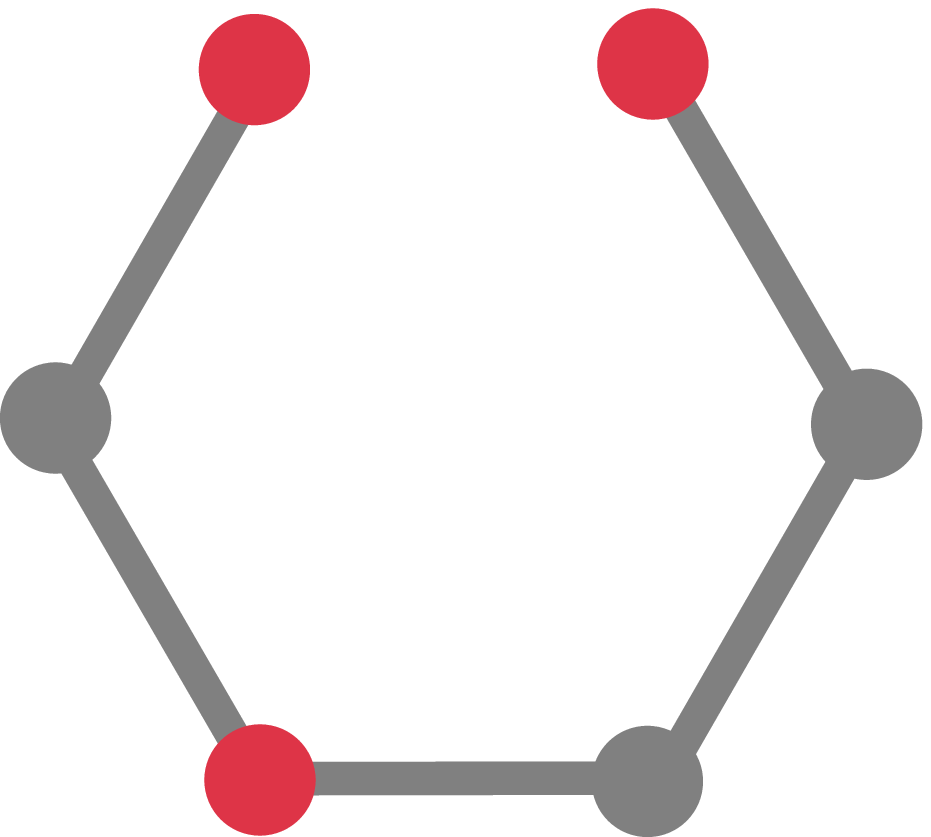}}\quad\quad
\subfigure[\label{Fig4_2}]{\includegraphics[width=0.07\textwidth]{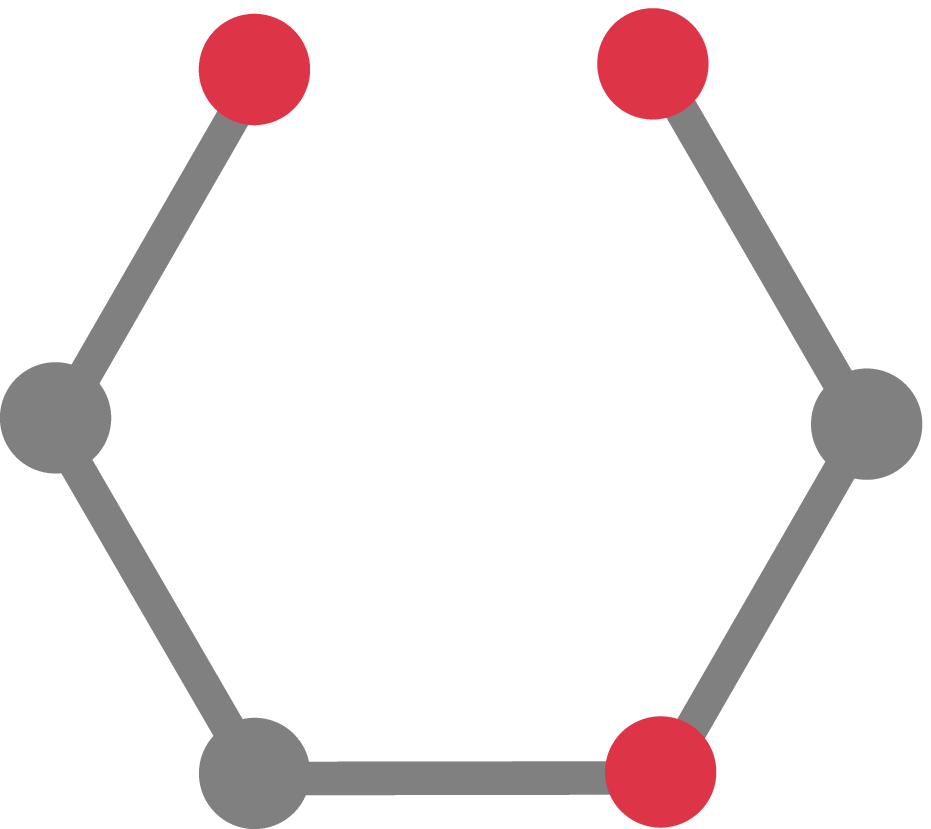}}\quad\quad
\subfigure[\label{Fig5}]{\includegraphics[width=0.07\textwidth]{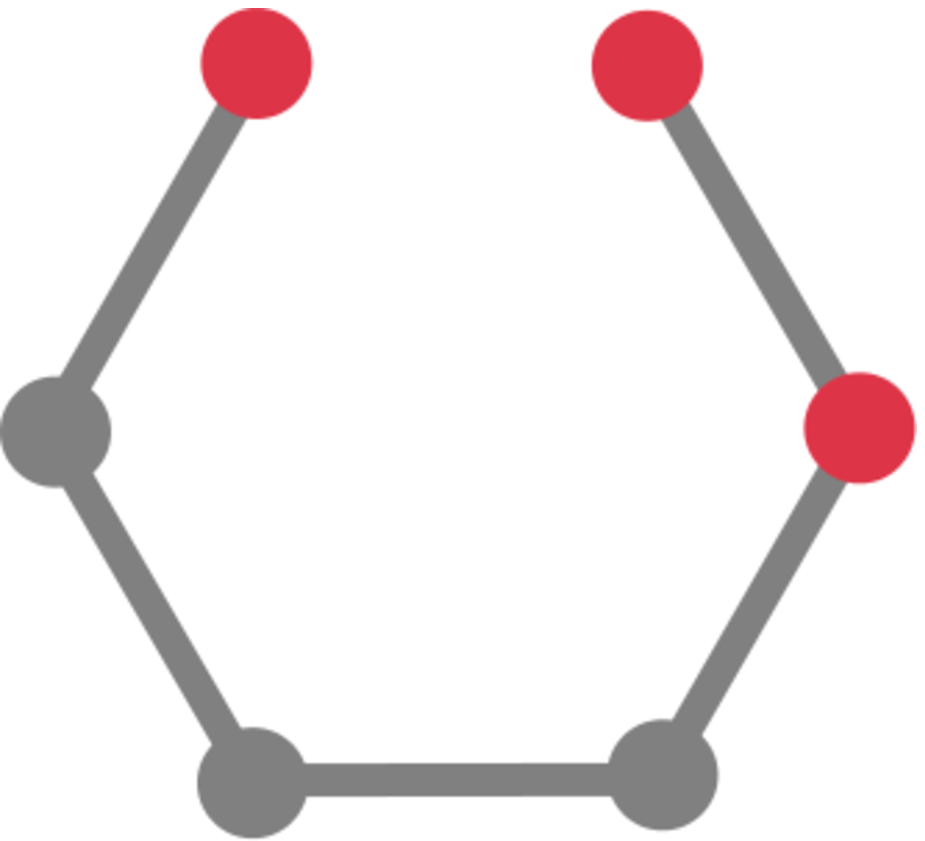}}\quad\quad
\subfigure[\label{Fig5_1}]{\includegraphics[width=0.07\textwidth]{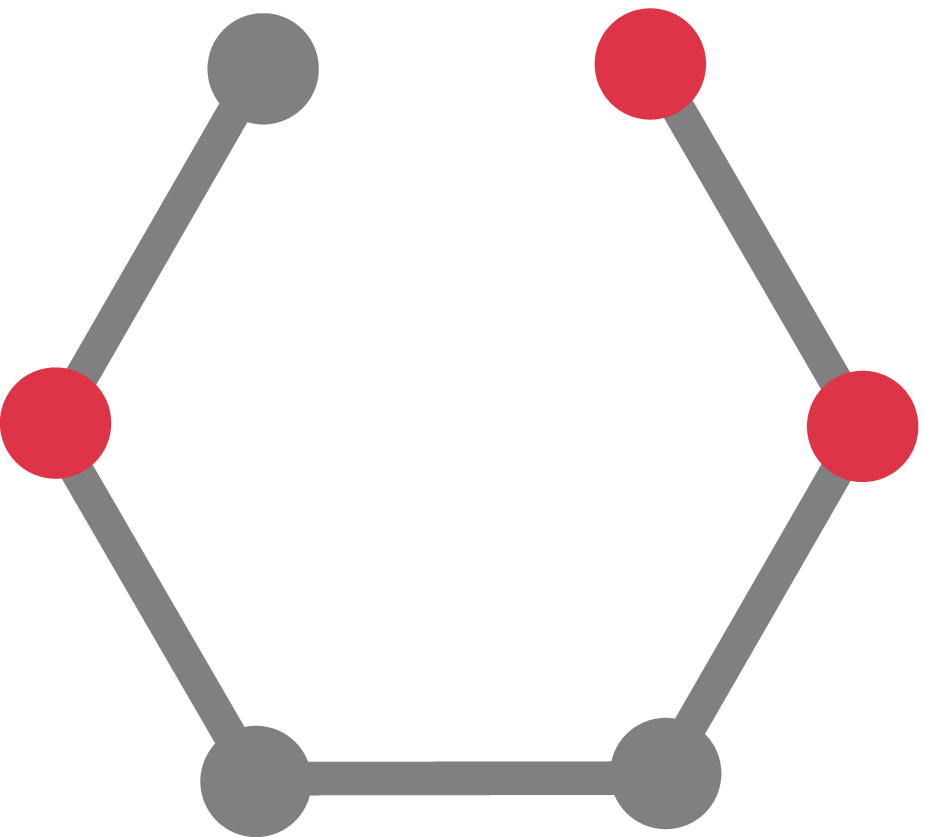}}\quad\quad
\subfigure[\label{Fig5_2}]{\includegraphics[width=0.07\textwidth]{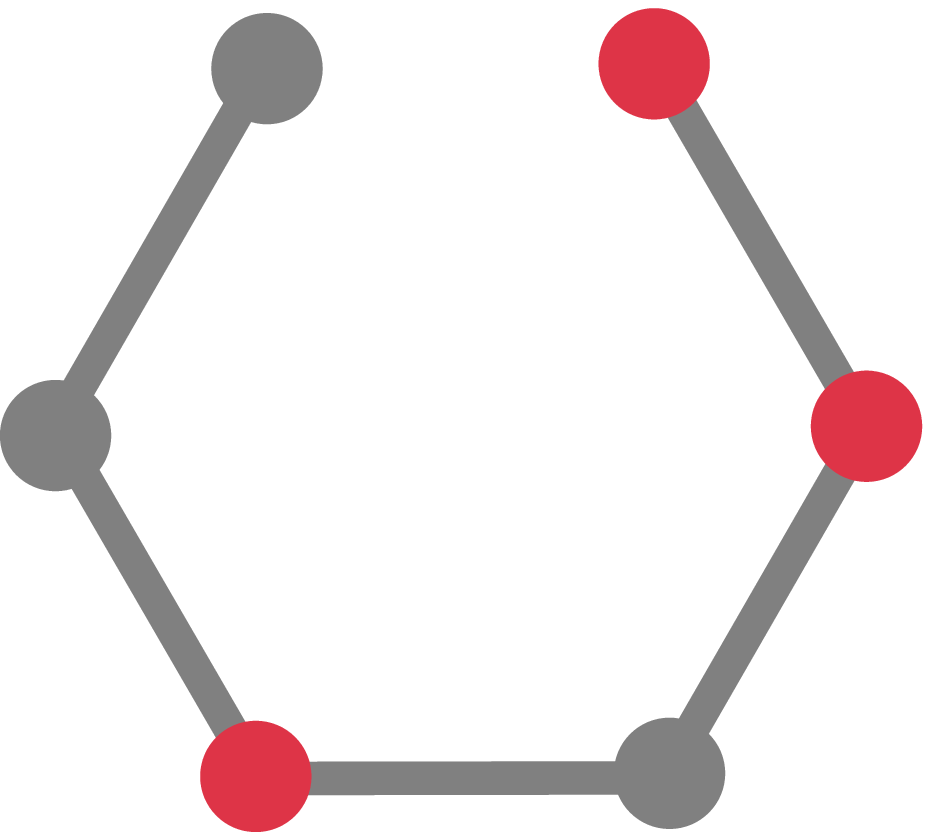}}\quad\quad
\subfigure[\label{Fig6}]{\includegraphics[width=0.08\textwidth]{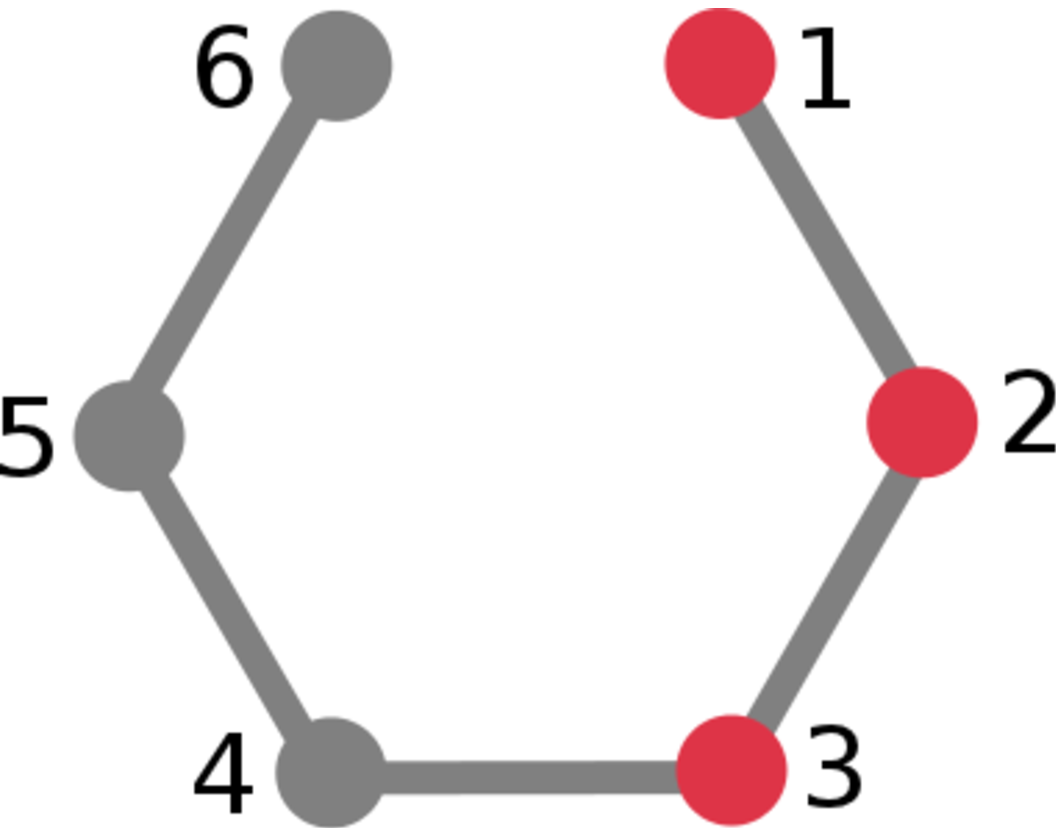}}
\caption{\ref{Fig3} through \ref{Fig6} represent a sequence of states in $O_g$ with $k_a=k_b=A=0$ and $B=3$ for the case of a connected one-dimensional lattice. The sequence connects states \ref{Fig3} and \ref{Fig6}.}
\end{figure}

In terms of the pictorial representation that we have described above, a sequence 
$\{\ket{\sigma,\lambda},\cdots,\ket{\chi_l,\theta_l},\cdots,\ket{\sigma',\lambda'}\}$ 
in $O_g$ connecting $\ket{\sigma,\lambda}$ to $\ket{\sigma',\lambda'}$ can be represented 
by a sequence of pictures~\footnote{Formally, it's a sequence of pairs of sets 
$\{(\sigma,\lambda),\cdots,(\chi_l,\theta_l),\cdots,(\sigma',\lambda')\}$.}. 
For example, Fig.~\ref{Fig3} through Fig.~\ref{Fig6} shows a sequence connecting states 
Fig.~\ref{Fig3} and Fig.~\ref{Fig6}, with matrix elements of $\mathbf{W_g}$ between 
any two adjacent pictures being nonzero.

Next step is the study of the degeneracy of $E_1$. Since Proposition~\ref{Prps1} and the PFT theorem also apply to $W_g$, it is sufficient to check for the existence 
of a sequence 
$\{\ket{\sigma,\lambda},\cdots,\ket{\chi_l,\theta_l},\cdots,\ket{\sigma',\lambda'}\}$ 
for any arbitrary $\ket{\sigma,\lambda}$ and $\ket{\sigma',\lambda'}$. In the following, 
we will study under which conditions arbitrary $\ket{\sigma,\lambda}$ and 
$\ket{\sigma',\lambda'}$ are connected. These conditions will differ
depending on the values of $A$ and $B$.

\subsection{One of the species has commensurate filling factor, i. e., $A=0$ or $B=0$}

It is obvious that when both species have commensurate filling factor, 
i.e. $A=0$ and $B=0$, $E_0$ is nondegenerate. In the strongly interacting regime one 
can apply the non-degenerate perturbation method. 
Let us therefore consider the case when only one of the species is at commensurate 
filling, e. g. $A= 0,B \ne 0$. In this case, besides grey sites, there are B red sites. 
The only assumption we are making on the lattice $\mathbf{G}$ is that it is connected 
(periodic boundary conditions do not necessarily need to be satisfied). 
Consider arbitrary $\ket{\sigma,\lambda}$ and $\ket{\sigma',\lambda'}$, e. g. states 
Fig.~\ref{Fig3} and Fig.~\ref{Fig6}. By using induction, one can show:
\medskip

\noindent 
\begin{proposition}
\label{Prps3} When $A=0,B\ne0$ (or $A\ne0,B=0$), any $\ket{\sigma,\lambda}$ and 
$\ket{\sigma',\lambda'}$ are connected if and only if $\mathbf{G}$ is connected.
\end{proposition}
\medskip

\noindent
Let us prove the sufficient condition by induction. 
Let us assume that $\mathbf{G}$ is connected. Then, it is always possible to 
label sites by $(i_1,i_2,\cdots,i_M)$ such that  if we remove the first $m$ sites 
in this sequence, the remaining $(i_{m+1},\cdots,i_M)$ sites still form a connected lattice 
for all $1 \le{m} <M$ \cite{Graph_Theory}. Fig.~\ref{Fig6} shows an example of labeling 
which satisfies this property. 
Constructing a state such that the color of $i_1$ (site $1$ in our example) is the same as in $\ket{\sigma',\lambda'}$ 
can be done by first locating a site $i_k$ (site $3$ in our example) in $\ket{\sigma,\lambda}$ 
with the same color as $i_1$ in $\ket{\sigma',\lambda'}$. Next, we successively exchange the 
colors along a path linking $i_1$ to $i_k$ (see e.g. Fig.~\ref{Fig3}-\ref{Fig4}). 
Let us assume that this can be done for an arbitrary $i_m$, with $1\le{m}<M$, and 
let us denote the constructed state by $\ket{\chi_{n},\theta_{n}}$. Then, since 
$(i_{m+1},\cdots,i_M)$ forms a connected lattice, by applying the same procedure 
as for $i_1$ we can fix the color on $i_{m+1}$ (see e.g. Fig.~\ref{Fig4}-\ref{Fig5}). 
Therefore, by induction, we have shown that $\ket{\sigma,\lambda}$ 
and $\ket{\sigma',\lambda'}$ are connected.

The necessary condition is proved by a similar argument as in Proposition~\ref{Prps2}, 
which implies that, if $\mathbf{G}$ is not connected, then there exist 
$\ket{\sigma,\lambda}$ and $\ket{\sigma',\lambda'}$ which are not connected. 

A direct consequence of Proposition~\ref{Prps1} and Proposition~\ref{Prps3} is the following:
\begin{theorem}
When $A=0,B\ne0$ or $A\ne0,B=0$, $E_1$ is nondegenerate if $\mathbf{G}$ is connected.
\end{theorem}

\subsection{\label{2-connected}Useful properties of a 2-connected lattice}

Let us first introduce the notion of 2-connectivity. A lattice $\mathbf{G}$ is said to 
be 2-connected if the removal of any site leaves the remaining sites connected. In the 
one-dimensional example of Fig.~\ref{Fig7}, this is equivalent to introducing periodic 
boundary conditions to get Fig.~\ref{Fig9}. In higher dimensions, the 2-connectivity conditions is satisfied by 
square, triangular, honeycomb, cubic, fcc lattices, etc. with any boundary conditions. 
Some useful properties of 2-connectivity are as follows: 
\begin{itemize}
\item[(a)] if $\mathbf{G}$ is 2-connected then it is also connected;
\item[(b)] $\mathbf{G}$ is 2-connected if and only if, for any two 
distinct sites, there exist two disjoint paths linking them
(two paths are disjoint if they only share the two ends).
This is the global version of Menger's theorem~\cite{Graph_Theory}; 
\item[(c)] for any distinct sites $i_1$, $i_2$ and $i_3$, 
there exists a path linking $i_1$ and $i_2$ which avoids $i_3$ (this is a direct consequence 
of property (b));
\item[(d)] in any state with at least one grey site, one can always move the blue 
or red color from an arbitrary site to another arbitrary site according to the rules given 
in Subsection~\ref{subsection4_1} (we prove this property in Appendix~\ref{App3}).
\end{itemize}

\subsection{$A=1$, $0<B<M-1$ and $A=M-1$, $1<B\le{M-1}$}
\label{sub4C}

\begin{figure}
\centering
\subfigure[\label{Fig7}]{\includegraphics[width=0.10\textwidth]{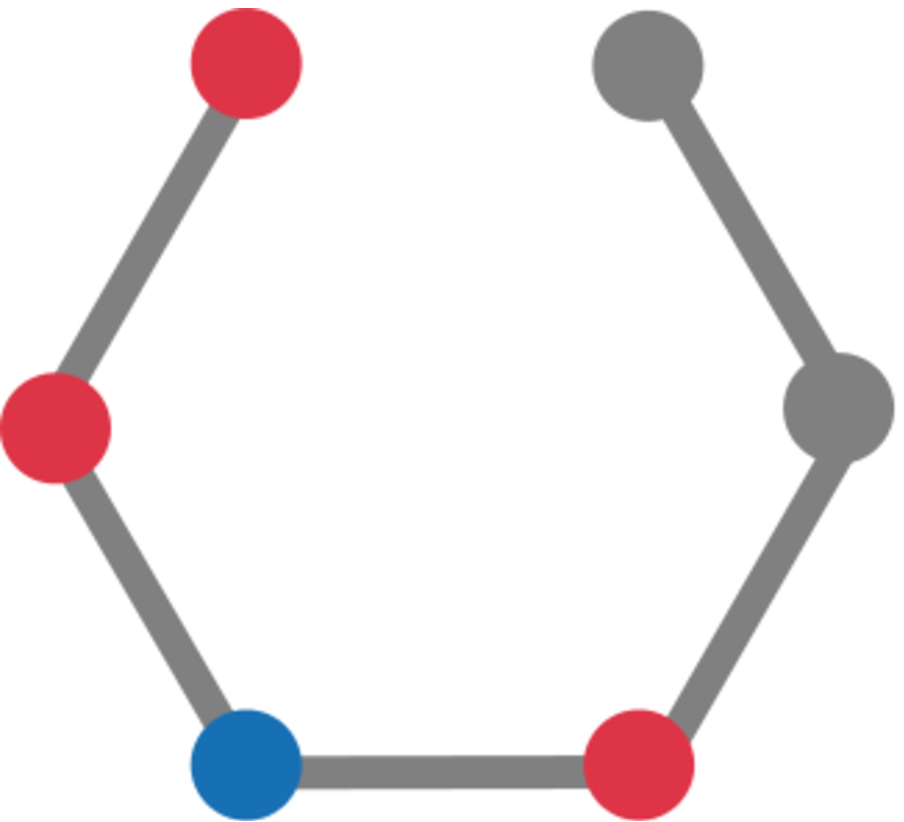}}\quad\quad
\subfigure[\label{Fig8}]{\includegraphics[width=0.10\textwidth]{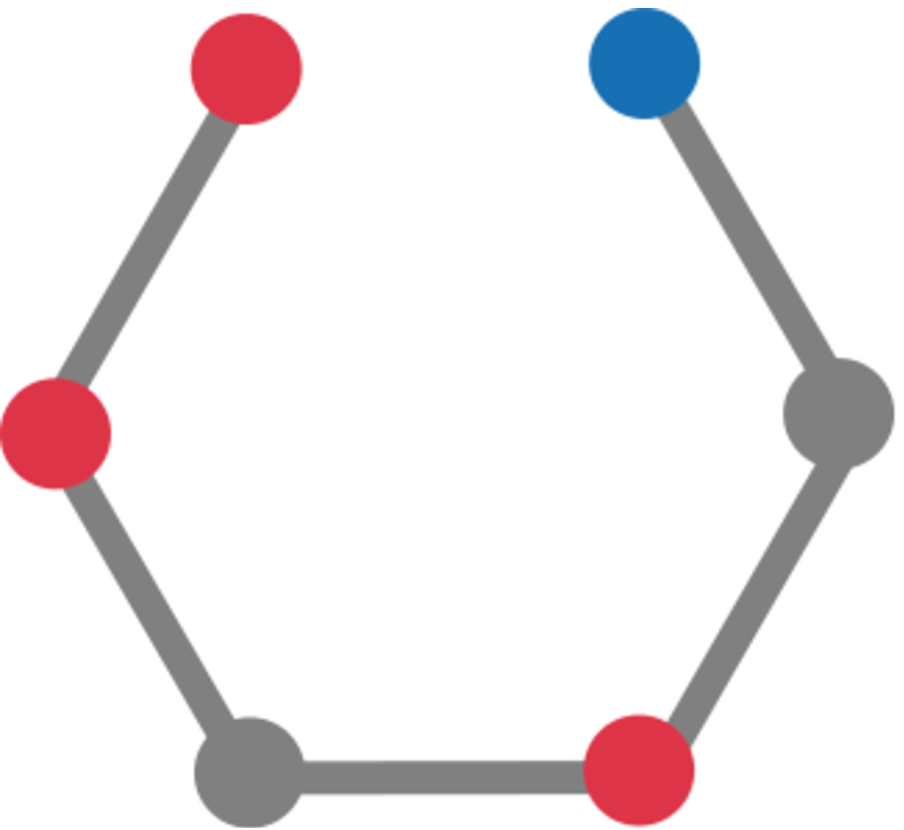}}\quad\quad
\subfigure[\label{Fig9}]{\includegraphics[width=0.10\textwidth]{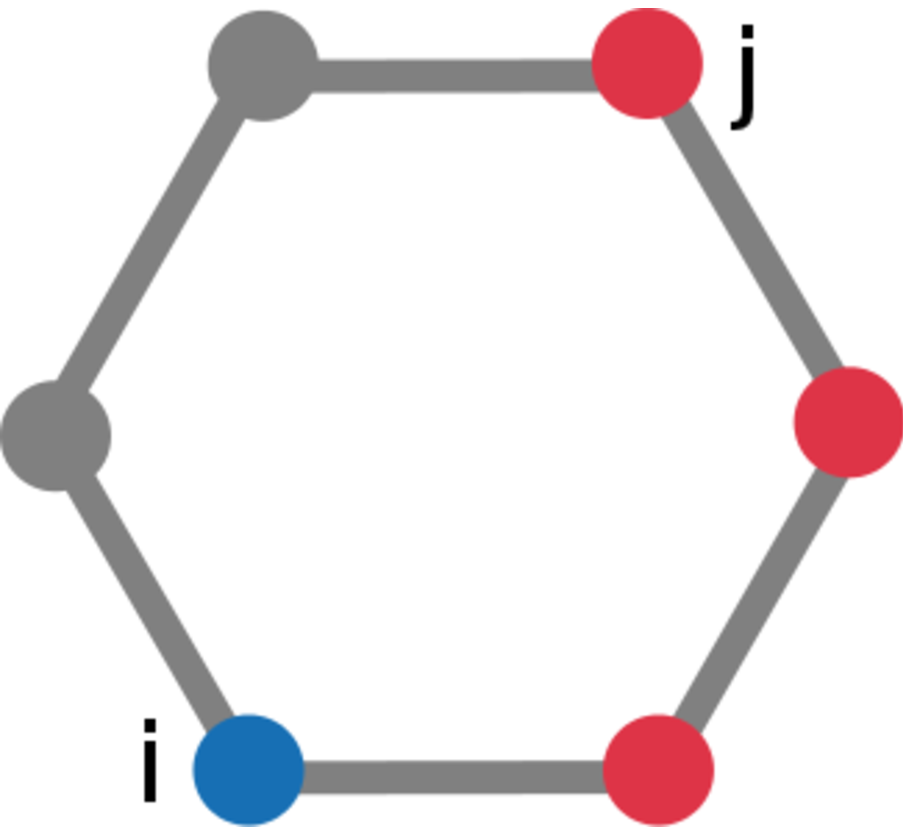}}\quad\quad
\subfigure[\label{Fig12}]{\includegraphics[width=0.10\textwidth]{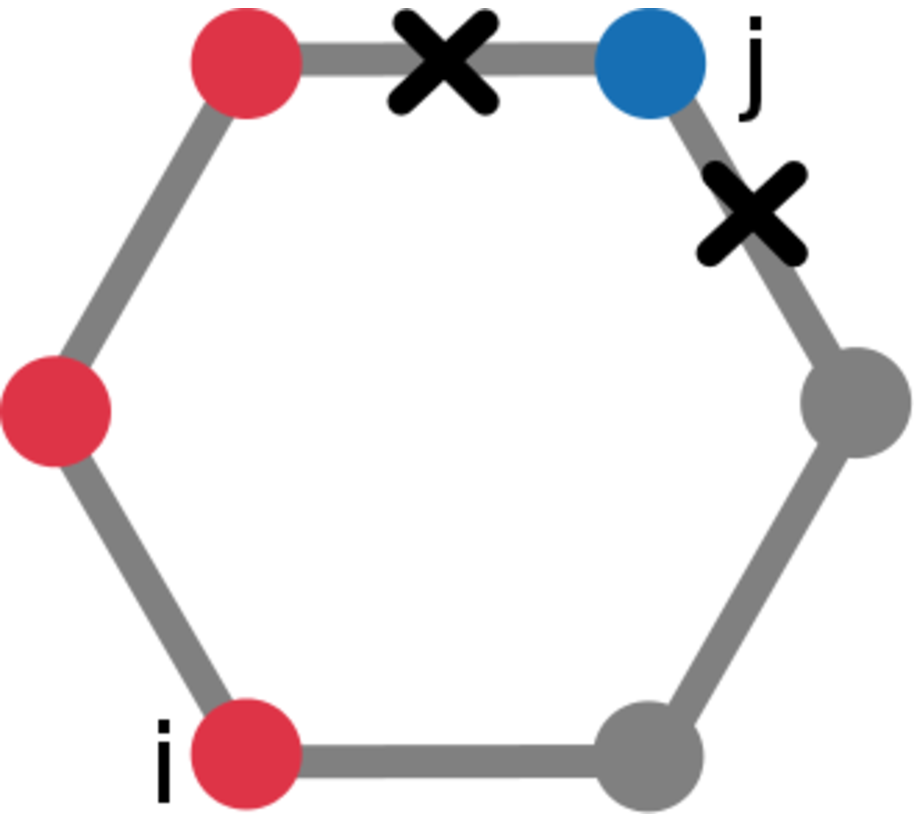}}\quad\quad
\subfigure[\label{Fig13}]{\includegraphics[width=0.10\textwidth]{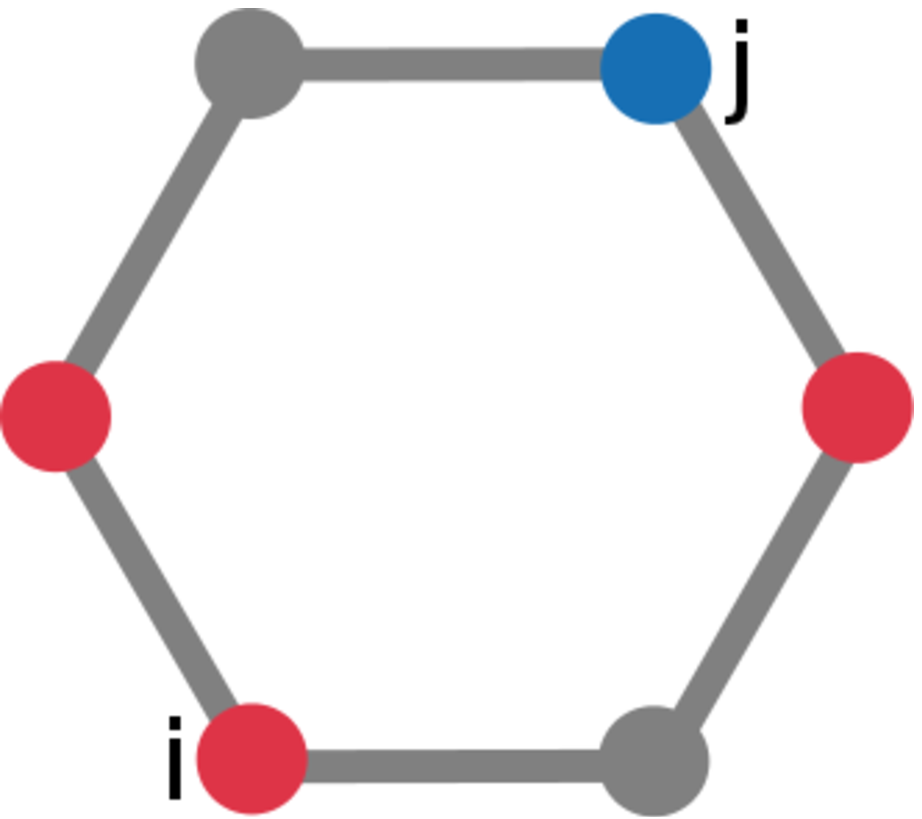}}
\caption{\label{FigTwo}\ref{Fig7} and \ref{Fig8} are two unconnected states in $O_g$ for the case of a connected one-dimensional lattice with $A=1,B=3$.  \ref{Fig9} and \ref{Fig13} are two connected states in $O_g$ for the case of a $2$-connected lattice with $A=1,B=3$. \ref{Fig12} is an intermediate state in the sequence connecting \ref{Fig9} to \ref{Fig13}. The two crosses in Fig.~\ref{Fig12} indicate that the removal of site $j$ leaves the remaining lattice still connected.}
\end{figure}

Although we will discuss our results explicitly for $A=1$, $0<B<M-1$, the case of 
$A=M-1$, $1<B\le{M-1}$ can be mapped onto $A=1,0<B<M-1$ by replacing blue with red and 
vice versa, and replacing purple with grey. Since purple sites can be moved in the same 
manner as grey sites the two cases are completely equivalent.

In the case $A=1$, $0<B<M-1$, i.e. one blue and B red sites are present, the requirement 
that $\mathbf{G}$ is connected is not sufficient for any two states to be connected. 
This is shown with an example in Fig.~\ref{Fig7} and \ref{Fig8}, where the two states 
represented are not connected because it is not possible to move the blue color in 
Fig.~\ref{Fig7} to its position in Fig.~\ref{Fig8} according to the rules given in 
Subsection~\ref{subsection4_1}. We need to impose that the lattice 
$\mathbf{G}$ is 2-connected. If this is the case, the following property holds:
\medskip

\noindent
\begin{proposition}
\label{Prps4}
Any two states $\ket{\sigma,\lambda}$, $\ket{\sigma',\lambda'}$ with $A=1$ and arbitrary $0<B<M-1$ (or $A=M-1$ and arbitrary $1<B\le{M-1}$) are connected if and only if $\mathbf{G}$ is 2-connected.
\end{proposition}

\begin{figure}
\centering
\subfigure[\label{Fig16}]{\includegraphics[width=0.11\textwidth]{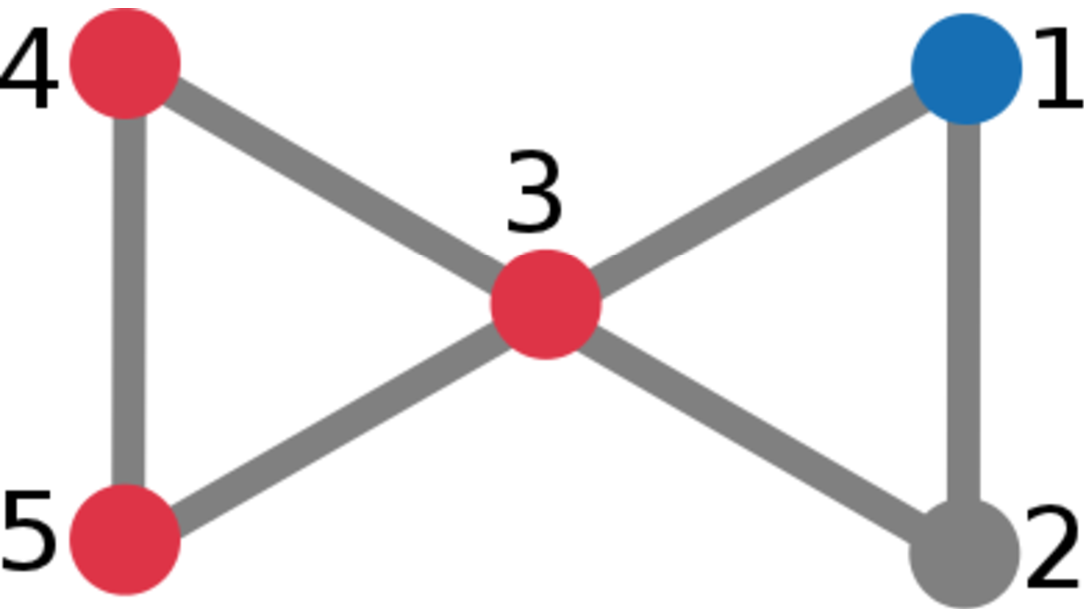}}\quad\quad
\subfigure[\label{Fig17}]{\includegraphics[width=0.11\textwidth]{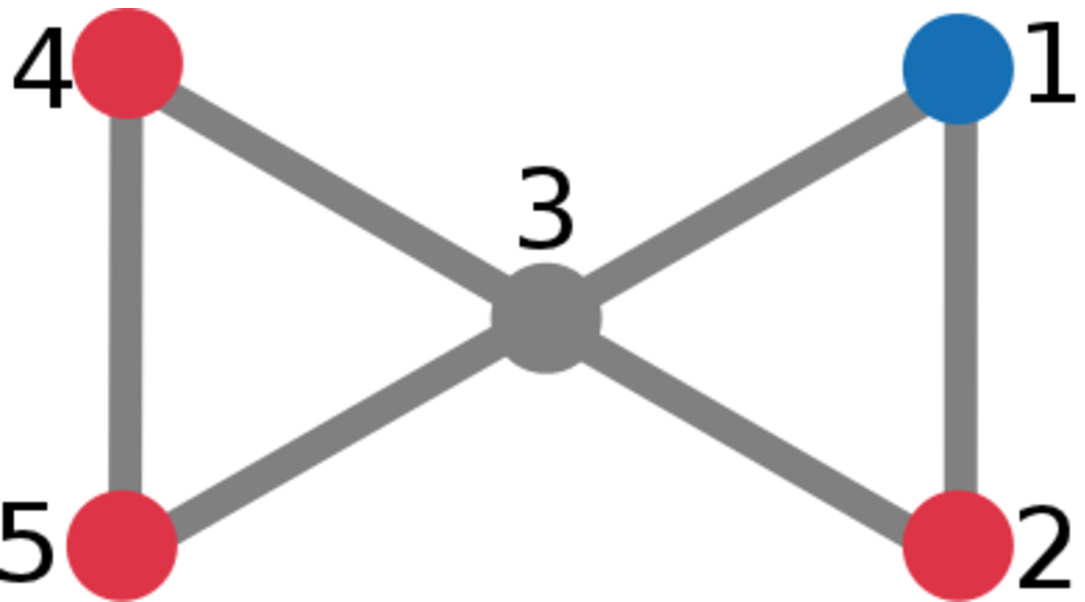}}\quad\quad
\subfigure[\label{Fig18}]{\includegraphics[width=0.11\textwidth]{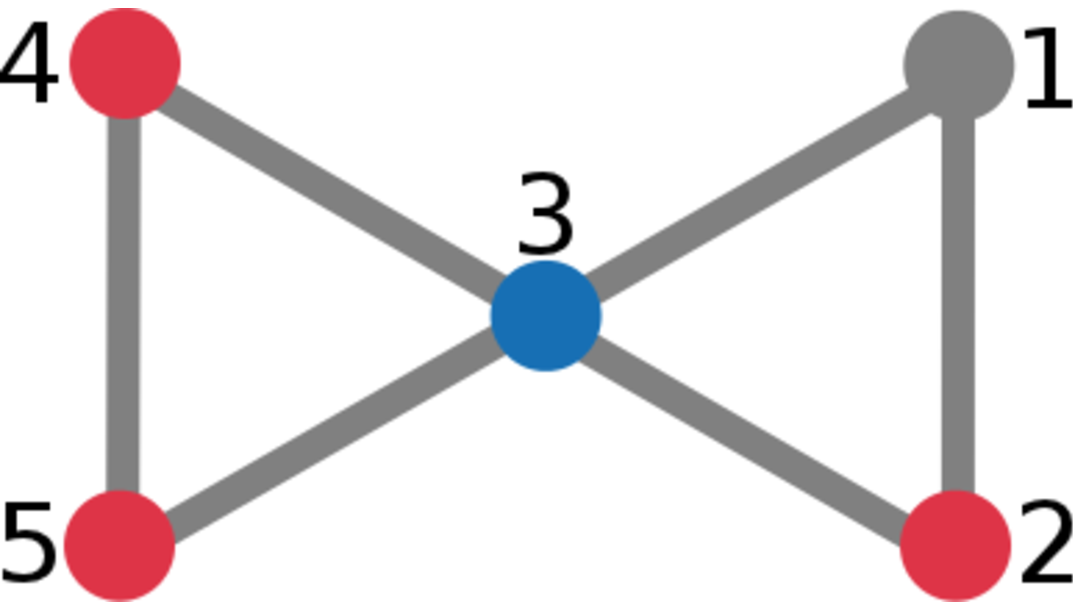}}\quad\quad
\subfigure[\label{Fig19}]{\includegraphics[width=0.11\textwidth]{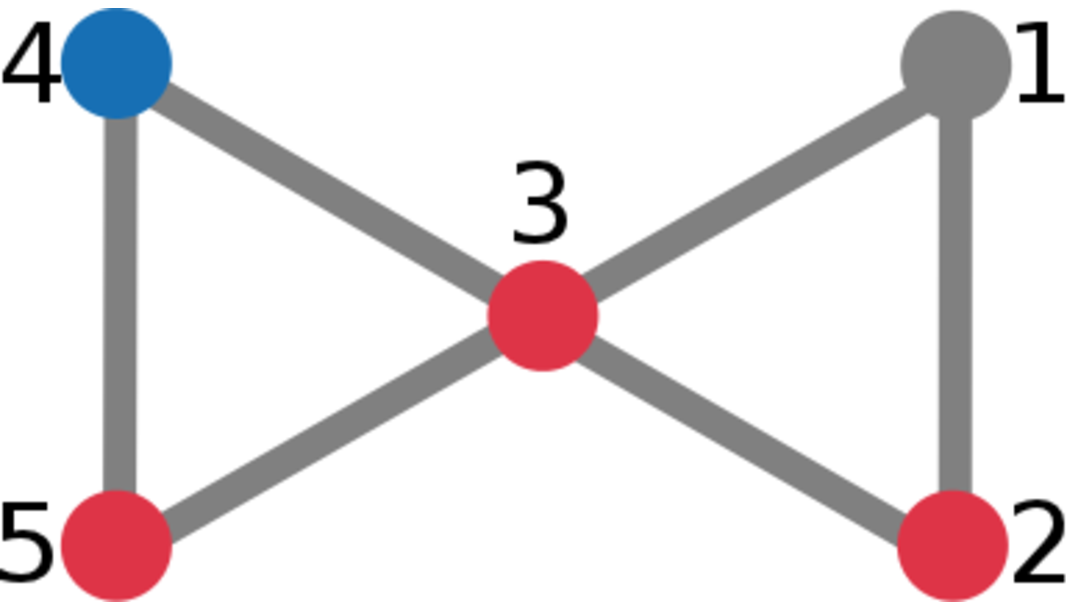}}
\caption{\ref{Fig16} through \ref{Fig19} represent states in $O_g$ with $A=1,B=3$ for the case of a lattice which is {\em{not}} 2-connected. The absence of 2-connectivity implies that the removal of the site $3$ leaves the remaining lattice unconnected. \ref{Fig16}, \ref{Fig17} and \ref{Fig18} are connected with each other, but they are not connected with \ref{Fig19}.}
\end{figure}
Let us start by proving the sufficient condition. 
Consider arbitrary $\ket{\sigma,\lambda}$ and $\ket{\sigma',\lambda'}$. Without loss 
of generality, assume that the blue color is on site $i$ in $\ket{\sigma,\lambda}$ and 
on site $j$ in $\ket{\sigma',\lambda'}$. Due to property (d) of 2-connectivity, the 
blue color can be moved from $i$ to $j$. In other words, we can construct a state 
$\ket{\chi_{n},\theta_{n}}$ connected with $\ket{\sigma,\lambda}$ in which $j$ is blue. 
An example of states $\ket{\sigma,\lambda}$, $\ket{\chi_{n},\theta_{n}}$ and 
$\ket{\sigma',\lambda'}$ is displayed in Fig.~\ref{Fig9}, \ref{Fig12} and \ref{Fig13} 
respectively. Moreover, the 2-connectivity of $\mathbf{G}$ implies that removal of site $j$ 
leaves the rest of the lattice still connected (see Fig.~\ref{Fig12}). 
Removing site $j$ leaves state $\ket{\sigma,\lambda}$ with red and grey sites only. 
Thus, following a similar argument as for the case $A=0,B\ne0$ in Proposition~\ref{Prps3}, 
we can show that $\ket{\chi_{n},\theta_{n}}$ is connected to $\ket{\sigma',\lambda'}$. 
Therefore, by transitivity of connectedness, $\ket{\sigma,\lambda}$ is connected to 
$\ket{\sigma',\lambda'}$.

The necessary condition is proved by contradiction. For simplicity but without loss of 
generality we choose a counter example with a single grey site. If $\mathbf{G}$ is not 
2-connected (e.g. the lattice shown in Fig.~\ref{Fig16}), then, there exists at least one site $i$ 
(site $3$ in our example) whose removal leaves the remaining sites partially 
unconnected. Let $J$ and $K$ denote the two unlinked sets (in our example $(1,2)$ and $(4,5)$ in 
Fig.~\ref{Fig16}). Let us consider two states, e. g. Fig.~\ref{Fig16} and Fig.~\ref{Fig19},
with the blue color on $J$ and $K$ respectively. In the attempt of transferring the blue 
color from $J$ to $K$ one can move the grey color as shown in Fig.~\ref{Fig17} and 
Fig.~\ref{Fig18}. At this point, though, the blue color cannot be moved to any site 
in $K$, because all sites in $K$ are red. So it is not possible to connect state 
Fig.~\ref{Fig18} to state Fig.~\ref{Fig19}). 

Finally we can conclude the following:
\begin{theorem}
In the case $A=1$, $0<B<M-1$ (or $A=M-1$, $1<B\le{M-1}$), $E_1$ is nondegenerate if 
$\mathbf{G}$ is 2-connected.
\end{theorem}


\section{\label{Sec5}Nondegeneracy of $E_1$ in the general cases $A+B<M-1$ and $A+B>M+1$.}

\begin{figure}
\centering
\subfigure[\label{Fig14}]{\includegraphics[width=0.10\textwidth]{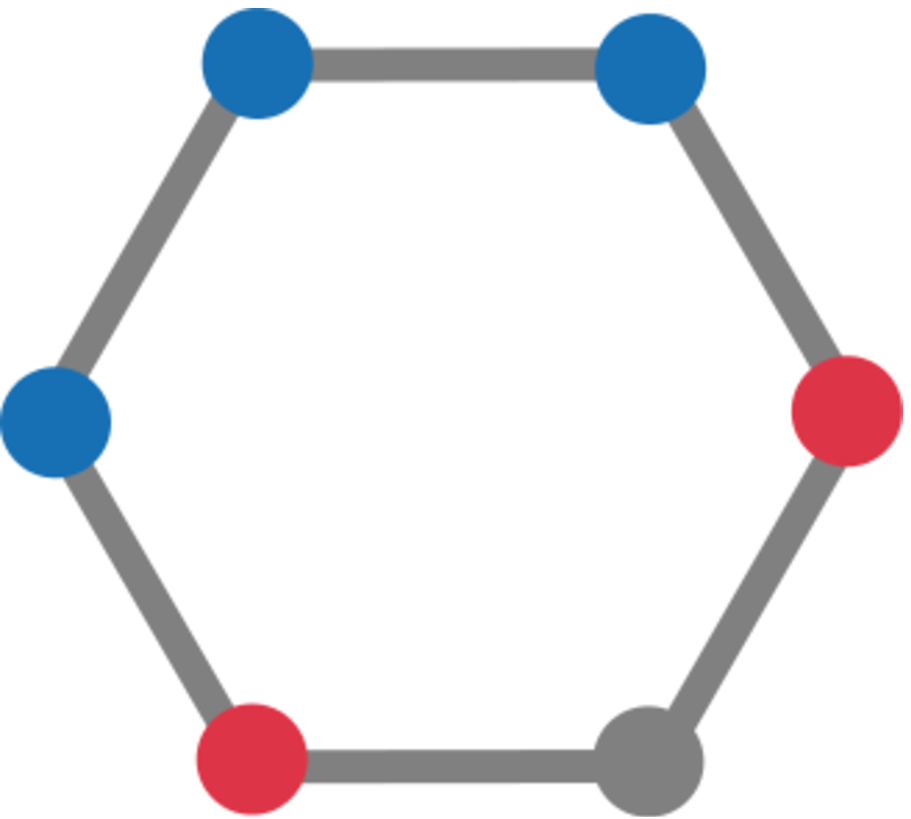}}\quad\quad
\subfigure[\label{Fig15}]{\includegraphics[width=0.10\textwidth]{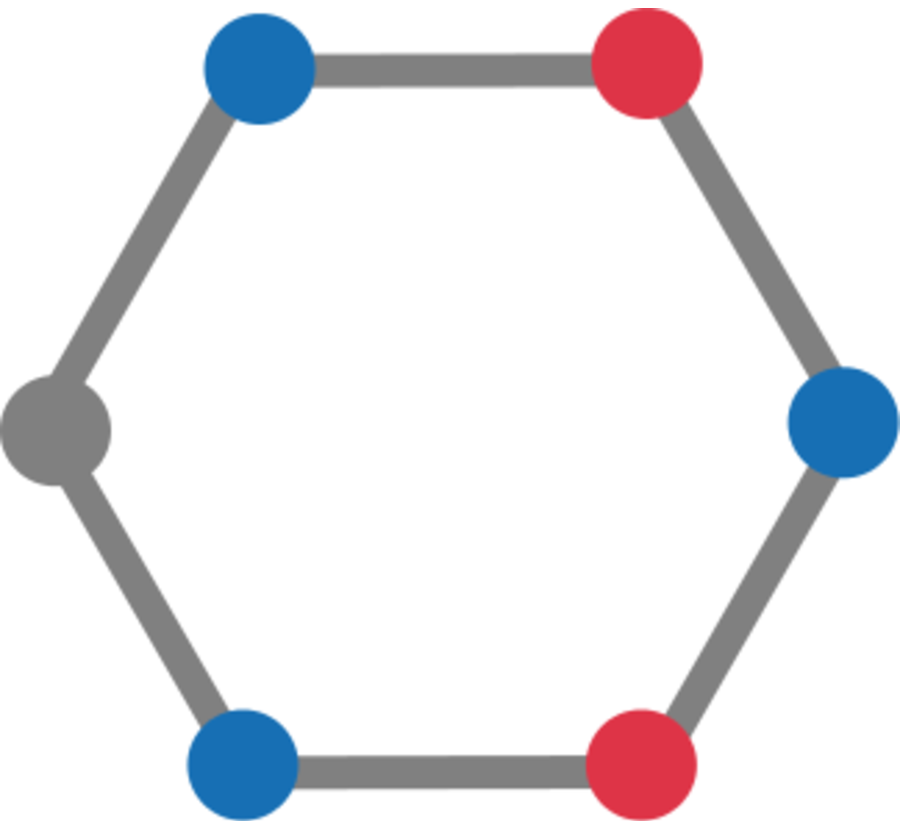}}
\caption{\label{FigThree}\ref{Fig14} and \ref{Fig15} are two unconnected states in $O_g$ for the case of a one-dimensional lattice of  $M=6$ sites with periodic boundary condition and $A=3,B=2$.}
\end{figure}

We extend the results of Subsection \ref{sub4C} to the general case $A+B<M-1$.
Replacing grey with purple, the case $A+B>M+1$ 
can be mapped onto $A+B<M-1$.

In general, for $A+B<M-1$, 2-connectivity is not a sufficient condition for any 
two states to be connected. This is shown by an example in Fig.~\ref{Fig14} and 
\ref{Fig15} for a one-dimensional system with periodic boundary conditions. For 
later convenience we refer to this type of lattice as 
circle~\footnote{A circle is a path where the two ends are the same.}. 
When there are at least two blue and two red sites, the order of the color on the 
circle becomes important. Note that the order of color only includes red and blue, 
since grey sites can be freely moved as explained previously. It is obvious that 
one cannot change the order of the color in Fig.~\ref{Fig14} to construct 
Fig.~\ref{Fig15}. However, it is easy to check that two states on a circle are 
connected if they have the same order of color. The sequence connecting them can 
be constructed by successively moving the grey color on the circle. Using this 
property one can show that:

\begin{proposition}
\label{Prps5} For $A+B<M-1$ (or $A+B>M+1$), if $\mathbf{G}$ is a circle with 
one added path linking two unbonded sites on the circle, then any $\ket{\sigma,\lambda}$, 
$\ket{\sigma',\lambda'}$ are connected.
\end{proposition}
 
\begin{figure}
\centering
\subfigure[\label{Fig300}]{\includegraphics[width=0.15\textwidth]{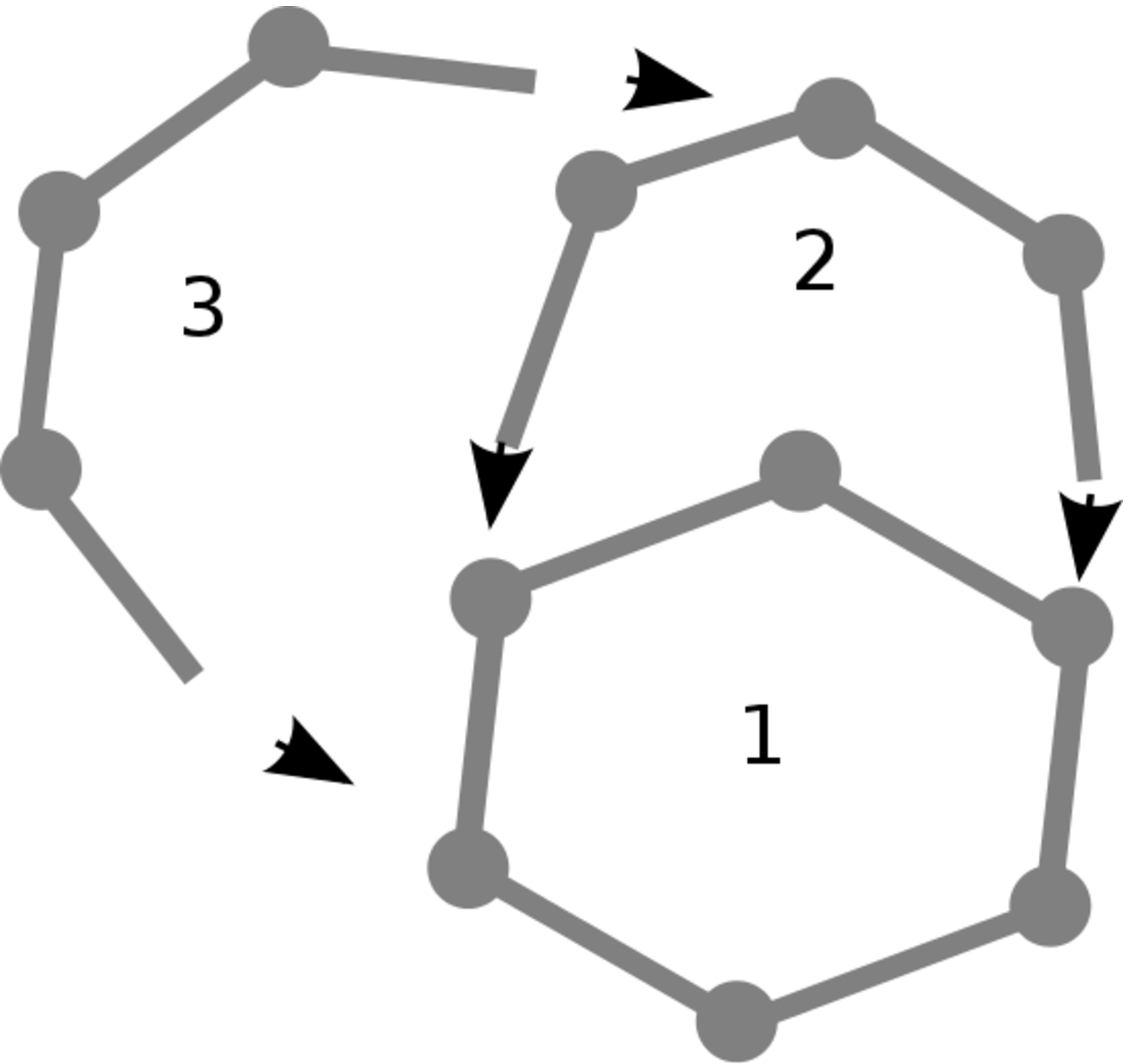}}\quad\quad
\subfigure[\label{Fig30}]{\includegraphics[width=0.12\textwidth]{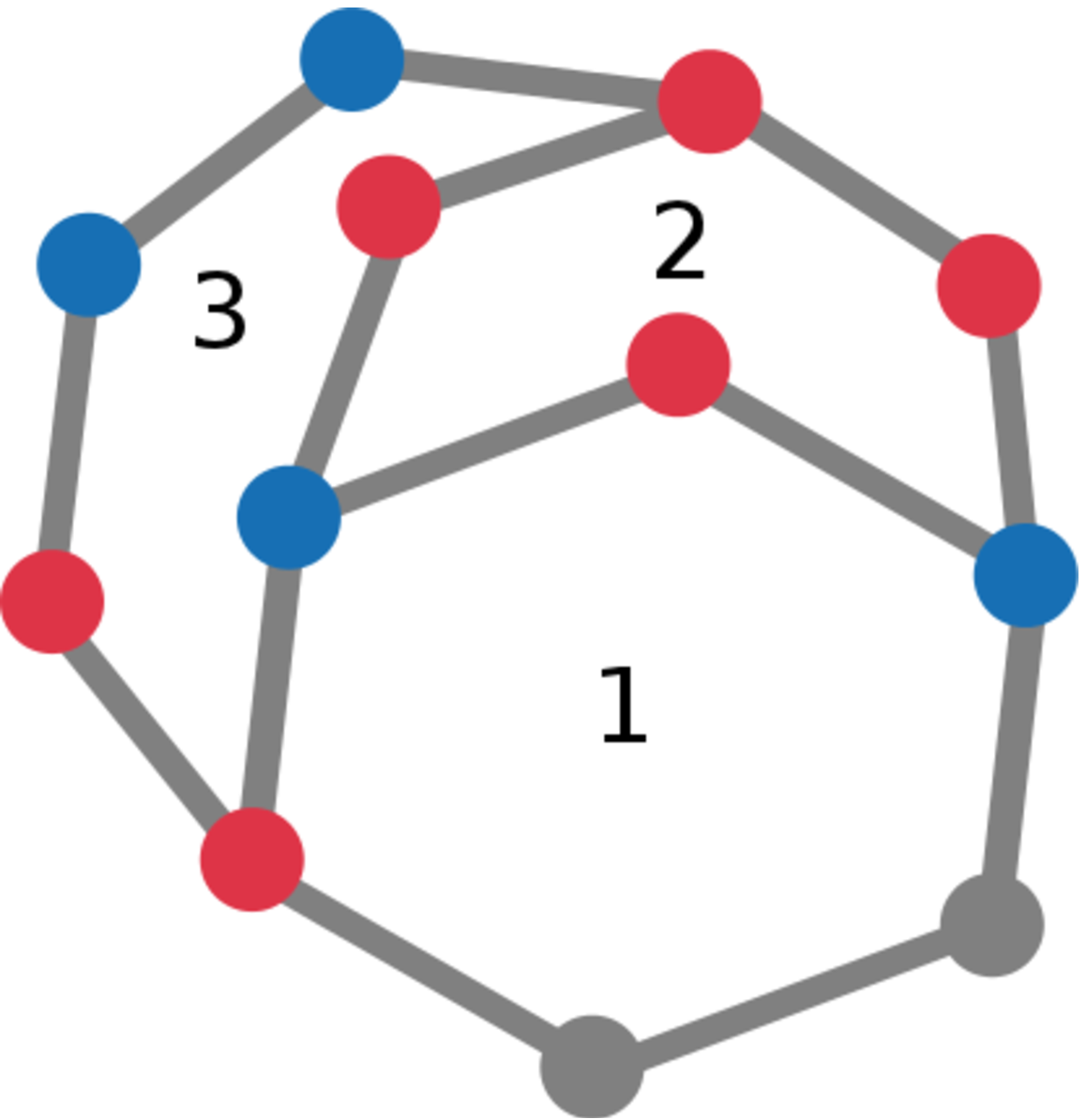}}\quad\quad
\subfigure[\label{Fig31}]{\includegraphics[width=0.12\textwidth]{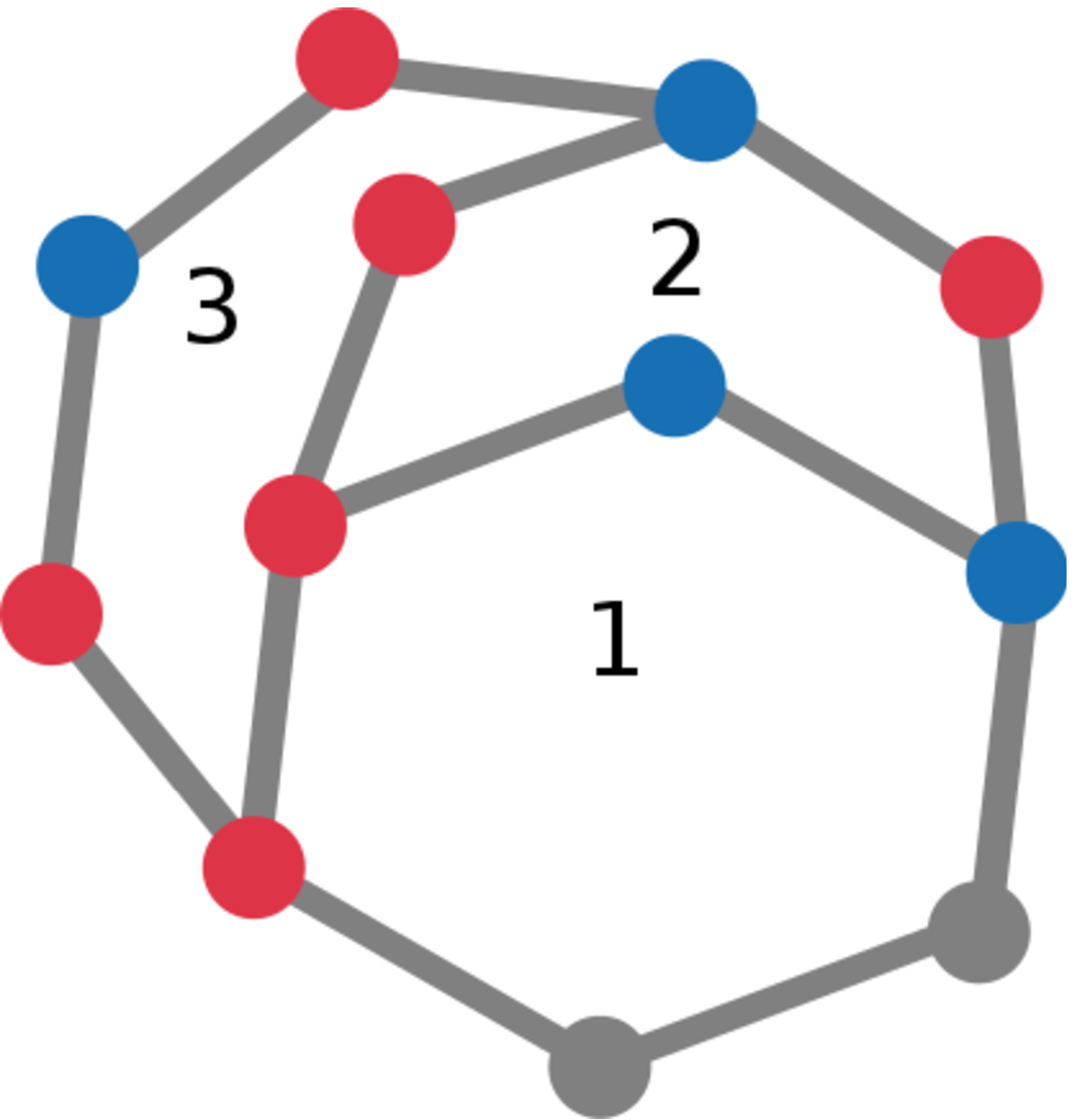}}\\
\subfigure[\label{Fig32}]{\includegraphics[width=0.12\textwidth]{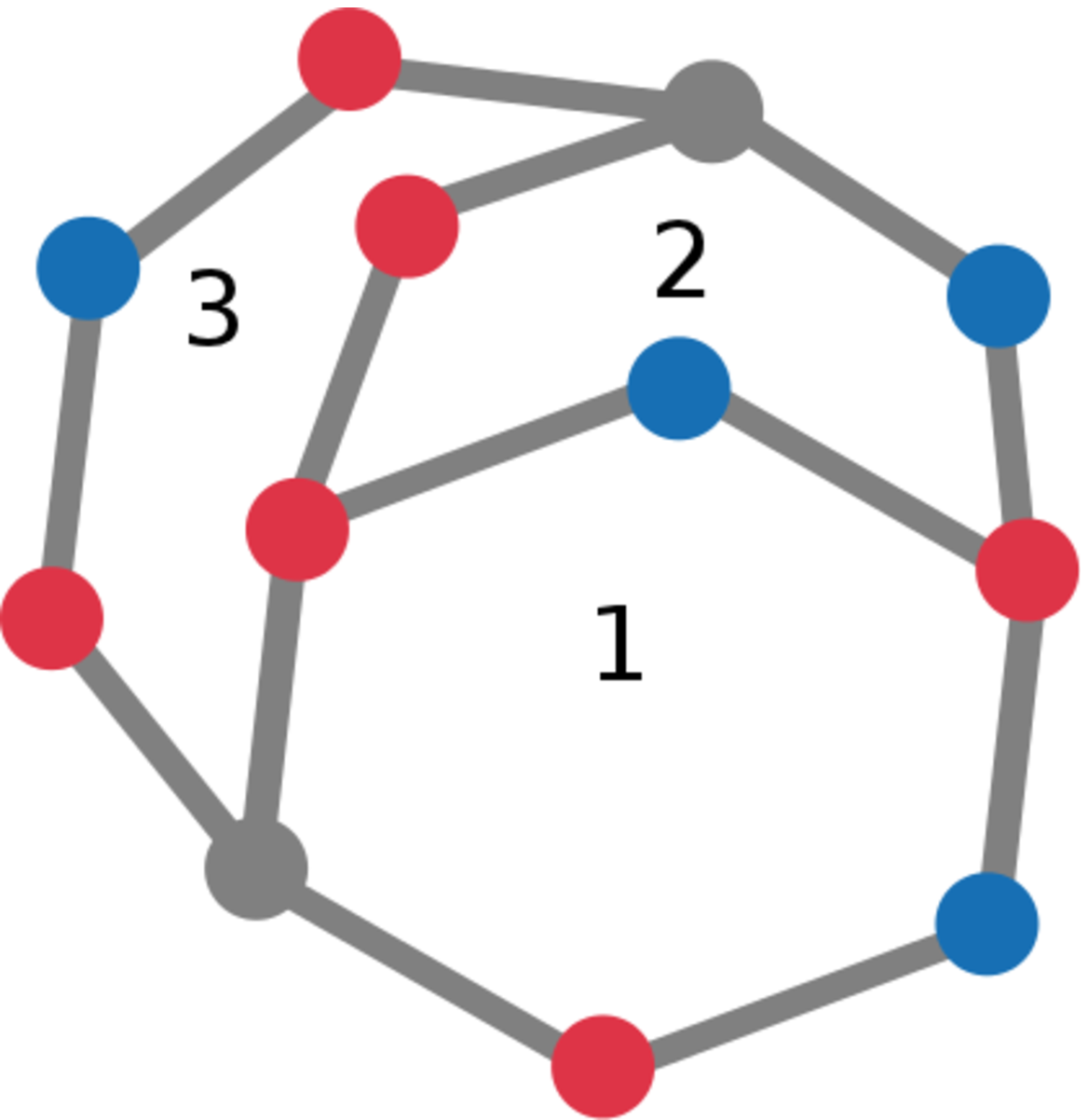}}\quad\quad
\subfigure[\label{Fig33_1}]{\includegraphics[width=0.12\textwidth]{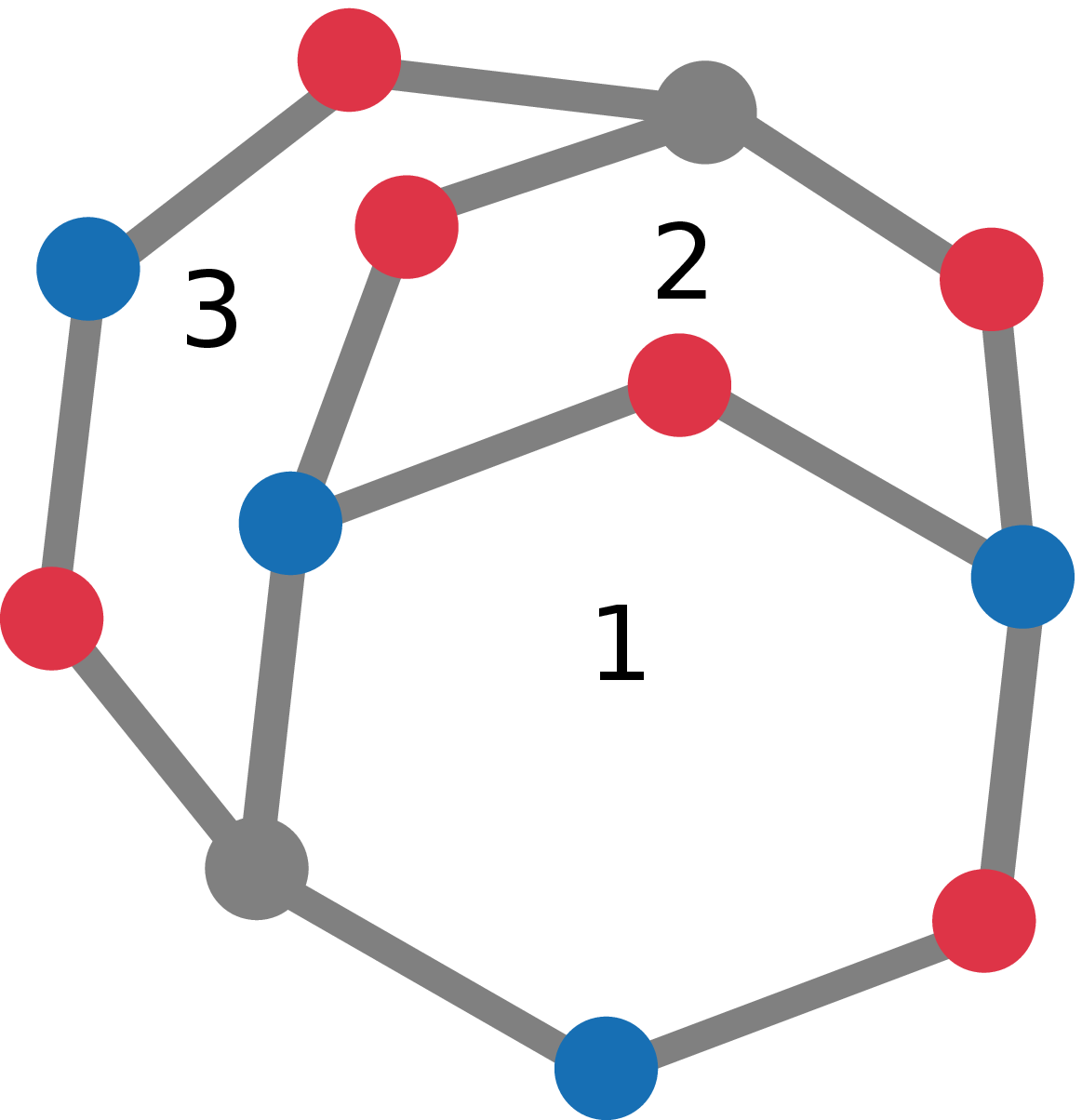}}\quad\quad
\subfigure[\label{Fig34}]{\includegraphics[width=0.12\textwidth]{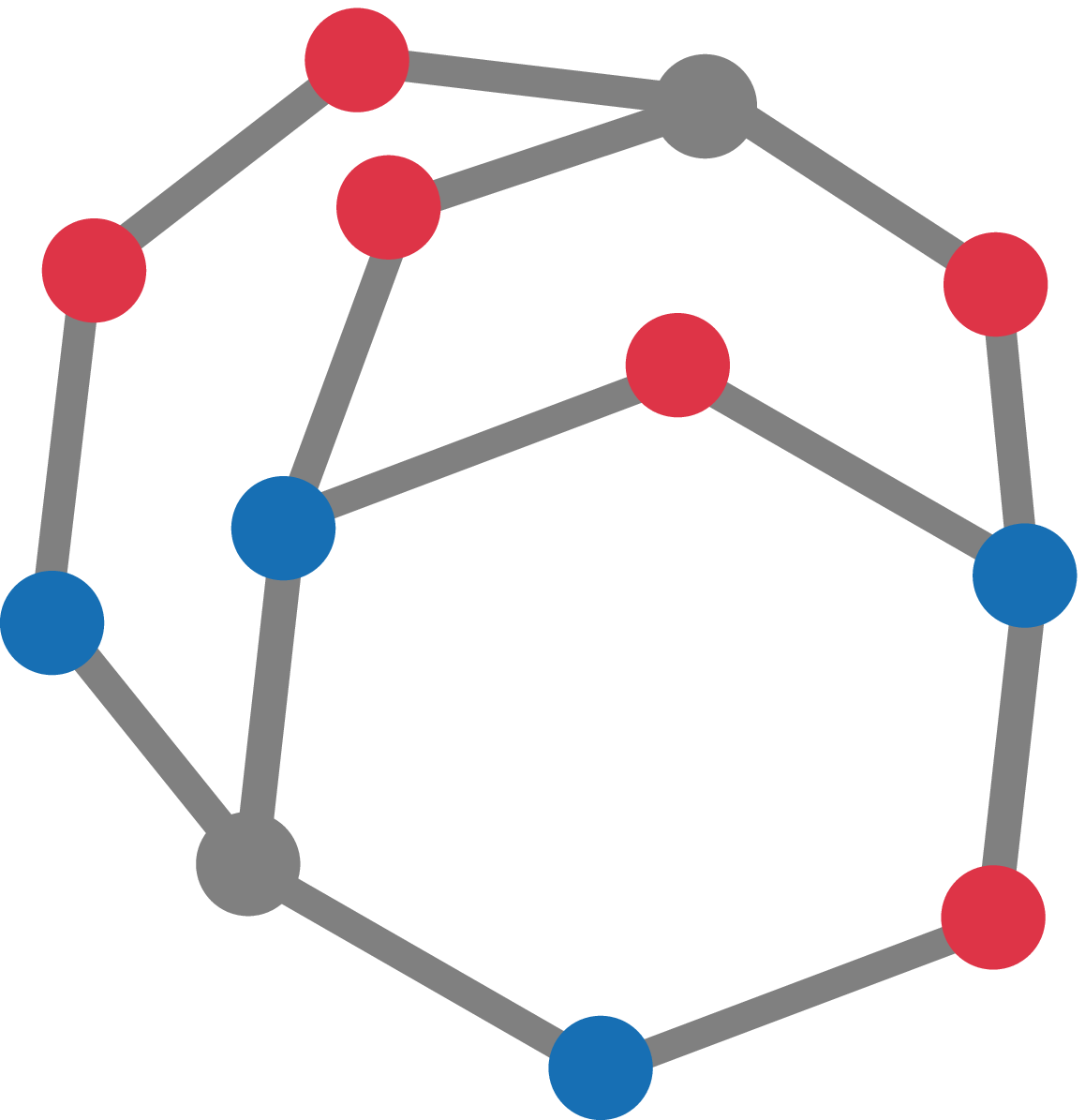}}\\
\subfigure[\label{Fig35}]{\includegraphics[width=0.12\textwidth]{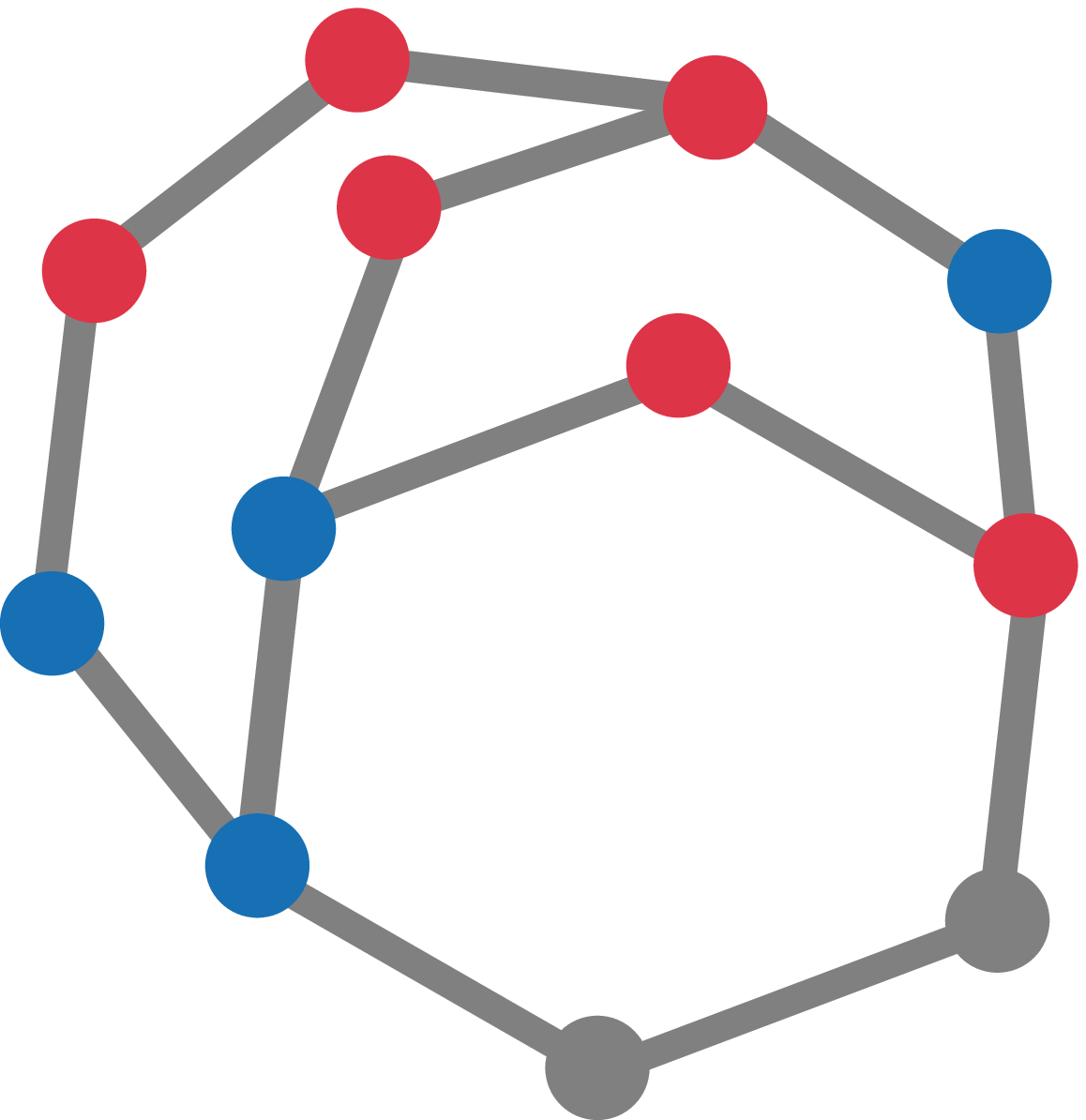}}\quad\quad
\subfigure[\label{Fig33}]{\includegraphics[width=0.11\textwidth]{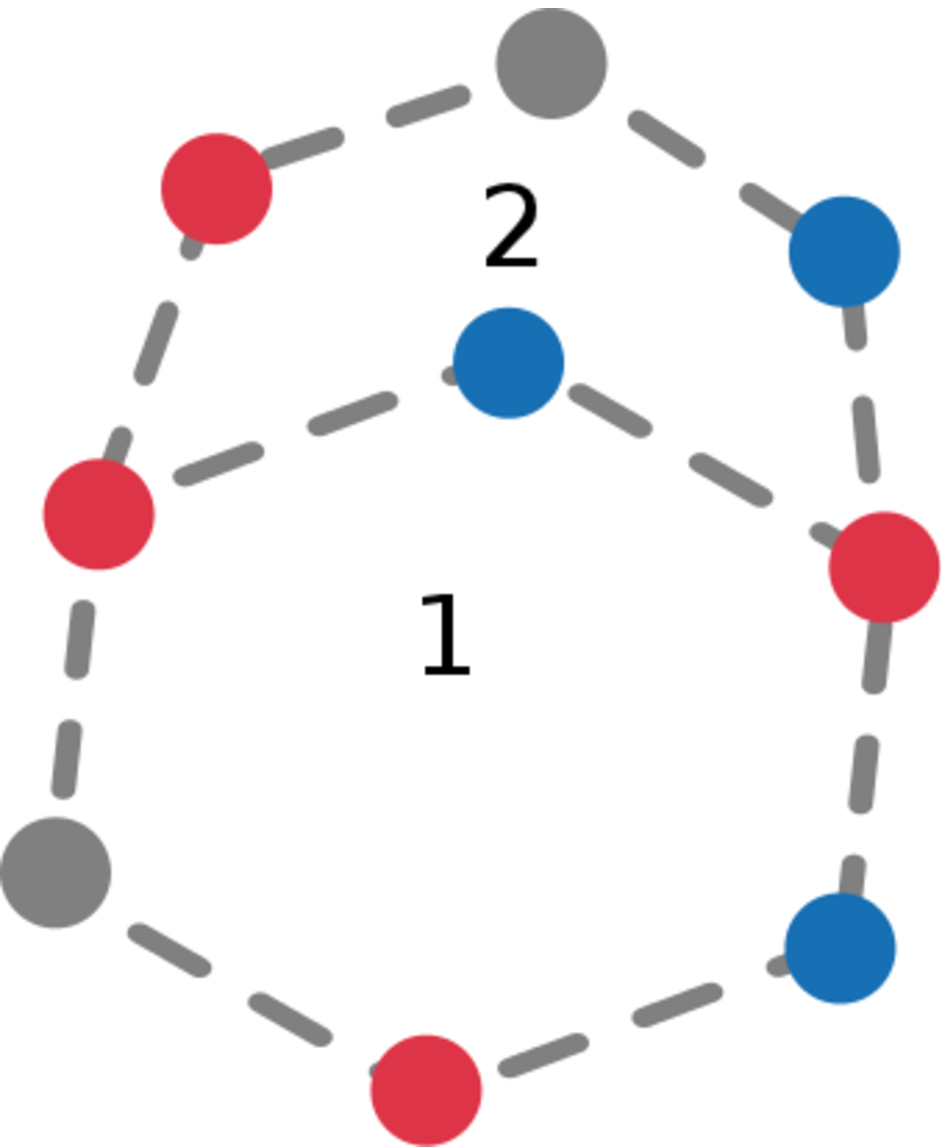}}\quad\quad
\subfigure[\label{Fig33_2}]{\includegraphics[width=0.12\textwidth]{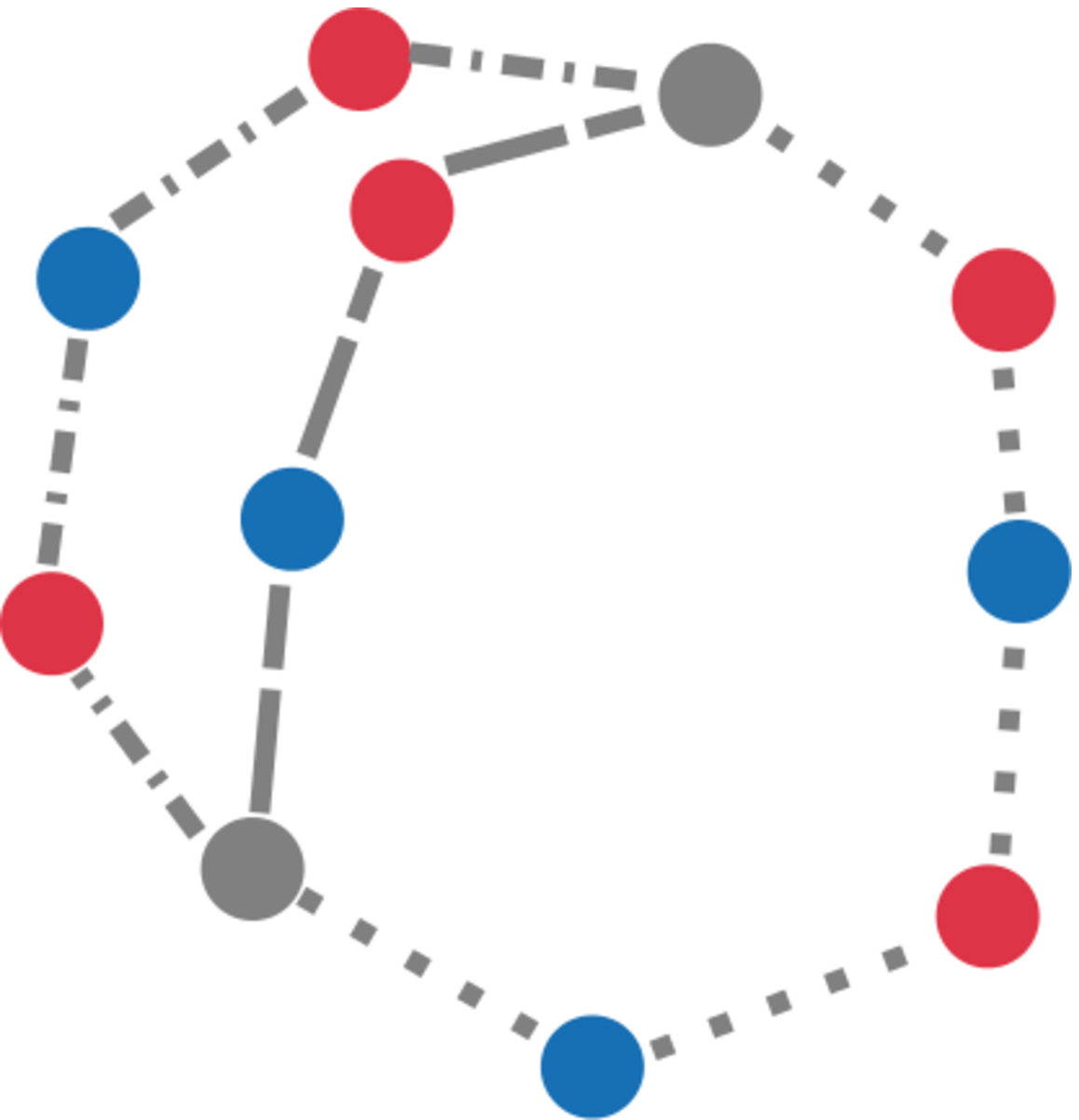}}\\
\caption{\label{FigFour}\ref{Fig300} displays a $2$-connected lattice constructed by adding path $2$ and $3$ to the original circle $1$. \ref{Fig30} and \ref{Fig35} represent two connected states in $O_g$ for the case of the lattice represented in \ref{Fig300}. Here $M=12$ and $A=4,B=6$. \ref{Fig31}-\ref{Fig34} are intermediate states in the sequence connecting \ref{Fig30} to \ref{Fig35}. \ref{Fig33} and \ref{Fig33_2} show states \ref{Fig32} and \ref{Fig33_1} on sublattices. \ref{Fig33_2} also shows three different paths (dashed, dot-dashed and dotted lines) linking the end points of path 3 (see text).}
\end{figure}

An example displaying the assumption of Proposition~\ref{Prps5} on the lattice is shown in Fig.~\ref{Fig28}. The proof of this proposition is given in the Appendix~\ref{App4}.
Note that, 
$\mathbf{G}$ being 2-connected is equivalent to $\mathbf{G}$ being constructed 
as follows: (1) start from a circle; (2) add a path which starts and ends on two 
distinct sites on the circle; (3) successively add paths to the already constructed 
lattice in the same manner as in (2)~\cite{Graph_Theory} 
(see e.g. Fig.~\ref{Fig300})~\footnote{Note that in the remainder of this Subsection
all added paths are non trivial in the sense that they add new sites to the already 
constructed lattice. Adding trivial bonds does not change our conclusions as it 
keeps the 2-connectivity properties of the lattice.}.  

By using this equivalence and Proposition~\ref{Prps5}, we give a necessary and 
sufficient condition in order for any two states to be connected for generic cases (including all cases we discussed above) with $A+B<M-1$ ( or $A+B>M+1$):
\begin{proposition}
\label{Prps6}
Any two states $\ket{\sigma,\lambda}$, $\ket{\sigma',\lambda'}$ with arbitrary 
$A+B<M-1$ (or arbitrary $A+B>M+1$) are connected if and only if $\mathbf{G}$ is 2-connected 
and is not a circle of $5$ 
or more sites~\footnote{Two states on a circle with $4$ or less sites are always connected, 
as shown in Lemma 4.6 of Tasaki's work~\cite{Tasaki1}. In Ref.~\cite{Tasaki1} the authors discuss the degeneracy problem of the ground-state energy of Fermi-Hubbard model with infinite $U$ at fixed number of spin-up(down) fermions and in the presence of a single hole. It is interesting noting that the basis $\ket{\sigma,\lambda}$ that we define here is equivalent to the corresponding basis defined in Ref.~\cite{Tasaki1}.}.
\end{proposition}
We will show that this is true with a specific example. The argument, though, can be straightforwardly generalized to the general case.
Let us start by proving the sufficient condition. Consider a 2-connected lattice 
(not a circle of 5 or more sites) as displayed in Fig.~\ref{Fig300}, and arbitrary $\ket{\sigma,\lambda}$, 
$\ket{\sigma',\lambda'}$ as displayed in Fig.~\ref{Fig30} and \ref{Fig35}. 
Note that, with $A+B<M-1$, there exists at least two grey sites. In this example 
we only consider two grey sites~\footnote{Nothing would change if more than 
two grey sites are present since the grey color can be freely moved on the lattice.}. 
Because the grey color can be moved to any site in the lattice, without loss of 
generality, we choose the grey sites to be the same for $\ket{\sigma,\lambda}$ 
and $\ket{\sigma',\lambda'}$ as shown Fig.~\ref{Fig30} and \ref{Fig35}. 
Starting from $\ket{\sigma,\lambda}$, we first construct an intermediate state 
$\ket{\chi_{n},\theta_{n}}$ (Fig.~\ref{Fig31}) connected to $\ket{\sigma,\lambda}$ 
such that the number of blue and red sites on the inner circle labelled by 1, 
and on the paths (in the context of this proof when we count the number of colors in the added paths we exclude the 
end points from the paths) labelled by $2$ and $3$, is the same as 
in $\ket{\sigma',\lambda'}$. 
This can be done according to property (d) of 
Section~\ref{2-connected}. Specifically, this is done by first fixing the 
colors on path $3$, next, since the remaining lattice is still 2-connected, 
on path $2$. Obviously, at this point, the colors on circle $1$ are 
automatically fixed.

\begin{figure}
\centering
\subfigure[\label{Fig39}]{\includegraphics[width=0.17\textwidth]{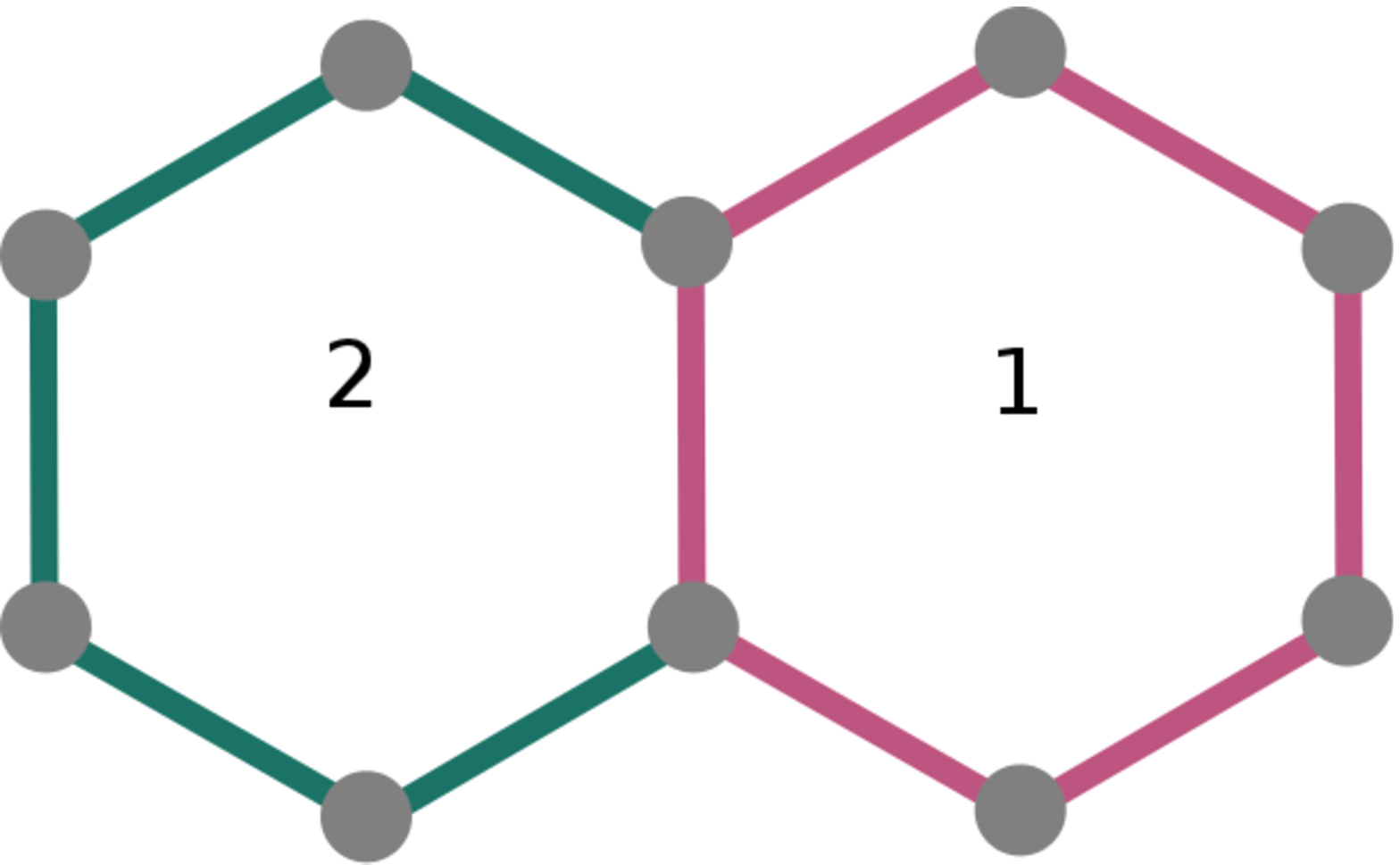}}\quad\quad
\subfigure[\label{Fig40}]{\includegraphics[width=0.17\textwidth]{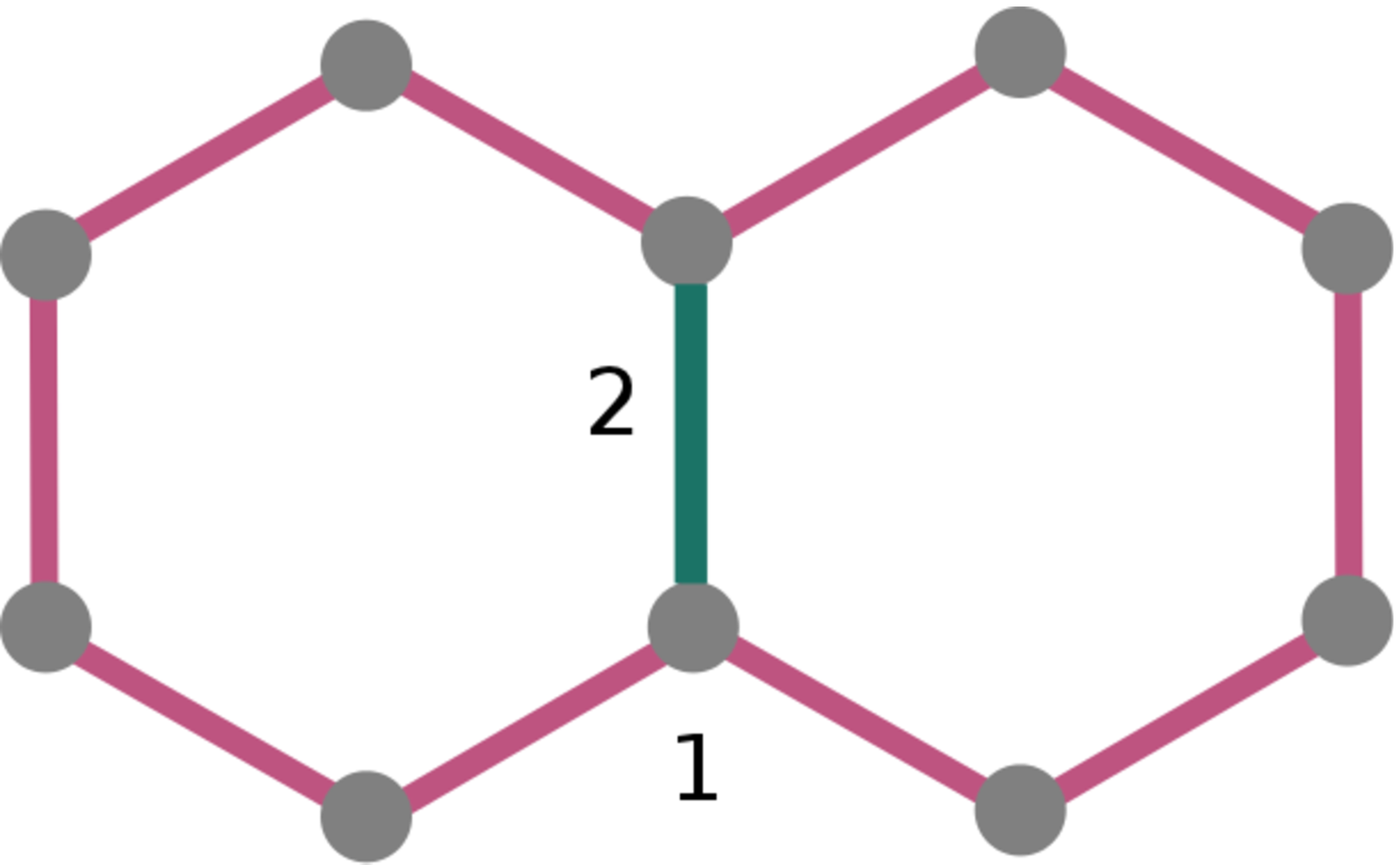}}
\caption{\ref{Fig39} and \ref{Fig40} display two different ways of viewing the same lattice. They can both be viewed as an already constructed lattice plus an added path. In \ref{Fig39}, two ends of the added path (green) form a bond on the already constructed lattice (pink). In \ref{Fig40}, two ends of the added path (green) doe not form a bond on the already constructed lattice (pink).}
\end{figure}

Next, we construct a sequence starting forward from $\ket{\chi_{n},\theta_{n}}$ 
(Fig.~\ref{Fig31}) and backward from $\ket{\sigma',\lambda'}$ 
(Fig.~\ref{Fig35}), by firstly moving the grey color from the original circle
to the two ends of path $3$, only through sites on circle $1$ and path $2$. 
This step generates state Fig.~\ref{Fig32} from Fig.~\ref{Fig31} and state 
Fig.~\ref{Fig34} from Fig.~\ref{Fig35}. In order to connect state~\ref{Fig32} to state~\ref{Fig34} we notice that path $2$ combined with circle $1$ 
(see Fig.~\ref{Fig33}) satisfies the assumption of 
Proposition~\ref{Prps5}~\footnote{If the two ends of any added path form a bond on 
the already constructed lattice as shown in Fig.~\ref{Fig39}, where pink indicates 
the already constructed lattice and green indicates the added path, then, since 
the path adds at least one new site, the ``new'' lattice we consider includes all sites
as shown in Fig.~\ref{Fig40} by pink bonds with an added green trivial path. Now, in order to apply Proposition~\ref{Prps5}, 
we regard the green bond in  Fig.~\ref{Fig40} as the added path.},
therefore we can construct state Fig.~\ref{Fig33_1} 
such that colors on circle $1$ and path $2$ are the same as in state 
Fig.~\ref{Fig34}. Next, we notice that both lattices in 
Fig.~\ref{Fig30} and Fig.~\ref{Fig33} are 2-connected, therefore there exists three disjoint 
paths linking the two ends of path $3$: path $3$ itself and two paths belonging 
to circle $1$ combined with path $2$. This is shown in Fig.~\ref{Fig33_2}. Hence, 
we can apply Proposition~\ref{Prps5} again to show that 
states \ref{Fig33_1} and \ref{Fig34} are connected. 
In conclusion, due to transitivity of connectedness, we have shown that 
$\ket{\sigma,\lambda}$ and $\ket{\sigma',\lambda'}$ are connected.

To prove the necessary condition, we simply observe that if $\mathbf{G}$ 
is not 2-connected or is a circle with $5$ or more sites, as shown in examples 
Fig.~\ref{Fig16}-\ref{Fig19} and Fig.~\ref{FigThree}, there exists some cases for 
which not every two states are connected.

In view of Proposition~\ref{Prps1} and Proposition~\ref{Prps6} we can formulate the 
following theorem:
\begin{theorem}
In case of arbitrary $A+B<M-1$ (or arbitrary $A+B>M+1$), $E_1$ is nondegenerate 
if $\mathbf{G}$ is 2-connected and not a circle with $5$ or more sites.
\end{theorem}
\medskip

\noindent
In the case $A+B=M-1$ (or $A+B=M+1$), finding a necessary and sufficient condition on 
the connectivity of a lattice for any two states to be connected is still an open question. 
Sufficient conditions for a specific model are provided by Tasaki~\cite{Tasaki2} and 
Katsura~\cite{Katsura2}~\footnote{In~\cite{Katsura2} the authors study the degeneracy of the ground-state energy
of the $SU(n)$ Fermi-Hubbard model with $U=\infty$ and with exactly 
one hole. Another sufficient condition for the $SU(2)$ Fermi-Hubbard model requiring the lattice to be constructed by ``exchange bond'' was given in~\cite{Tasaki2}.}


\section{\label{Sec6}Determination of $\ket{\psi^0}$ with $N_a$, $N_b$ 
such that $A=1,B=M-1$}

In the general case of $A+B=M$, there are neither grey nor purple sites. 
Hence, according to the rules given in Subection~\ref{subsection4_1}, any 
two different states are {\em{not}} connected. This statement is valid independently on 
the connectivity properties of $\mathbf{G}$. 
Therefore, all matrix elements in $\mathbf{W_g}$ are zero, which results in $E_1=0$ and 
degenerate. 
We are interested in the case $A=1,B=M-1$ ($A=M-1,B=1$) which correspond to doping 
species-$\cal A$ ($\cal B$) with one particle and species-$\cal B$ ($\cal A$) with one hole. 
In this case,  $\ket{\psi^0}$ is not uniquely determined by solving Eq.~\ref{Eq3}. 
In the following we will take advantage of the symmetry properties of the lattice 
to uniquely determine $\ket{\psi^0}$. 

\begin{figure}
\subfigure[\label{Fig36}]{\includegraphics[width=0.27\textwidth]{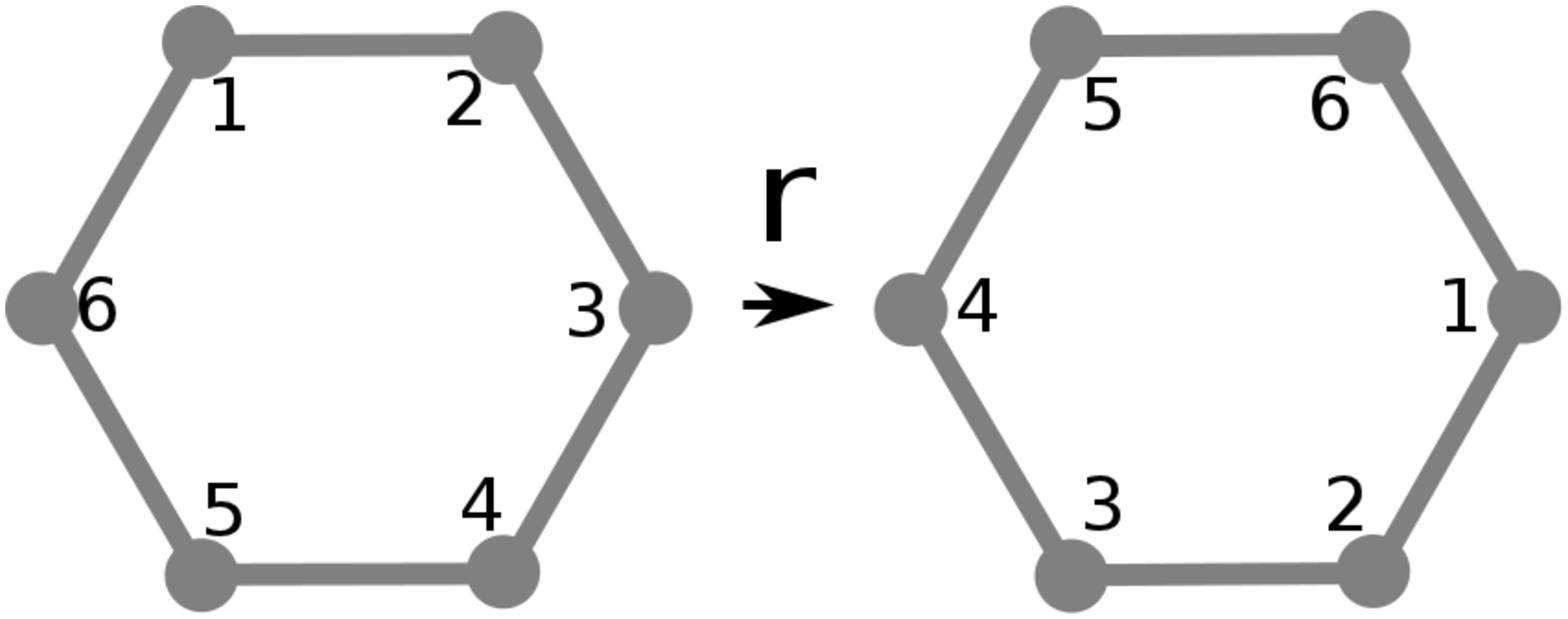}}\quad\quad
\subfigure[\label{Fig37}]{\includegraphics[width=0.30\textwidth]{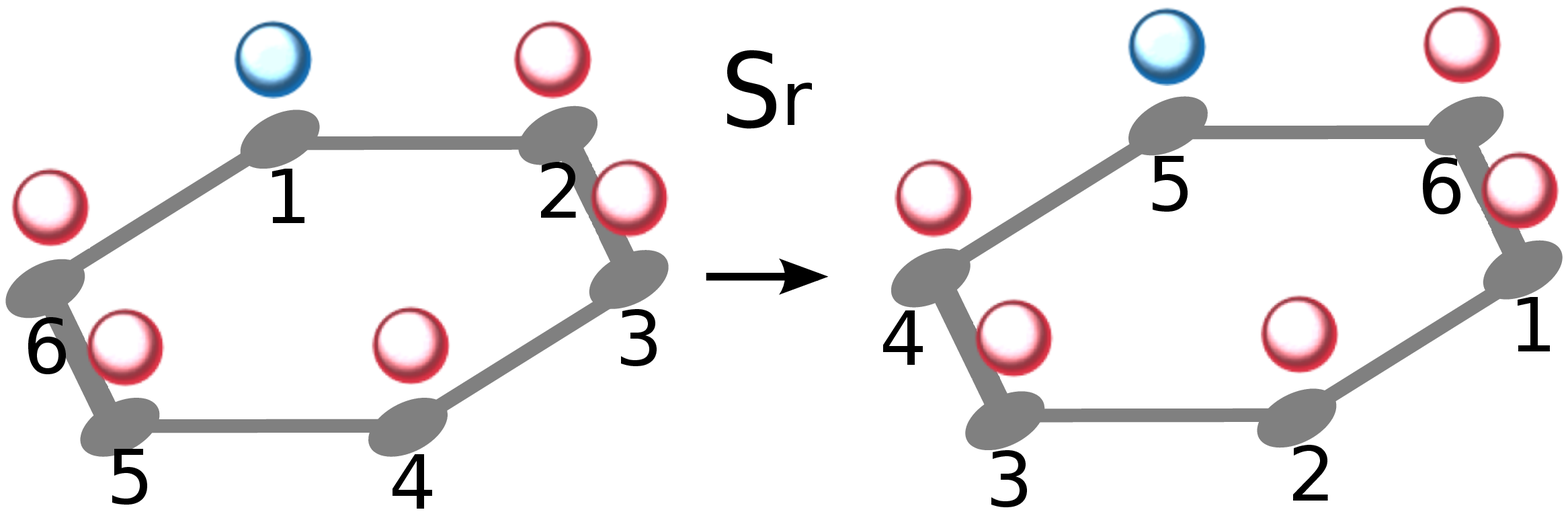}}
\caption{\label{FigSix}\ref{Fig36} displays a lattice automorphism $r$ on an hexagon. \ref{Fig37} displays the action on Fock states of  the corresponding operator $S_r$:  the lattice is rotated while the physical position 
of particles is unchanged.}
\end{figure}

Let us start by defining a symmetry operation $r$ on the lattice and its corresponding 
operator $S_r$. 
We say $r$ is a (bond-weighted) lattice automorphism
of $\mathbf{G}$ if $r$ maps $\mathcal{V}(\mathbf{G})$ one-to-one onto itself and satisfies 
(i) $\{i,j\}$ is a bond if and only if $\{r(i),r(j)\}$ is a bond, 
(ii) $I_{(i,j)}=I_{(r(i),r(j))}$. 
The inverse of $r$, $r^{-1}$, is also a lattice automorphism. 
Given a lattice automorphism $r$, one can define a linear operator $S_r$ 
on $\mathscr{H}$ such that $S_r\ket{\xi,\gamma}=\ket{\xi',\gamma'}$, 
where $\xi'_i=\xi_{r(i)}$ and $\gamma'_i=\gamma_{r(i)}$.
If we take the example of Fig.~\ref{FigSix} with equal weight on all bonds ($I_{(i,j)}=1$ for every 
$\{i,j\}$), the lattice automorphism $r$ is a $2\pi/3$ clockwise rotation 
(see Fig.~\ref{Fig36}). The action of the corresponding $S_r$ is shown in 
Fig.~\ref{Fig37}, where the lattice is rotated while the physical position 
of particles is unchanged.

Since $r$ is invertible, $S_r$ also has an inverse, ${S_r}^{-1}$. It is easy to show 
that $S_{r^{-1}}=S_{r}^{-1}$. Moreover, by definition 
$\ket{\xi',\gamma'}=S_r\ket{\xi,\gamma}$ is also a normalized Fock state. 
Therefore $S_r$ preserves 
the norm $\sqrt{\braket{\psi|\psi}}$ for any arbitrary state $\ket{\psi}$
in a finite-dimensional $\mathscr{H}$. 
Hence,
$S_r$ is a unitary operator, i.e. ${S_r}^{\dagger}={S_r}^{-1}$, and thus a bounded operator~
\footnote{A unitary operator is always bounded. Then, if an operator $S_r$ is bounded,
$\left(\ket{\psi^0}+\sum_{n=1}^\infty{\epsilon}^n\ket{\psi^n}\right) 
\rightarrow \ket{\Psi}$ implies 
$\left(S_r\ket{\psi^0}+\sum_{n=1}^\infty{\epsilon}^n{S_r}\ket{\psi^n}\right) 
\rightarrow S_r\ket{\Psi}$.}.

Note that, by definition, the state $S_r\ket{\xi,\gamma}$ has exactly the same 
spatial configuration of bosons as $\ket{\xi,\gamma}$. 
Then the interaction-dependent terms in $H$ are unchanged.
Hence, state $S_r\ket{\xi,\gamma}$ has the same eigenvalue of $H_0$ as $\ket{\xi,\gamma}$, 
and thus $S_r$ commutes with $H_0$. Moreover, according to Eq.~\ref{Eq2} 
and the definition of $r$,
\begin{widetext}
\begin{multline}
\braket{\xi,\gamma|S_{r}^{-1}WS_r|\xi',\gamma'}=-\delta_{\gamma,\gamma'} t_a
\sum_{(r(i),r(j))}
\Big[
I_{{r(i)},r(j)}
\sqrt{\xi_{r(j)}+1}\sqrt{\xi_{r(i)}'+1}\; 
\delta_{\xi_{r(j)}+1,\xi'_{r(j)}}
\delta_{\xi_{r(i)},\xi_{r(i)}'+1}
\prod_{l\ne{r(i),r(j)}}^M\delta_{\xi_{r(l)}\xi_{r(l)}'}\Big] \\
-\delta_{\xi,\xi'} t_b
\sum_{(r(i),r(j))}
\Big[I_{{r(i)},r(j)}\sqrt{\gamma_{r(j)}+1}\sqrt{\gamma'_{r(i)}+1}\; 
\delta_{\gamma_{r(j)}+1,\gamma'_{r(j)}}\delta_{\gamma_{r(i)},\gamma'_{r(i)}+1}
\prod_{m\ne{r(i),r(j)}}^M\delta_{\gamma_m,\gamma'_m}\Big]=\braket{\xi,\gamma|W|\xi',\gamma'}.
\end{multline}
\end{widetext}
Therefore, $S_r$ also commutes with $H$.

The boundedness of $S_r$ implies:
\begin{equation}
S_r\ket{\Psi}= S_r \ket{\psi^0}
+
\sum_{n=1}^\infty  {\epsilon}^n {S_r}\ket{\psi^n}.
\end{equation}
Since $E$ is nondegenerate and 
$HS_r\ket{\Psi}=ES_r\ket{\Psi}$, then $S_r\ket{\Psi}=e^{i\theta}\ket{\Psi}$. 
More specifically,
\begin{equation}
\label{eq_S_r}
S_r\ket{\psi^0}+\sum_{n=1}^\infty {\epsilon}^n{S_r}\ket{\psi^n}
=e^{i\theta}\ket{\psi^0}+ \sum_{n=1}^\infty{\epsilon}^n{e^{i\theta}}\ket{\psi^n}.
\end{equation}
Taking the limit $\epsilon \rightarrow{0}$, 
we have $S_r\ket{\psi^0} = e^{i\theta}\ket{\psi^0}$. Furthermore, multiplying by 
$\bra{\xi,\gamma}$ Eq.~\ref{eq_S_r}, we obtain two power series of $\epsilon$:
\begin{multline}
\braket{\xi,\gamma|S_r|\psi^0}
+\sum_{n=1}^\infty {\epsilon}^n \braket{\xi,\gamma|{S_r}|\psi^n}
\\
= e^{i\theta}\braket{\xi,\gamma|\psi^0}+\sum_{n=1}^\infty{\epsilon}^n{e^{i\theta}}
\braket{\xi,\gamma|\psi^n}.
\end{multline}
Because the series are analytic in a small neighborhood of $\epsilon =0$, we can equate 
the coefficients at each order $n$ to get 
$\braket{\xi,\gamma|{S_r}|\psi^n}={e^{i\theta}}\braket{\xi,\gamma|\psi^n}$. 
In other words, $\ket{\psi^n}$ is $S_r$-invariant (apart from a phase factor) for 
any lattice automorphism $r$.

In the following we will use these properties to determine the expansion coefficients 
of the first order correction to the ground state Eq.~\ref{Eq3}. 
Let us consider arbitrary states $\ket{\sigma,\lambda}$ and $\ket{\sigma',\lambda'}$. 
We denote the unique blue site (recall $A=1$ so all sites but one are red) on 
these states by $i$ and $j$ respectively. If there exists a lattice automorphism $r$ 
such that $r(j)=i$, then $\ket{\sigma',\lambda'}=S_r\ket{\sigma,\lambda}$. 
Moreover, as shown above, 
$\braket{\sigma',\lambda'|S_r|\psi^0}
=\braket{S_r^{\dagger}\sigma',\lambda'|\psi^0}
=\braket{S_r^{-1}\sigma',\lambda'|\psi^0}
=\braket{S_{r^{-1}}\sigma',\lambda'|\psi^0}
=\braket{\sigma,\lambda|\psi^0}=e^{i\theta}\braket{\sigma',\lambda'|\psi^0}$. 
If we choose $\ket{\Psi}$ to be positive (see Theorem~\ref{The1}), 
in the limit of $\epsilon$ arbitrarily small, all $\braket{\sigma,\lambda|\psi^0}$ 
are also positive. This implies 
$\braket{\sigma, \lambda |\psi^0}=\braket{\sigma',\lambda'|\psi^0}$. 
Therefore we can conclude the following:

\begin{theorem}
If $\mathbf{G}$ is connected and for any two sites $i$ and $j$ there exists 
a lattice automorphism mapping $j$ to $i$, then $\braket{\sigma,\lambda|\psi^0}=1/\sqrt{M}$.
\end{theorem}

The assumption made on the lattice is very easily satisfied by any 
regular lattice with periodic boundary 
conditions (e.g. hypercubic, triangular, honeycomb....) as long as 
$I_{(i,j)}=I_{\bf{i}-\bf{j}}$, where $\bf{i},\bf{j}$ 
refer to the position of sites $i,j$. We also note that this assumption seems to be independent
from the size of the lattice.

\section{Conclusion}
We have studied the degeneracy of the ground-state energy $E$ of the two-component
Bose-Hubbard model and of the perturbative correction $E_1$ in terms of connectivity properties of the optical lattice.
We have shown that the degeneracy properties of $E$ and $E_1$ are closely related to the
connectivity properties of the lattice. We can summarize our main results as follows:
\begin{itemize}
\item The ground-state energy E is nondegenerate if the lattice is connected.
\item When $A=0,B\ne{0}$ ($B=0,A\ne{0}$), $E_1$ is nondegenerate if the lattice is connected.
\item When $A=1,0<B<M-1$ or $A=M-1,\\
1<B\le{M-1}$, $E_1$ is nondegenerate if the lattice is $2$-connected.
\item In generic cases with $A+B<M-1$ or\\
 $A+B>M+1$, $E_1$ is nondegenerate if the lattice is $2$-connected and not a circle with $5$ or more sites.
\item When $A+B=M$, $E_1$ is degenerate independently on the connectivity of the optical lattice. In the case of $A=M-1,B=1$ ($A=1,B=M-1$), we\\
 have determined the 0th order correction of state $\psi^0$. We have shown that $\psi^0$ possesses equal expansion coefficient provided that there exists a lattice automorphism mapping a generic site of the lattice into another one. 
\end{itemize}
These results are used to ensure a valid perturbative
approach of the two-component Bose-Hubbard model also in the case of degenerate $E_1$. 
We expect that the analysis developed in this paper and the
results about the ground-state degeneracy provide an effective tool
to study the asymmetric character of the Mott-insulator to superfluid
transition between the particle and hole side and, more in
general, the entanglement properties that appear to
characterize this process.

\begin{acknowledgments}
The work of one of the authors (VP) has been partially supported by the M.I.U.R. project Collective
quantum phenomena: From strongly correlated systems to quantum simulators (PRIN 2010LLKJBX).
\end{acknowledgments}


\appendix

\section{Proof of Proposition~\ref{Prps1}\label{App1} }
We only prove the equivalence between (a) and (b).
A proof based on the connectivity of the underlying graph of 
matrices is given in \cite{Irreducible_Graph}.
The equivalence bewteen (b) and (c) is a direct 
consequence of Theorem 4.3 in Ref.~\cite{Tasaki1}.

Let us first prove the necessary condition by contradiction. 
Let us assume that $\ket{\xi,\gamma}$ and $\ket{\xi',\gamma'}$ are not connected by $X$. 
Then $\ket{\xi,\gamma}/\mathfrak{R}_X\ne\ket{\xi',\gamma'}/\mathfrak{R}_X$ and they both 
belong to $O/\mathfrak{R}_X$. So $O/\mathfrak{R}_X$ is a nontrivial partition of $O$, 
i.e. it includes more than one subset of $O$, and thus $\mathbf{X}$ is reducible. 
We get contradiction. Therefore we proved the necessary condition.

Let us now prove the sufficient condition also by contradiction. 
Let us assume $\mathbf{X}$ is reducible. Then, there exists a nontrivial partition 
of $O$ containing at least two disjoint nonempty subsets $O_1$ and $O_2$ of $O$, and 
the blocks $\mathbf{X}_{O_1\times{O_1}^c}$ (${O_1}^c$ is the complement of $O_1$ in $O$), 
$\mathbf{X}_{{O_1}^c\times{O_1}}$, 
$\mathbf{X}_{O_2\times{O_2}^c}$ 
and $\mathbf{X}_{{O_2}^c\times{O_2}}$ are zero. 
On the other hand, by hypothesis, for any $\ket{\xi,\gamma}$ and $\ket{\xi',\gamma'}$ 
which belong respectively to $O_1$ and $O_2$, there exists a finite sequence in the basis 
$\{\ket{\alpha_1,\beta_1},\ket{\alpha_2,\beta_2},\cdots,\ket{\alpha_N,\beta_N}\}$ 
such that $\ket{\alpha_1,\beta_1}=\ket{\xi,\gamma}$, 
$\ket{\alpha_N,\beta_N}\}=\ket{\xi',\gamma'}$ and for any $1\le{i}<N$, 
$\braket{\alpha_i,\beta_i|X|\alpha_{i+1},\beta_{i+1}}\ne{0}$. 
Hence, for some $1\le{i}<N$, $\ket{\alpha_i,\beta_i}\in{O_2}^c$ and 
$\ket{\alpha_{i+1},\beta_{i+1}}\in{O_2}$ 
with $\braket{\alpha_i,\beta_i|X|\alpha_{i+1},\beta_{i+1}}\ne{0}$ which implies 
$\mathbf{X}_{O_2\times{O_2}^c}\ne{0}$. We get contradiction, hence $\mathbf X$ is irreducible.


\section{Proof of Corollary~\ref{Cor2}\label{App2}}
Let us consider $t_b=0$, $t_a\ne{0}$ 
($\leftrightarrow T_b=0$, $T_a\ne{0}$). The basic idea is to show that $\mathbf{W}$ can be block 
diagonalized in terms of $M^{N_b}$ identical blocks.
Let us start by noticing that matrix elements of $W$:
\begin{widetext}
\begin{equation}
\braket{\xi,\gamma|W|\xi',\gamma'}
=-t_a \delta_{\gamma,\gamma'}
\sum_{\{i,j\}\in\mathcal{E}(\mathbf{G})}
\Big[I_{(i,j)}\sqrt{\xi_{j}+1}\sqrt{\xi'_{i}+1}
\delta_{\xi_{j}+1,\xi'_j}\delta_{\xi_{i},\xi'_{i}+1}
\prod_{l\ne{i,j}}^M\delta_{\xi_l\xi'_l}\Big].
\end{equation}
\end{widetext}
are nonzero only when $\gamma=\gamma'$. Moreover, if $\gamma=\gamma'$, the value of 
matrix elements is independent of $\gamma$.

Let us define a function $f$ which provides a one-to-one
mapping from the set $P$ of all $\ket{\gamma}$s onto
$O/\mathfrak{R}_W$, such that for any $\ket{\gamma}$, 
$f(\ket{\gamma})=\ket{\xi,\gamma}/\mathfrak{R}_W$. The mapping $f$ can be easily defined and 
one just needs to show that $f$ is one-to-one and onto. 

Let $\ket{\gamma}$ and 
$\ket{\gamma'}$ be different. It's obvious that any member in $f(\ket{\gamma})$ 
is not connected with any member in $f(\ket{\gamma'})$ by $W$, hence 
$f(\ket{\gamma})\ne{f(\ket{\gamma'})}$, i.e. $f$ is one-to-one. 
Next, let $x\in{O/\mathfrak{R}_W}$, i.e. $x=\ket{\xi'',\gamma''}/\mathfrak{R}_W$ 
for some $\ket{\xi'',\gamma''}$. 
Let us now consider $f(\ket{\gamma''})=\ket{\xi''',\gamma''}/\mathfrak{R}_W$ 
for some $\ket{\xi'''}\ne \ket{\xi''}$. By using connection properties of $W$, 
one can show that $\ket{\xi'',\gamma''}\mathfrak{R}_W\ket{\xi''',\gamma''}$. 
So $\ket{\xi'',\gamma''}/\mathfrak{R}_W=\ket{\xi''',\gamma''}/\mathfrak{R}_W$ 
and thus $x=f(\ket{\gamma''})$, i.e. $f$ is onto. In conclusion, $f$ is a 
one-to-one mapping from $P$ onto $O/\mathfrak{R}_W$.

The total number of elements in $P$ is $M^{N_b}$, hence $O/\mathfrak{R}_W$ 
is a nontrivial partition of $O$. Then, it is obvious that for any 
$O_i\in{O/\mathfrak{R}_W}$, $\mathbf{W}_{O_i\times{O_i}^c}$ and 
$\mathbf{W}_{{O_i}^c\times{O_i}}$ are zero matrices. In other words, 
$O/\mathfrak{R}_W$ block-diagonalizes $\mathbf{W}$. 

Next, we show that each block is an irreducible nonnegative matrix and all
blocks have the same set of eigenvalues. Let $\ket{\gamma}\in{P}$, then 
$f(\ket{\gamma})\in{O/\mathfrak{R}_W}$. We can express $f(\ket{\gamma})$ as 
$f(\ket{\gamma})=Q\otimes\{\ket{\gamma}\}$, where $Q$ is the set of all 
$\ket{\xi}$'s. From Frobenius theorem, one can conclude that 
$\mathbf{W}_{f(\ket{\gamma})\times{f(\ket{\gamma})}}$ has a nondegenerate 
ground-state energy. Since $f(\ket{\gamma})= Q \otimes\{\ket{\gamma}\}$, one can use 
the identity map
on $Q$, so that for any $\ket{\gamma}\ne\ket{\gamma'}$, a one-to-one 
mapping from $\mathbf{W}_{f(\ket{\gamma})\times{f(\ket{\gamma})}}$ onto 
$\mathbf{W}_{f(\ket{\gamma'})\times{f(\ket{\gamma'})}}$ can be constructed.
By construction, this mapping keeps matrix element identical, i.e. the 
two matrices have the same set of eigenvalues.

In conclusion, we have shown that the ground-state energy of $W$ is 
$M^{N_b}$-degenerate.


\begin{figure}
\centering
\subfigure[\label{Fig51}]{\includegraphics[width=0.16\textwidth]{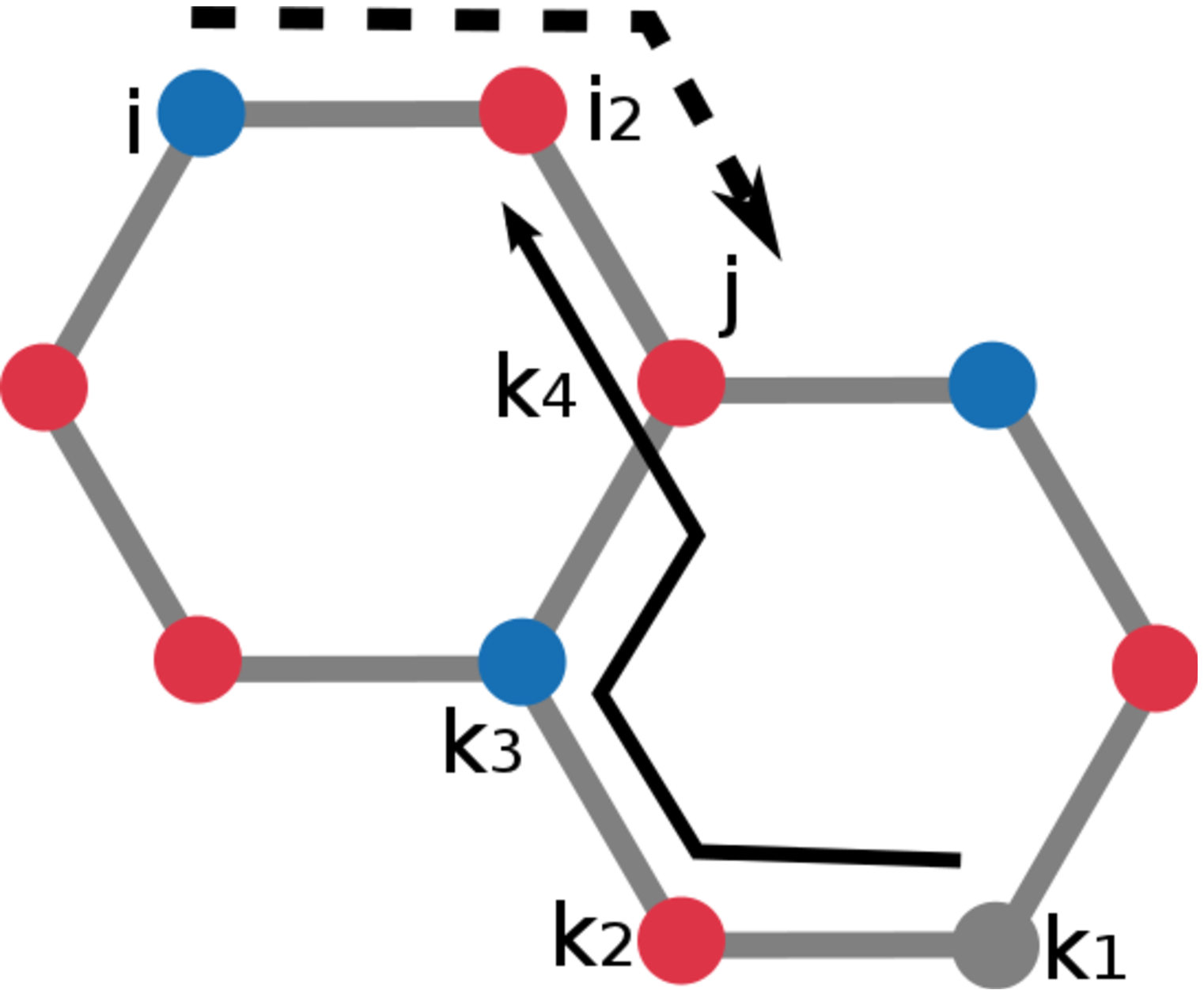}}\quad
\subfigure[\label{Fig52}]{\includegraphics[width=0.16\textwidth]{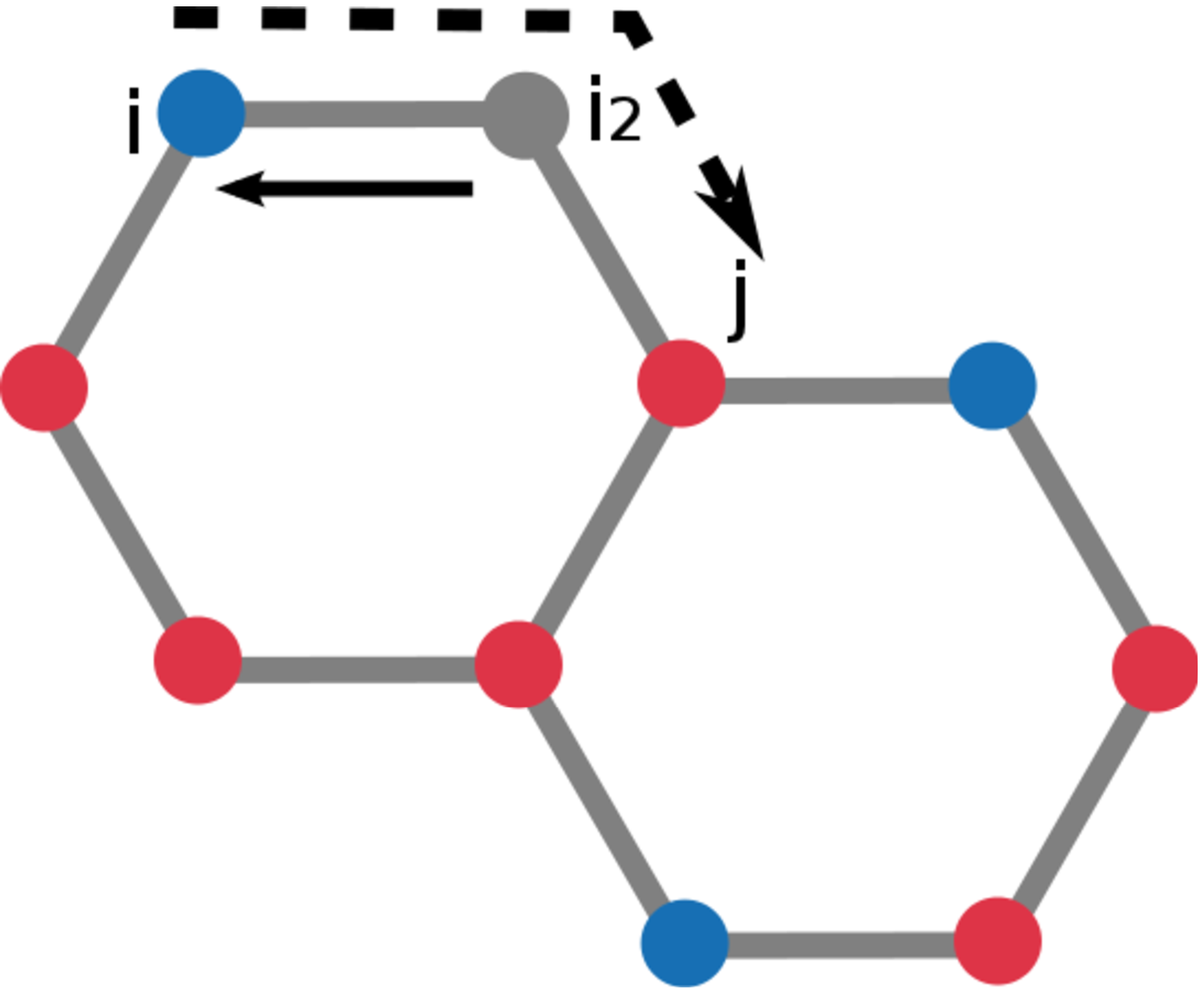}}\quad
\subfigure[\label{Fig53}]{\includegraphics[width=0.18\textwidth]{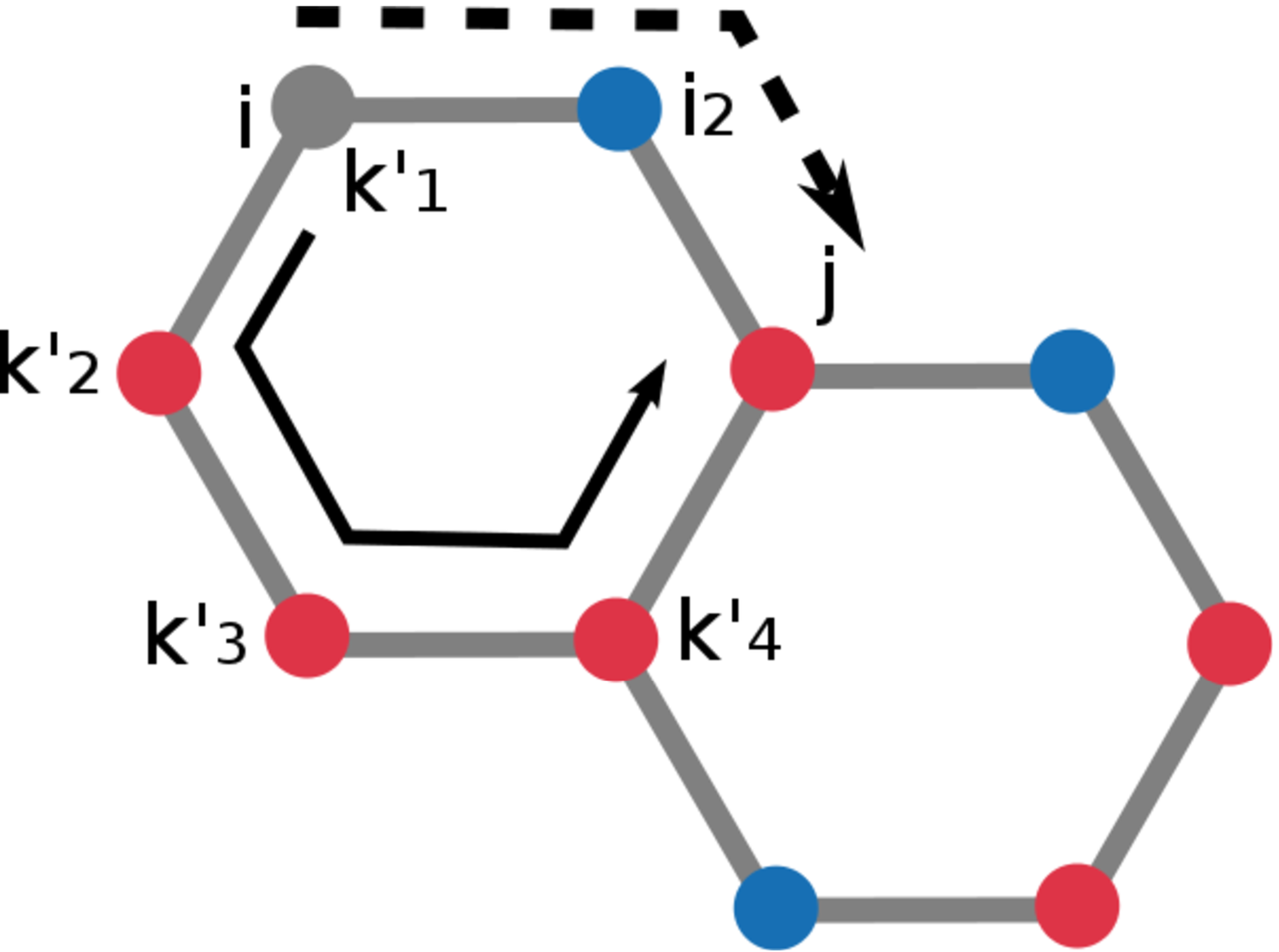}}\quad
\subfigure[\label{Fig54}]{\includegraphics[width=0.16\textwidth]{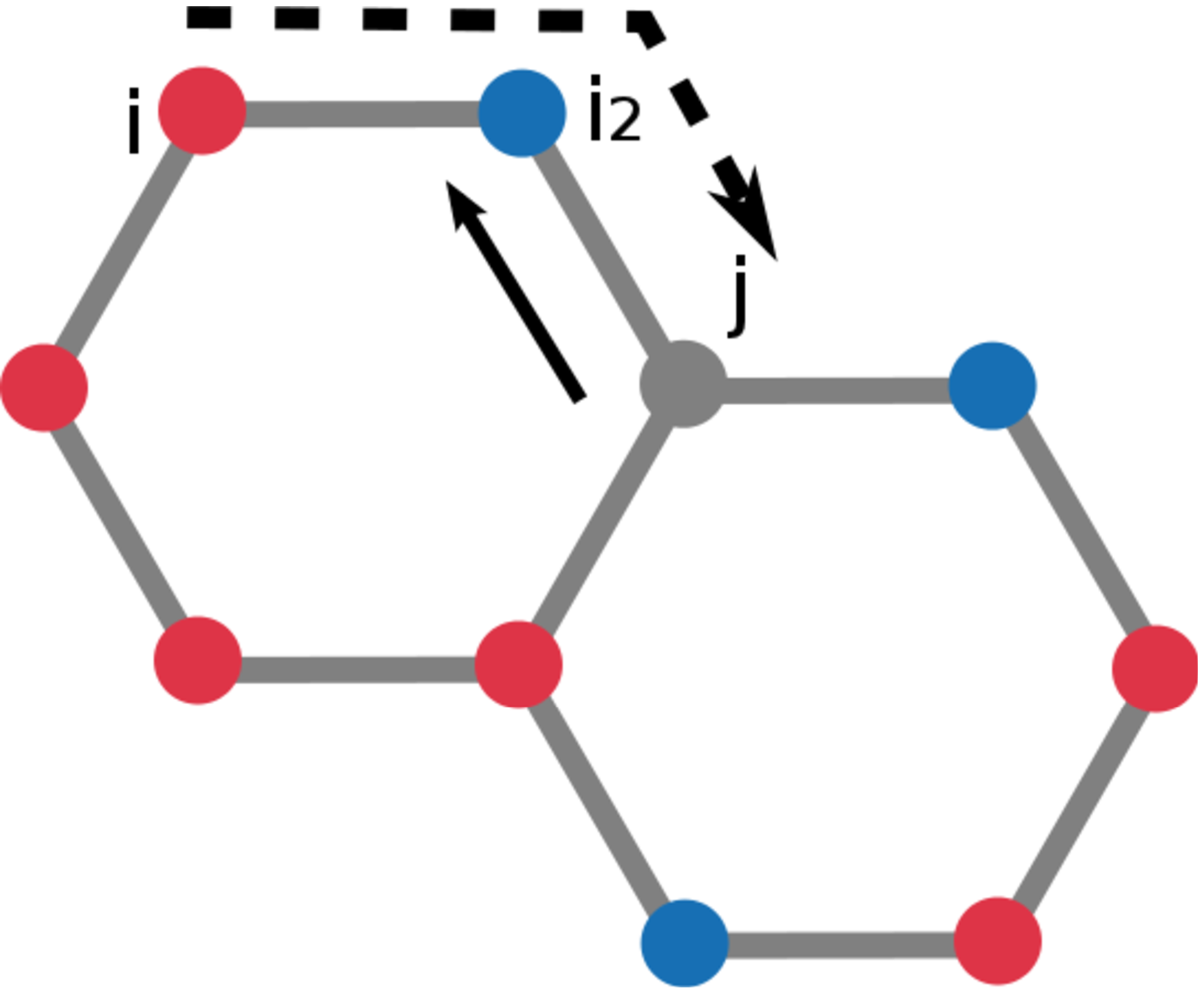}}\quad
\subfigure[\label{Fig55}]{\includegraphics[width=0.16\textwidth]{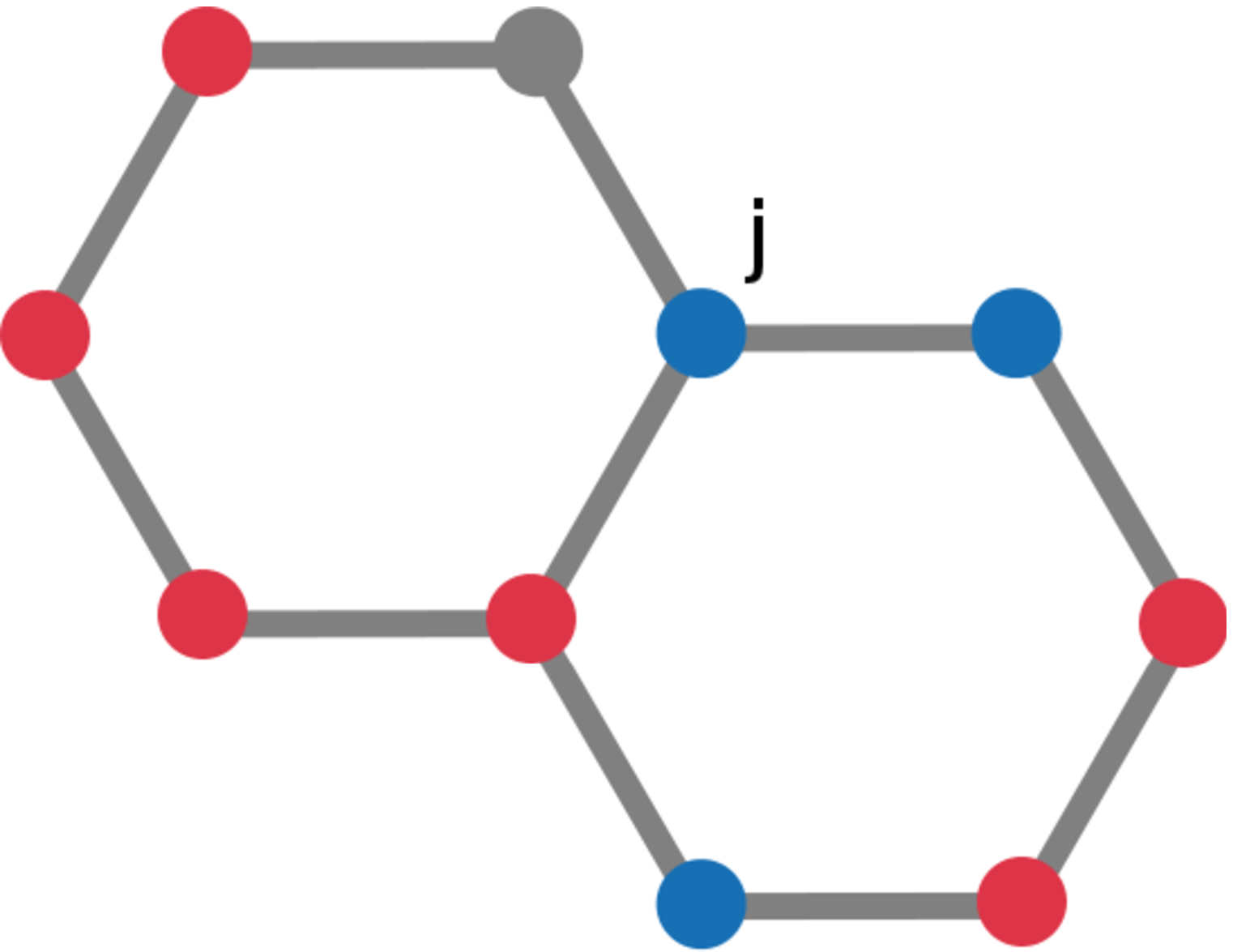}}
\caption{\label{FigSeven} \ref{Fig51} through \ref{Fig55} are examples of states in $O_g$ on a $2$-connected lattice with $A=3,B=6$. The steps for moving the blue color from site $i$ to $j$ are explained in the text and displayed pictorially in the sequence \ref{Fig51}-\ref{Fig55}. The dashed arrow indicates a path connecting $i$ and $j$. Black arrows indicate the path along which the grey color is moved at each step.}
\end{figure}

\section{Proof of property (d) in Subsection~\ref{2-connected} \label{App3}}

We only consider the case of blue color. The proof for the red color is trivially equal.

Consider an arbitrary state and arbitrary sites $i$ and $j$, where $i$ is blue. 
We want to move the blue color from $i$ to $j$ according to the rules given in 
Subsection~\ref{subsection4_1}. Because $\mathbf{G}$ is connected, there exists 
a path $\{i,i_2,\cdots,j\}$ linking $i$ to $j$. The idea of following proof is to move the blue color along this path.

To illustrate the proof, we give an example in Fig.~\ref{FigSeven}, where \ref{Fig51} is the starting state,
and the path $\{i,i_2,\cdots,j\}$ is indicated by a dashed arrow. To move the blue color along the path,
e.g. from $i_n$ to $i_{n+1}$, we need to firstly move a grey color to $i_{n+1}$ and then exchange the color
on the bond $\{i_n,i_{n+1}\}$.
In order to do so we observe that due to the 2-connectivity 
of $\mathbf{G}$, there also exists a path $\{k_1,k_2,\cdots,i_2\}$ which avoids $i$
but links a grey site $k_1$ to $i_2$. The grey color can be moved successively on 
$\{k_1,k_2,\cdots,i_2\}$ so that $i_2$ becomes grey. Note that the fact that the path $\{k_1,k_2,\cdots,i_2\}$ avoids $i$ is important,
because it allows us to keep the blue color on $i$ while moving the grey color to $i_2$. Next step consists of exchanging 
the color on the bond $\{i,i_2\}$ so that $i_2$ becomes blue. The last two steps 
can be repeated successively (i.e. finding a path $\{{k'}_1,{k'}_2,\cdots,i_3\}$ 
linking a grey site ${k'}_1$ to $i_3$ and avoiding $i_{2}$, moving the grey color
along this path until $i_3$ becomes grey, exchanging the color on bond $\{i_2,i_3\}$ 
so that $i_3$ becomes blue and so on) until $j$ acquires the blue color. This process is illustrated in Fig.~\ref{Fig51} through \ref{Fig55}
where solid black arrows indicate the path along which the grey color is moved at each step.

\section{Proof of Proposition~\ref{Prps5}\label{App4}}
\begin{figure}
\centering
\subfigure[\label{Fig20}]{\includegraphics[width=0.12\textwidth]{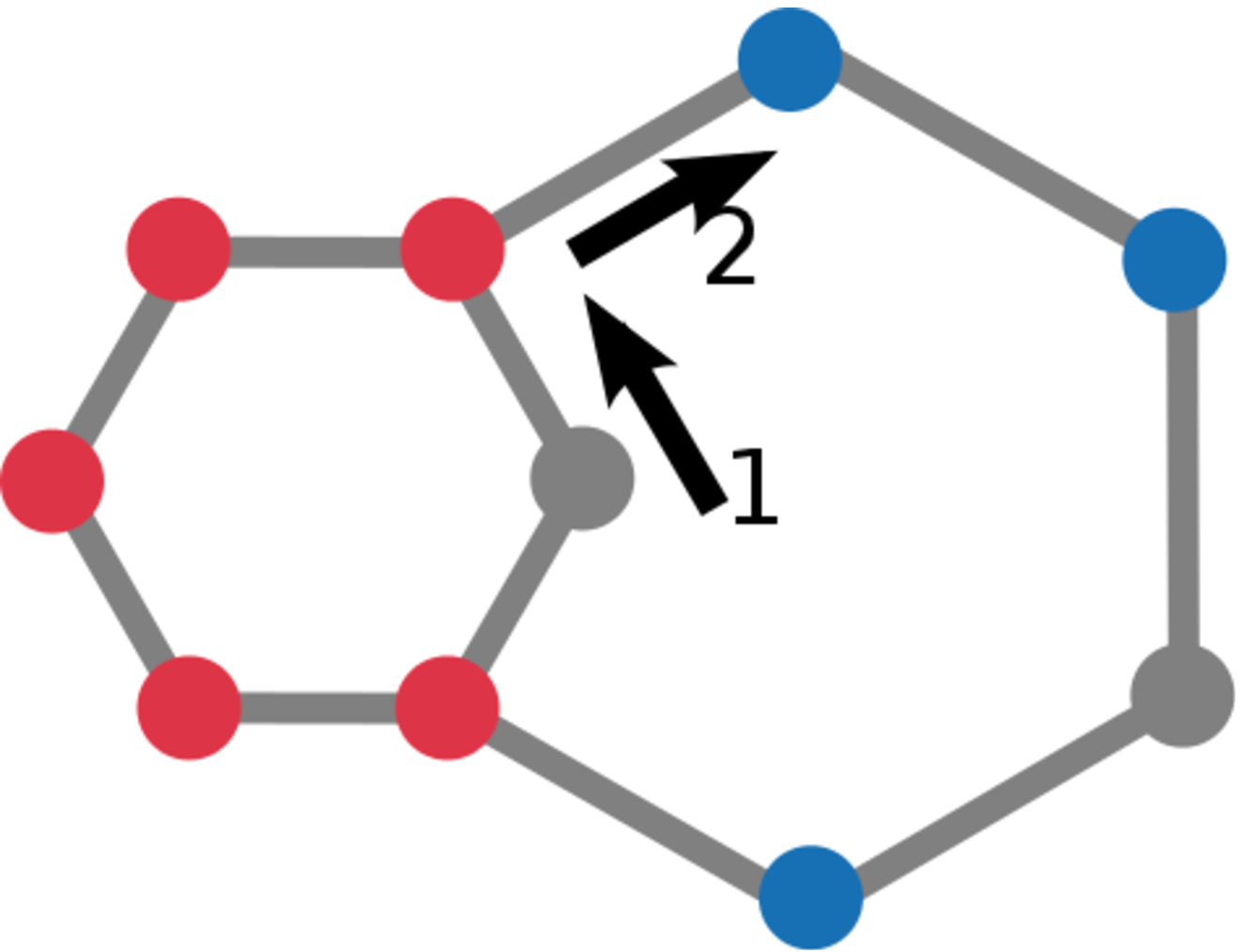}}\quad\quad
\subfigure[\label{Fig21}]{\includegraphics[width=0.12\textwidth]{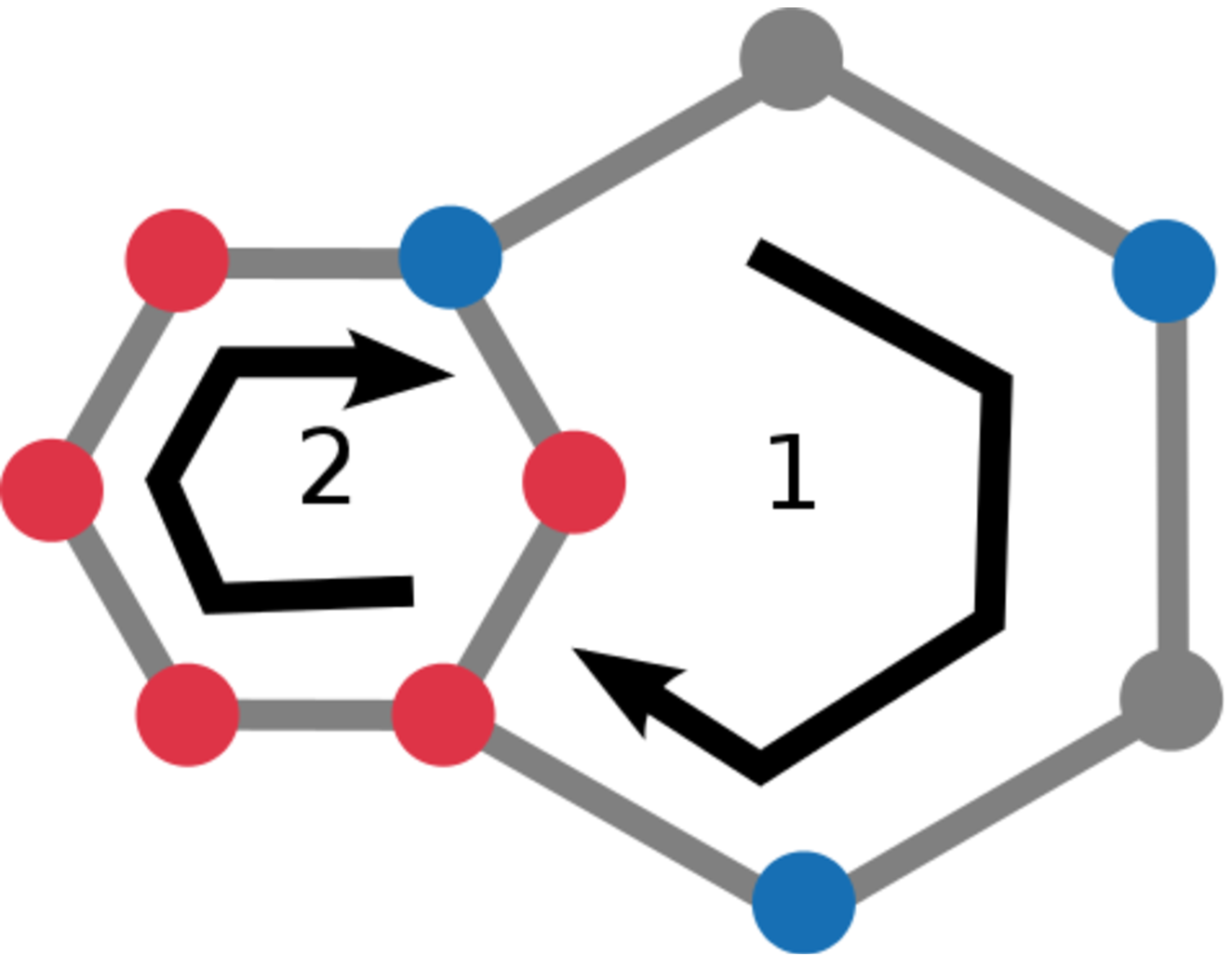}}\quad\quad
\subfigure[\label{Fig22}]{\includegraphics[width=0.12\textwidth]{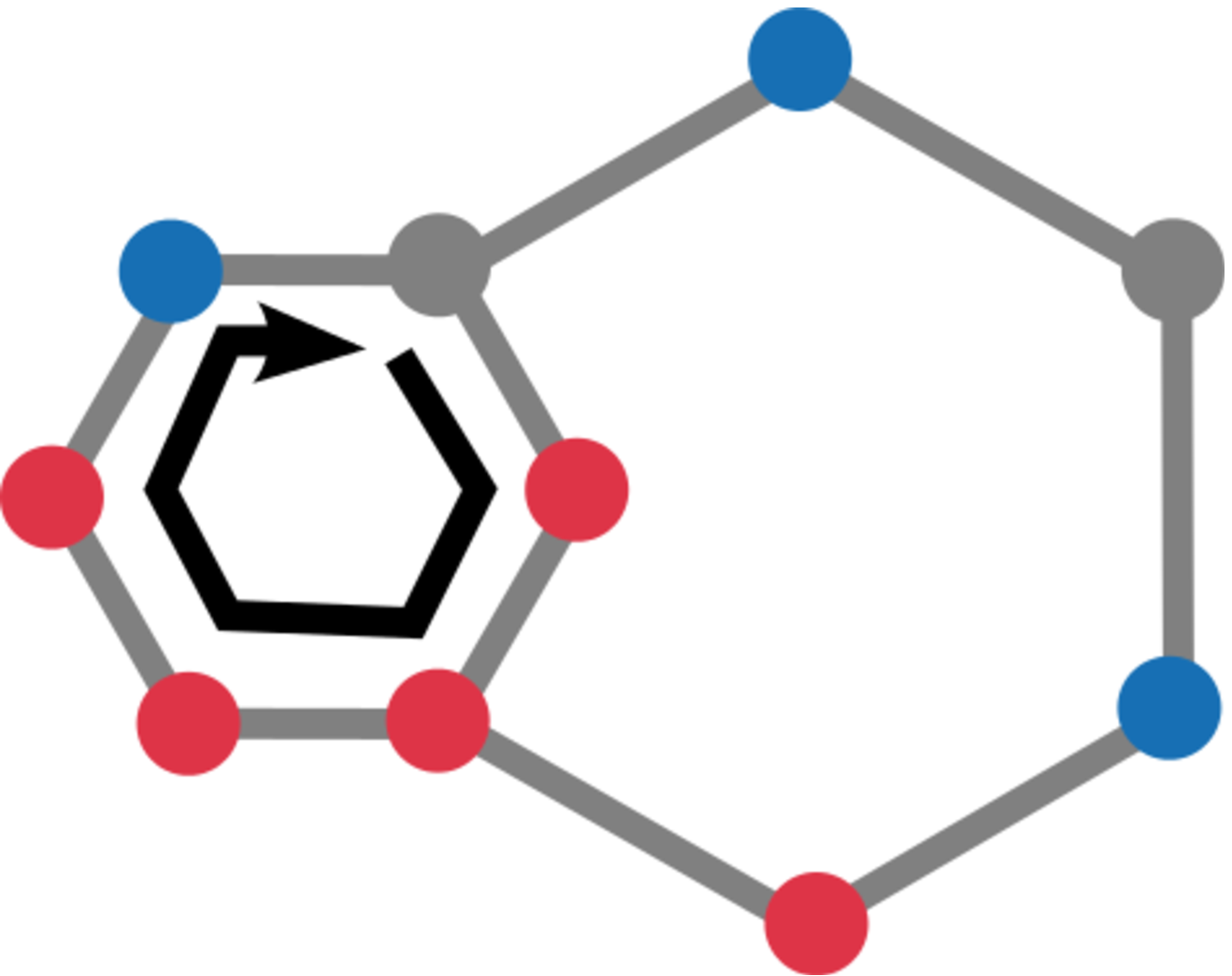}}\quad\quad
\subfigure[\label{Fig23}]{\includegraphics[width=0.12\textwidth]{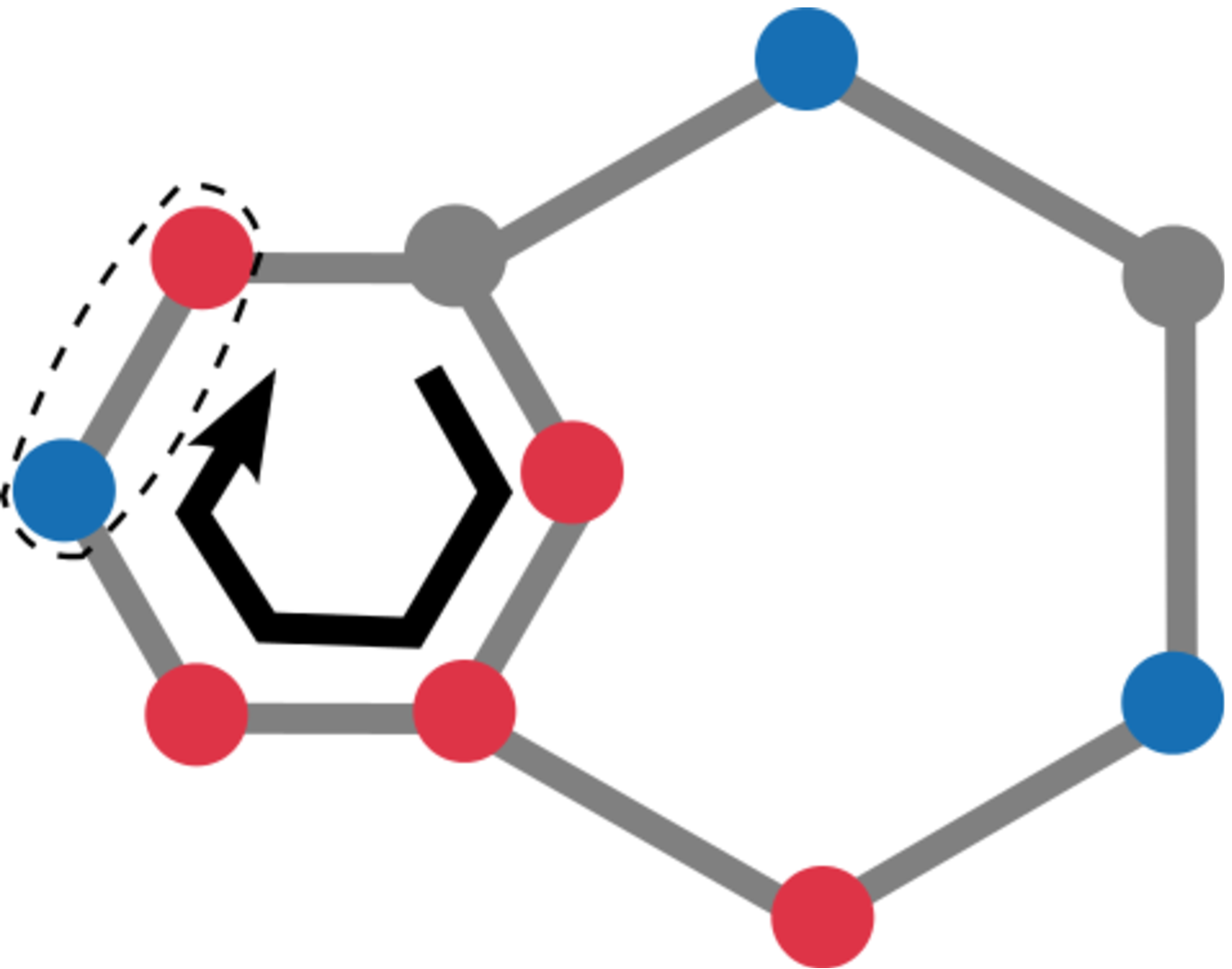}}\quad\quad
\subfigure[\label{Fig24}]{\includegraphics[width=0.12\textwidth]{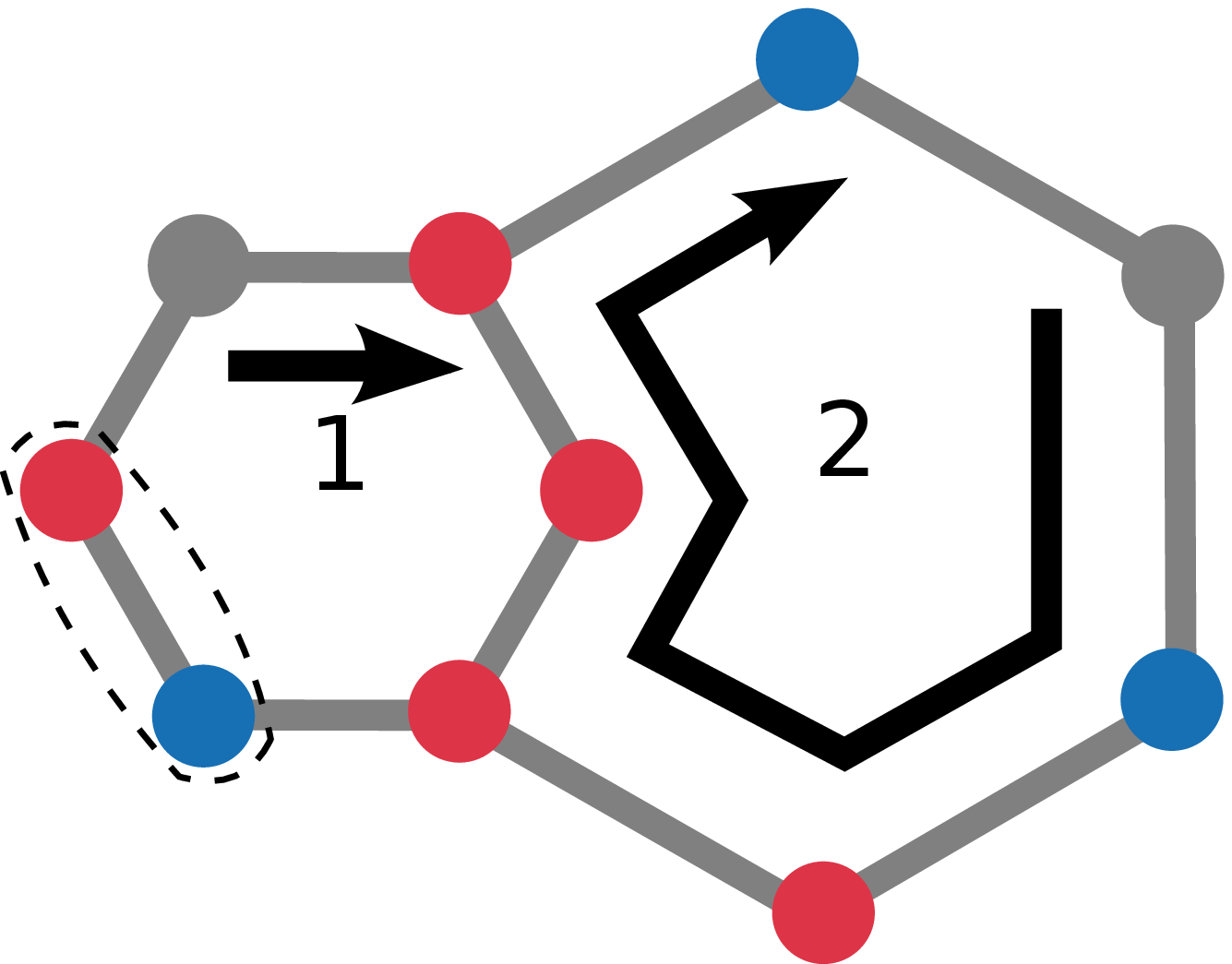}}\\
\subfigure[\label{Fig25}]{\includegraphics[width=0.12\textwidth]{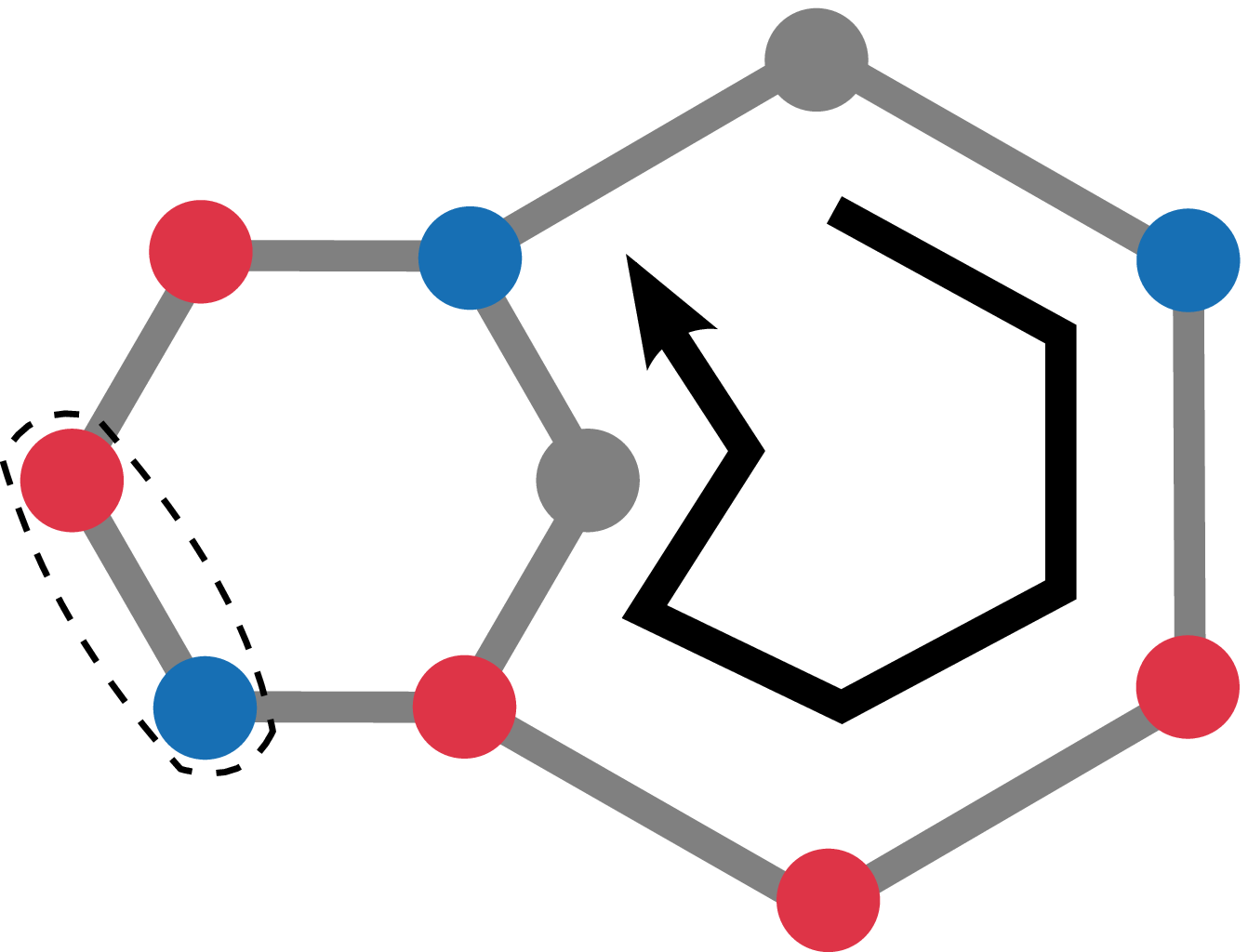}}\quad\quad
\subfigure[\label{Fig26}]{\includegraphics[width=0.12\textwidth]{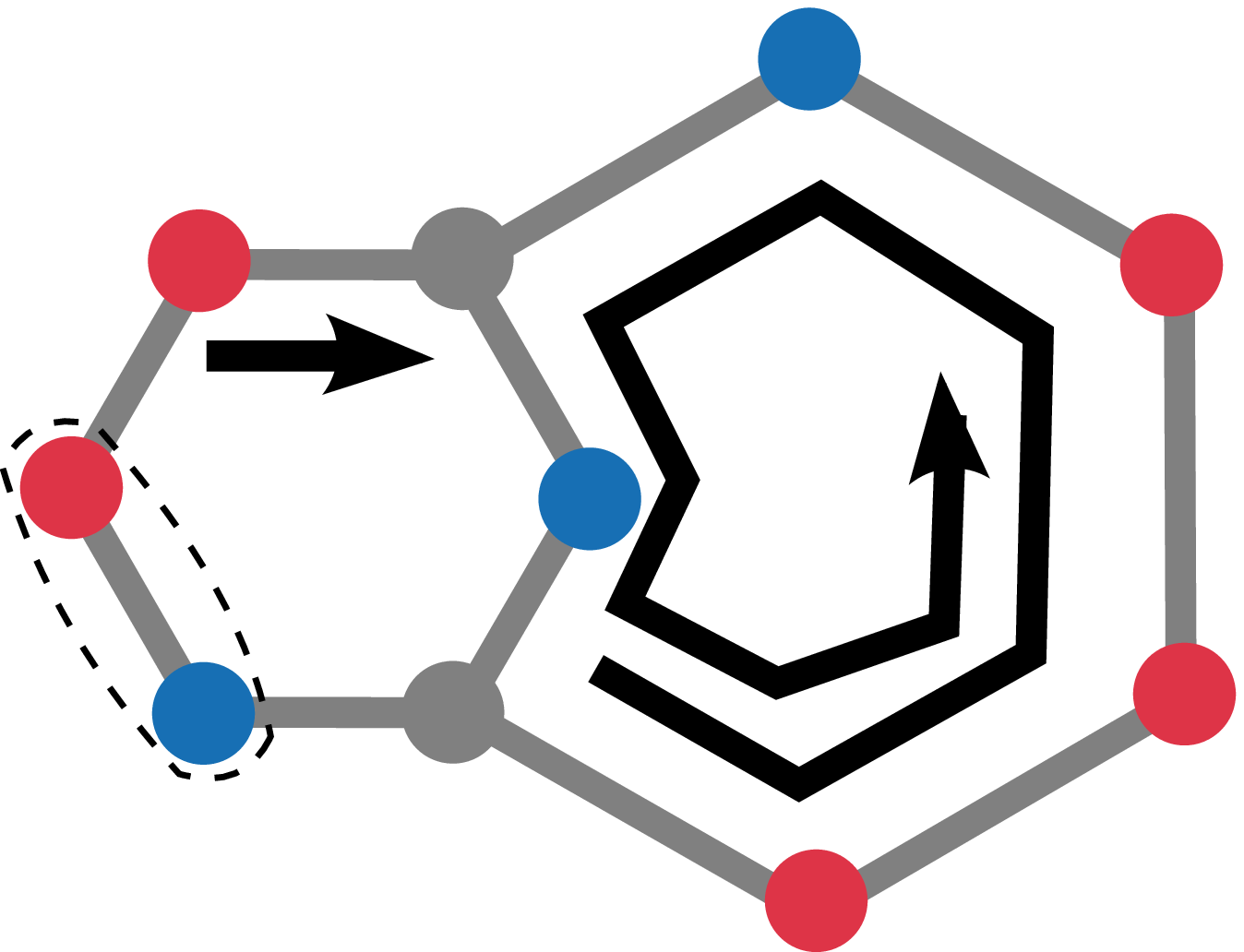}}\quad\quad
\subfigure[\label{Fig27}]{\includegraphics[width=0.12\textwidth]{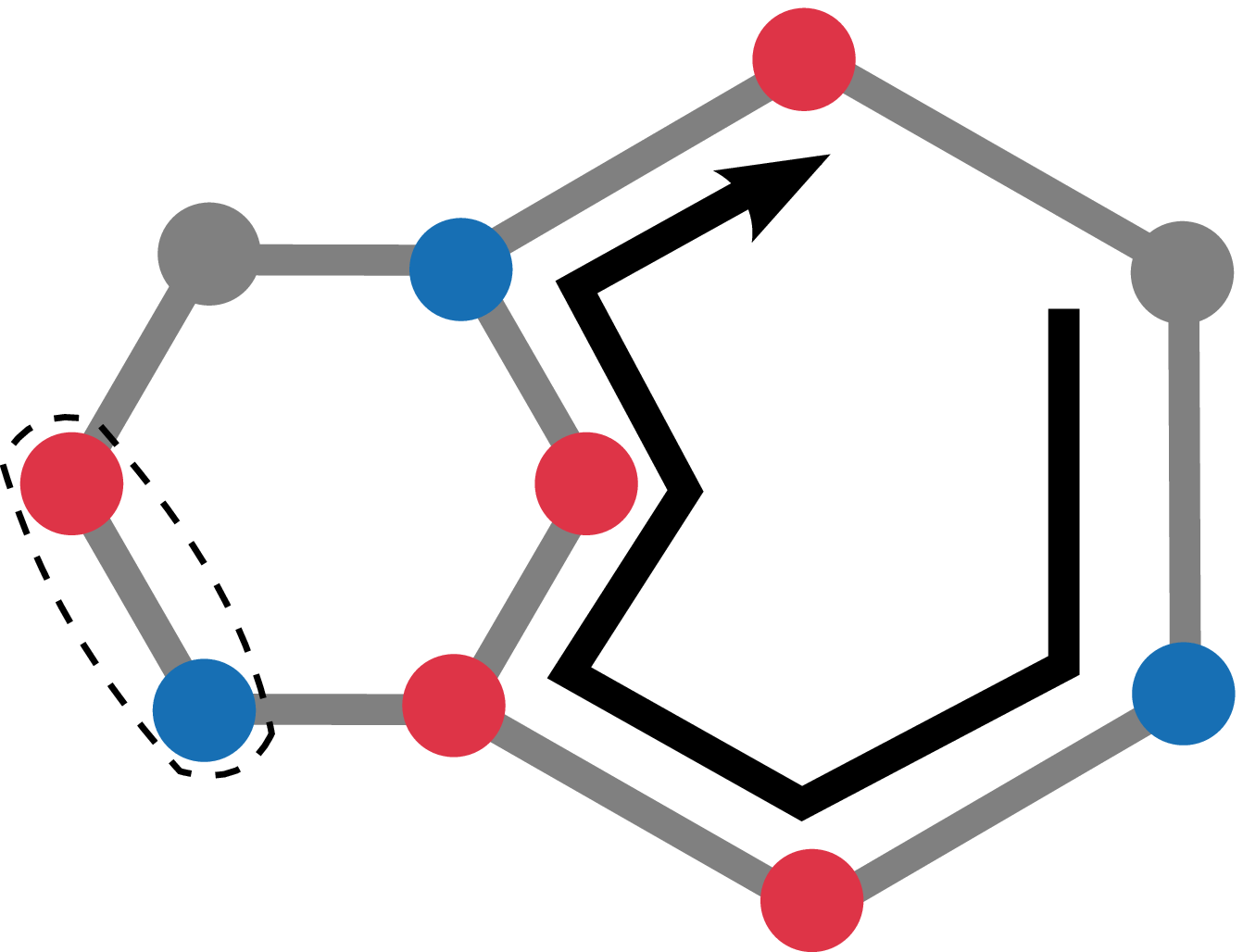}}\quad\quad
\subfigure[\label{Fig28}]{\includegraphics[width=0.12\textwidth]{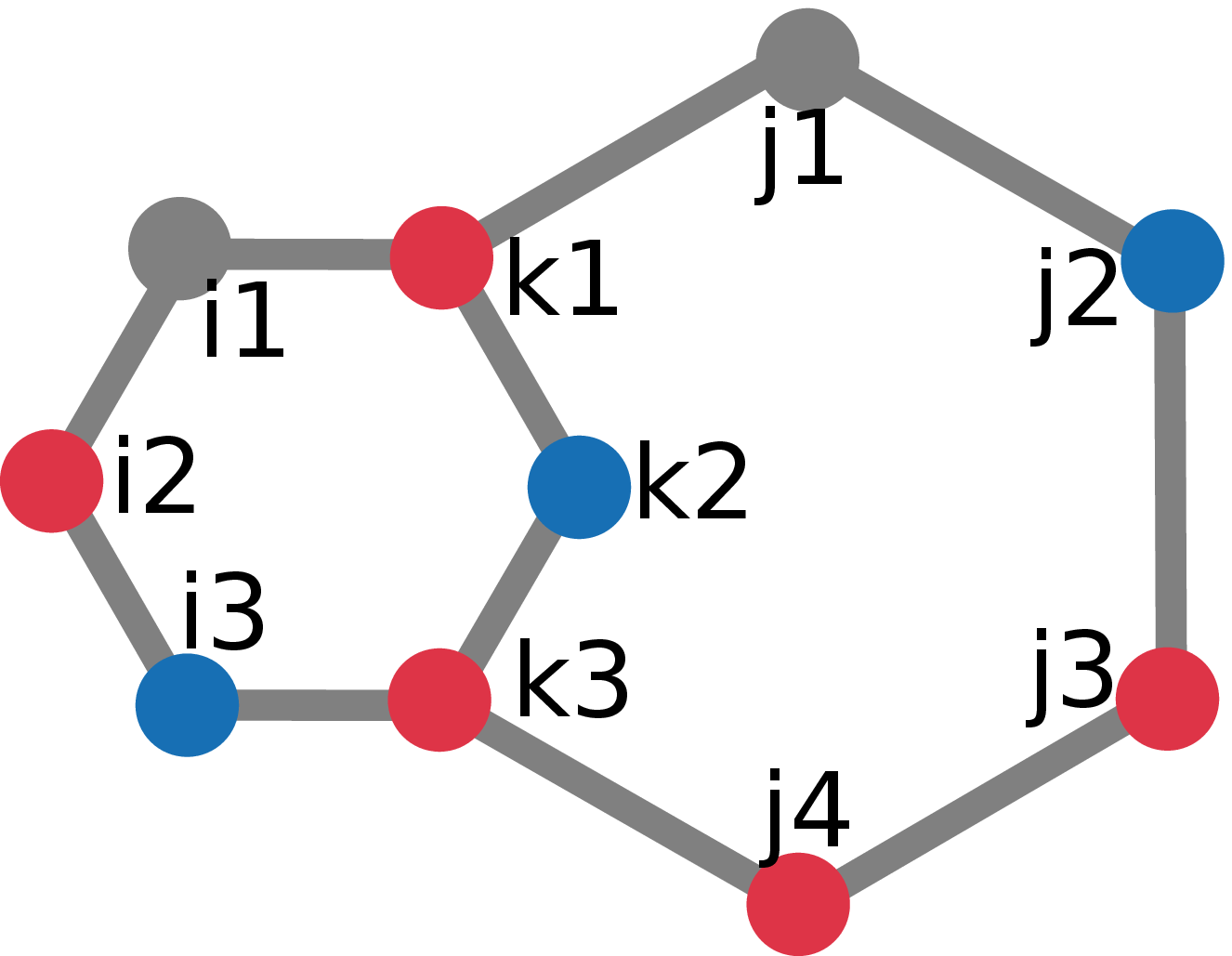}}
\caption{\label{FigFive}\ref{Fig20} through \ref{Fig28} represents states in $O_g$ for the case of a lattice consisting of a circle with one added path linking to unbonded sites. In this example $A=3,B=5$. States \ref{Fig20} and \ref{Fig28} are connected through the sequence \ref{Fig21}-\ref{Fig27}. Black arrows indicate how the grey color is moved at each step. Dashed circle shows the order of the color which needs to be keep fixed (see text).}
\end{figure}

For simplicity, but without loss of generality, we prove the proposition for the specific example shown in Fig.~\ref{FigFive}. The general case only differs in the number of sites on the circle and the extra path connecting the two sites which are unbounded in the original circle, and in the color of sites. 

Let us consider the lattice shown in Fig.~\ref{Fig28}. 
We denote by $i_l$, $j_l$, $k_1$, and $k_3$ the sites belonging to the original circle. 
Sites on the extra path connecting the initially unbounded sites $k_1$ and $k_3$ 
are denoted by $k_l$ (in this case we only have $k_2$). This path separates the 
original circle into the left and right circles. Let us now consider arbitrary 
$\ket{\sigma,\lambda}$ and $\ket{\sigma',\lambda'}$ displayed by, e.g., 
Fig.~\ref{Fig20} and \ref{Fig28} respectively. Because the grey color 
can be moved to any site of the lattice according to the rules given in
Subsection~\ref{subsection4_1}, we choose $\ket{\sigma',\lambda'}$ such that 
both, left and right circles, have one grey site. Note that the proof below 
does not depend on the number of grey sites.

The proof is based on the fact that we can first construct a generic state 
connected with $\ket{\sigma,\lambda}$ and such that the order of color on $i$-sites 
is the same as in $\ket{\sigma',\lambda'}$. In our particular example, because one 
of the $i$ sites is grey, this reduces to fixing the color on the bond specified by 
the dashed line in Figs.~\ref{Fig23}-\ref{Fig27}. In order to do so, we first construct 
a state connected to $\ket{\sigma,\lambda}$ where $k_1$ is blue, as shown in Fig.~\ref{Fig21}.
This process is depicted in Fig.~\ref{Fig20} by black arrows indicating how the 
grey color moves. This process is always possible due to the fact that $\mathbf{G}$ 
is 2-connected (see a similar argument given to prove Proposition.~\ref{Prps4}).
We keep moving the grey color as depicted by black arrows in Fig.~\ref{Fig21}-\ref{Fig23}, 
in order to construct the sequence Fig.~\ref{Fig22}-\ref{Fig24}. We have finally 
constructed a state such that the order of the color on $i$ sites is the same as 
$\ket{\sigma',\lambda'}$. Similar procedures can be followed if the color of more
than two $i$ sites need to be fixed.

Next, we need to fix the order of color on the right circle. In a general case, this is 
equivalent to switching the order of color on a certain number of bonds. In our case, 
we only need to do so for the color on bond $\{j_1,k_1\}$ of state Fig.~\ref{Fig24}. The procedure is depicted by 
black arrows in \ref{Fig24}-\ref{Fig26}, so that we end up with state Fig.~\ref{Fig27}. 
The idea of the procedure is to transfer the pair of colors on bond $\{j_1,k_1\}$ to bond 
$\{k_1,i_1\}$ (see Fig.~\ref{Fig25}) and then move grey sites in order to transfer the pair 
of colors back to the original bond $\{j_1,k_1\}$ (see Fig.~\ref{Fig27}) but with the order of the color inverted. 
In general, this procedure will ensure that the order of the color on $\{j_1,k_1\}$ is inverted. 
Now the order of color on the right circle is the same as in $\ket{\sigma',\lambda'}$. 
The last step consists of moving the grey color (which does not change the order of color) 
on the right circle in order to reach the state $\ket{\sigma',\lambda'}$. This is depicted 
by the black arrow in Fig.~\ref{Fig27}.

In a more general case that the one described here, one simply needs to repeat 
similar procedures to switch the color on bonds on the right circle as needed.

\bibliography{jun132014}

\begin{thebibliography}{42}%
\makeatletter
\providecommand \@ifxundefined [1]{%
 \@ifx{#1\undefined}
}%
\providecommand \@ifnum [1]{%
 \ifnum #1\expandafter \@firstoftwo
 \else \expandafter \@secondoftwo
 \fi
}%
\providecommand \@ifx [1]{%
 \ifx #1\expandafter \@firstoftwo
 \else \expandafter \@secondoftwo
 \fi
}%
\providecommand \natexlab [1]{#1}%
\providecommand \enquote  [1]{``#1''}%
\providecommand \bibnamefont  [1]{#1}%
\providecommand \bibfnamefont [1]{#1}%
\providecommand \citenamefont [1]{#1}%
\providecommand \href@noop [0]{\@secondoftwo}%
\providecommand \href [0]{\begingroup \@sanitize@url \@href}%
\providecommand \@href[1]{\@@startlink{#1}\@@href}%
\providecommand \@@href[1]{\endgroup#1\@@endlink}%
\providecommand \@sanitize@url [0]{\catcode `\\12\catcode `\$12\catcode
  `\&12\catcode `\#12\catcode `\^12\catcode `\_12\catcode `\%12\relax}%
\providecommand \@@startlink[1]{}%
\providecommand \@@endlink[0]{}%
\providecommand \url  [0]{\begingroup\@sanitize@url \@url }%
\providecommand \@url [1]{\endgroup\@href {#1}{\urlprefix }}%
\providecommand \urlprefix  [0]{URL }%
\providecommand \Eprint [0]{\href }%
\providecommand \doibase [0]{http://dx.doi.org/}%
\providecommand \selectlanguage [0]{\@gobble}%
\providecommand \bibinfo  [0]{\@secondoftwo}%
\providecommand \bibfield  [0]{\@secondoftwo}%
\providecommand \translation [1]{[#1]}%
\providecommand \BibitemOpen [0]{}%
\providecommand \bibitemStop [0]{}%
\providecommand \bibitemNoStop [0]{.\EOS\space}%
\providecommand \EOS [0]{\spacefactor3000\relax}%
\providecommand \BibitemShut  [1]{\csname bibitem#1\endcsname}%
\let\auto@bib@innerbib\@empty
\bibitem [{\citenamefont {Altman}\ \emph {et~al.}(2003)\citenamefont {Altman},
  \citenamefont {Hofstetter}, \citenamefont {Demler},\ and\ \citenamefont
  {Lukin}}]{quantummagneticphase2}%
  \BibitemOpen
  \bibfield  {author} {\bibinfo {author} {\bibfnamefont {E.}~\bibnamefont
  {Altman}}, \bibinfo {author} {\bibfnamefont {W.}~\bibnamefont {Hofstetter}},
  \bibinfo {author} {\bibfnamefont {E.}~\bibnamefont {Demler}}, \ and\ \bibinfo
  {author} {\bibfnamefont {M.~D.}\ \bibnamefont {Lukin}},\ }\href {<Go to
  ISI>://WOS:000185630900001} {\bibfield  {journal} {\bibinfo  {journal} {New
  J. Phys.}\ }\textbf {\bibinfo {volume} {5}},\ \bibinfo {pages} {113}
  (\bibinfo {year} {2003})}\BibitemShut {NoStop}%
\bibitem [{\citenamefont {Catani}\ \emph {et~al.}(2008)\citenamefont {Catani},
  \citenamefont {De~Sarlo}, \citenamefont {Barontini}, \citenamefont
  {Minardi},\ and\ \citenamefont {Inguscio}}]{bosonbosonexperiment1}%
  \BibitemOpen
  \bibfield  {author} {\bibinfo {author} {\bibfnamefont {J.}~\bibnamefont
  {Catani}}, \bibinfo {author} {\bibfnamefont {L.}~\bibnamefont {De~Sarlo}},
  \bibinfo {author} {\bibfnamefont {G.}~\bibnamefont {Barontini}}, \bibinfo
  {author} {\bibfnamefont {F.}~\bibnamefont {Minardi}}, \ and\ \bibinfo
  {author} {\bibfnamefont {M.}~\bibnamefont {Inguscio}},\ }\href {<Go to
  ISI>://WOS:000252862000011} {\bibfield  {journal} {\bibinfo  {journal} {Phys.
  Rev. A}\ }\textbf {\bibinfo {volume} {77}},\ \bibinfo {pages} {011603}
  (\bibinfo {year} {2008})}\BibitemShut {NoStop}%
\bibitem [{\citenamefont {Thalhammer}\ \emph {et~al.}(2008)\citenamefont
  {Thalhammer}, \citenamefont {Barontini}, \citenamefont {De~Sarlo},
  \citenamefont {Catani}, \citenamefont {Minardi},\ and\ \citenamefont
  {Inguscio}}]{bosonbosonexperiment2}%
  \BibitemOpen
  \bibfield  {author} {\bibinfo {author} {\bibfnamefont {G.}~\bibnamefont
  {Thalhammer}}, \bibinfo {author} {\bibfnamefont {G.}~\bibnamefont
  {Barontini}}, \bibinfo {author} {\bibfnamefont {L.}~\bibnamefont {De~Sarlo}},
  \bibinfo {author} {\bibfnamefont {J.}~\bibnamefont {Catani}}, \bibinfo
  {author} {\bibfnamefont {F.}~\bibnamefont {Minardi}}, \ and\ \bibinfo
  {author} {\bibfnamefont {M.}~\bibnamefont {Inguscio}},\ }\href {<Go to
  ISI>://WOS:000256585500002} {\bibfield  {journal} {\bibinfo  {journal} {Phys.
  Rev. Lett.}\ }\textbf {\bibinfo {volume} {100}},\ \bibinfo {pages} {210402}
  (\bibinfo {year} {2008})}\BibitemShut {NoStop}%
\bibitem [{\citenamefont {Gadway}\ \emph {et~al.}(2010)\citenamefont {Gadway},
  \citenamefont {Pertot}, \citenamefont {Reimann},\ and\ \citenamefont
  {Schneble}}]{bosonbosonexperiment3}%
  \BibitemOpen
  \bibfield  {author} {\bibinfo {author} {\bibfnamefont {B.}~\bibnamefont
  {Gadway}}, \bibinfo {author} {\bibfnamefont {D.}~\bibnamefont {Pertot}},
  \bibinfo {author} {\bibfnamefont {R.}~\bibnamefont {Reimann}}, \ and\
  \bibinfo {author} {\bibfnamefont {D.}~\bibnamefont {Schneble}},\ }\href {<Go
  to ISI>://WOS:000280234400015} {\bibfield  {journal} {\bibinfo  {journal}
  {Phys. Rev. Lett.}\ }\textbf {\bibinfo {volume} {105}},\ \bibinfo {pages}
  {045303} (\bibinfo {year} {2010})}\BibitemShut {NoStop}%
\bibitem [{\citenamefont {Iskin}(2010)}]{Iskin1}%
  \BibitemOpen
  \bibfield  {author} {\bibinfo {author} {\bibfnamefont {M.}~\bibnamefont
  {Iskin}},\ }\href {http://link.aps.org/doi/10.1103/PhysRevA.82.033630}
  {\bibfield  {journal} {\bibinfo  {journal} {Phys. Rev. A}\ }\textbf {\bibinfo
  {volume} {82}},\ \bibinfo {pages} {033630} (\bibinfo {year}
  {2010})}\BibitemShut {NoStop}%
\bibitem [{\citenamefont {Kuklov}\ and\ \citenamefont
  {Svistunov}(2003)}]{Kuklov1}%
  \BibitemOpen
  \bibfield  {author} {\bibinfo {author} {\bibfnamefont {A.~B.}\ \bibnamefont
  {Kuklov}}\ and\ \bibinfo {author} {\bibfnamefont {B.~V.}\ \bibnamefont
  {Svistunov}},\ }\href {http://link.aps.org/doi/10.1103/PhysRevLett.90.100401
  http://prl.aps.org/abstract/PRL/v90/i10/e100401} {\bibfield  {journal}
  {\bibinfo  {journal} {Phys. Rev. Lett.}\ }\textbf {\bibinfo {volume} {90}},\
  \bibinfo {pages} {100401} (\bibinfo {year} {2003})}\BibitemShut {NoStop}%
\bibitem [{\citenamefont {Kuklov}\ \emph {et~al.}(2004)\citenamefont {Kuklov},
  \citenamefont {Prokof'ev},\ and\ \citenamefont {Svistunov}}]{Kuklov2}%
  \BibitemOpen
  \bibfield  {author} {\bibinfo {author} {\bibfnamefont {A.}~\bibnamefont
  {Kuklov}}, \bibinfo {author} {\bibfnamefont {N.}~\bibnamefont {Prokof'ev}}, \
  and\ \bibinfo {author} {\bibfnamefont {B.}~\bibnamefont {Svistunov}},\ }\href
  {<Go to ISI>://WOS:000188785200002} {\bibfield  {journal} {\bibinfo
  {journal} {Phys. Rev. Lett.}\ }\textbf {\bibinfo {volume} {92}},\ \bibinfo
  {pages} {050402} (\bibinfo {year} {2004})}\BibitemShut {NoStop}%
\bibitem [{\citenamefont {Isacsson}\ \emph {et~al.}(2005)\citenamefont
  {Isacsson}, \citenamefont {Cha}, \citenamefont {Sengupta},\ and\
  \citenamefont {Girvin}}]{Isacsson1}%
  \BibitemOpen
  \bibfield  {author} {\bibinfo {author} {\bibfnamefont {A.}~\bibnamefont
  {Isacsson}}, \bibinfo {author} {\bibfnamefont {M.~C.}\ \bibnamefont {Cha}},
  \bibinfo {author} {\bibfnamefont {K.}~\bibnamefont {Sengupta}}, \ and\
  \bibinfo {author} {\bibfnamefont {S.~M.}\ \bibnamefont {Girvin}},\ }\href
  {<Go to ISI>://WOS:000233603600071} {\bibfield  {journal} {\bibinfo
  {journal} {Phys. Rev. B}\ }\textbf {\bibinfo {volume} {72}},\ \bibinfo
  {pages} {184507} (\bibinfo {year} {2005})}\BibitemShut {NoStop}%
\bibitem [{\citenamefont {Pai}\ \emph {et~al.}(2012)\citenamefont {Pai},
  \citenamefont {Kurdestany}, \citenamefont {Sheshadri},\ and\ \citenamefont
  {Pandit}}]{Pai1}%
  \BibitemOpen
  \bibfield  {author} {\bibinfo {author} {\bibfnamefont {R.~V.}\ \bibnamefont
  {Pai}}, \bibinfo {author} {\bibfnamefont {J.~M.}\ \bibnamefont {Kurdestany}},
  \bibinfo {author} {\bibfnamefont {K.}~\bibnamefont {Sheshadri}}, \ and\
  \bibinfo {author} {\bibfnamefont {R.}~\bibnamefont {Pandit}},\ }\href
  {http://link.aps.org/doi/10.1103/PhysRevB.85.214524} {\bibfield  {journal}
  {\bibinfo  {journal} {Phys. Rev. B}\ }\textbf {\bibinfo {volume} {85}},\
  \bibinfo {pages} {214524} (\bibinfo {year} {2012})}\BibitemShut {NoStop}%
\bibitem [{\citenamefont {Ozaki}\ \emph {et~al.}()\citenamefont {Ozaki},
  \citenamefont {Danshita},\ and\ \citenamefont {Nikuni}}]{Ozaki1}%
  \BibitemOpen
  \bibfield  {author} {\bibinfo {author} {\bibfnamefont {T.}~\bibnamefont
  {Ozaki}}, \bibinfo {author} {\bibfnamefont {I.}~\bibnamefont {Danshita}}, \
  and\ \bibinfo {author} {\bibfnamefont {T.}~\bibnamefont {Nikuni}},\ }\href
  {http://arxiv.org/abs/1210.1370} {\bibinfo  {journal} {arXiv:1210.1370}\
  }\BibitemShut {NoStop}%
\bibitem [{\citenamefont {Nakano}\ \emph {et~al.}(2012)\citenamefont {Nakano},
  \citenamefont {Ishima}, \citenamefont {Kobayashi}, \citenamefont {Yamamoto},
  \citenamefont {Ichinose},\ and\ \citenamefont {Matsui}}]{Nakano1}%
  \BibitemOpen
\bibfield  {journal} {  }\bibfield  {author} {\bibinfo {author} {\bibfnamefont
  {Y.}~\bibnamefont {Nakano}}, \bibinfo {author} {\bibfnamefont
  {T.}~\bibnamefont {Ishima}}, \bibinfo {author} {\bibfnamefont
  {N.}~\bibnamefont {Kobayashi}}, \bibinfo {author} {\bibfnamefont
  {T.}~\bibnamefont {Yamamoto}}, \bibinfo {author} {\bibfnamefont
  {I.}~\bibnamefont {Ichinose}}, \ and\ \bibinfo {author} {\bibfnamefont
  {T.}~\bibnamefont {Matsui}},\ }\href
  {http://link.aps.org/doi/10.1103/PhysRevA.85.023617} {\bibfield  {journal}
  {\bibinfo  {journal} {Phys. Rev. A}\ }\textbf {\bibinfo {volume} {85}},\
  \bibinfo {pages} {023617} (\bibinfo {year} {2012})}\BibitemShut {NoStop}%
\bibitem [{\citenamefont {Capogrosso-Sansone}\ \emph
  {et~al.}(2011)\citenamefont {Capogrosso-Sansone}, \citenamefont
  {Guglielmino},\ and\ \citenamefont {Penna}}]{Barbara1}%
  \BibitemOpen
  \bibfield  {author} {\bibinfo {author} {\bibfnamefont {B.}~\bibnamefont
  {Capogrosso-Sansone}}, \bibinfo {author} {\bibfnamefont {M.}~\bibnamefont
  {Guglielmino}}, \ and\ \bibinfo {author} {\bibfnamefont {V.}~\bibnamefont
  {Penna}},\ }\href {http://dx.doi.org/10.1134/S1054660X11150023
  http://download.springer.com/static/pdf/814/art%253A10.1134%252FS1054660X11150023.pdf?auth66=1379625905_42801338b90427a6016e7529b9bef292&ext=.pdf}
  {\bibfield  {journal} {\bibinfo  {journal} {Laser Phys.}\ }\textbf {\bibinfo
  {volume} {21}},\ \bibinfo {pages} {1443} (\bibinfo {year}
  {2011})}\BibitemShut {NoStop}%
\bibitem [{\citenamefont {Guglielmino}\ \emph {et~al.}(2010)\citenamefont
  {Guglielmino}, \citenamefont {Penna},\ and\ \citenamefont
  {Capogrosso-Sansone}}]{Barbara2}%
  \BibitemOpen
  \bibfield  {author} {\bibinfo {author} {\bibfnamefont {M.}~\bibnamefont
  {Guglielmino}}, \bibinfo {author} {\bibfnamefont {V.}~\bibnamefont {Penna}},
  \ and\ \bibinfo {author} {\bibfnamefont {B.}~\bibnamefont
  {Capogrosso-Sansone}},\ }\href {<Go to ISI>://WOS:000280575300001} {\bibfield
   {journal} {\bibinfo  {journal} {Phys. Rev. A}\ }\textbf {\bibinfo {volume}
  {82}},\ \bibinfo {pages} {021601} (\bibinfo {year} {2010})}\BibitemShut
  {NoStop}%
\bibitem [{\citenamefont {Fisher}\ \emph {et~al.}(1989)\citenamefont {Fisher},
  \citenamefont {Weichman}, \citenamefont {Grinstein},\ and\ \citenamefont
  {Fisher}}]{Fisher1}%
  \BibitemOpen
  \bibfield  {author} {\bibinfo {author} {\bibfnamefont {M.~P.~A.}\
  \bibnamefont {Fisher}}, \bibinfo {author} {\bibfnamefont {P.~B.}\
  \bibnamefont {Weichman}}, \bibinfo {author} {\bibfnamefont {G.}~\bibnamefont
  {Grinstein}}, \ and\ \bibinfo {author} {\bibfnamefont {D.~S.}\ \bibnamefont
  {Fisher}},\ }\href {http://link.aps.org/doi/10.1103/PhysRevB.40.546}
  {\bibfield  {journal} {\bibinfo  {journal} {Phys. Rev. B}\ }\textbf {\bibinfo
  {volume} {40}},\ \bibinfo {pages} {546} (\bibinfo {year} {1989})}\BibitemShut
  {NoStop}%
\bibitem [{\citenamefont {Jaksch}\ \emph {et~al.}(1998)\citenamefont {Jaksch},
  \citenamefont {Bruder}, \citenamefont {Cirac}, \citenamefont {Gardiner},\
  and\ \citenamefont {Zoller}}]{Jaksch1}%
  \BibitemOpen
  \bibfield  {author} {\bibinfo {author} {\bibfnamefont {D.}~\bibnamefont
  {Jaksch}}, \bibinfo {author} {\bibfnamefont {C.}~\bibnamefont {Bruder}},
  \bibinfo {author} {\bibfnamefont {J.~I.}\ \bibnamefont {Cirac}}, \bibinfo
  {author} {\bibfnamefont {C.~W.}\ \bibnamefont {Gardiner}}, \ and\ \bibinfo
  {author} {\bibfnamefont {P.}~\bibnamefont {Zoller}},\ }\href
  {http://link.aps.org/doi/10.1103/PhysRevLett.81.3108} {\bibfield  {journal}
  {\bibinfo  {journal} {Phys. Rev. Lett.}\ }\textbf {\bibinfo {volume} {81}},\
  \bibinfo {pages} {3108} (\bibinfo {year} {1998})}\BibitemShut {NoStop}%
\bibitem [{\citenamefont {Wang}\ \emph {et~al.}()\citenamefont {Wang},
  \citenamefont {Penna},\ and\ \citenamefont
  {Capogrosso-Sansone}}]{inprogress}%
  \BibitemOpen
  \bibfield  {author} {\bibinfo {author} {\bibfnamefont {W.}~\bibnamefont
  {Wang}}, \bibinfo {author} {\bibfnamefont {V.}~\bibnamefont {Penna}}, \ and\
  \bibinfo {author} {\bibfnamefont {B.}~\bibnamefont {Capogrosso-Sansone}},\
  }\href@noop {} {\ }\bibinfo {note} {In progress}\BibitemShut {NoStop}%
\bibitem [{\citenamefont {Tasaki}(1998)}]{Tasaki1}%
  \BibitemOpen
  \bibfield  {author} {\bibinfo {author} {\bibfnamefont {H.}~\bibnamefont
  {Tasaki}},\ }\href {<Go to ISI>://WOS:000073659300001} {\bibfield  {journal}
  {\bibinfo  {journal} {Progr. Theoret. Phys.}\ }\textbf {\bibinfo {volume}
  {99}},\ \bibinfo {pages} {489} (\bibinfo {year} {1998})}\BibitemShut
  {NoStop}%
\bibitem [{\citenamefont {Katsura}\ and\ \citenamefont
  {Tasaki}(2013)}]{Katsura1}%
  \BibitemOpen
  \bibfield  {author} {\bibinfo {author} {\bibfnamefont {H.}~\bibnamefont
  {Katsura}}\ and\ \bibinfo {author} {\bibfnamefont {H.}~\bibnamefont
  {Tasaki}},\ }\href {http://link.aps.org/doi/10.1103/PhysRevLett.110.130405
  http://prl.aps.org/pdf/PRL/v110/i13/e130405} {\bibfield  {journal} {\bibinfo
  {journal} {Phys. Rev. Lett.}\ }\textbf {\bibinfo {volume} {110}},\ \bibinfo
  {pages} {130405} (\bibinfo {year} {2013})}\BibitemShut {NoStop}%
\bibitem [{\citenamefont {Nagaoka}(1966)}]{Nagaoka1}%
  \BibitemOpen
  \bibfield  {author} {\bibinfo {author} {\bibfnamefont {Y.}~\bibnamefont
  {Nagaoka}},\ }\href {<Go to ISI>://WOS:A19667956200054} {\bibfield  {journal}
  {\bibinfo  {journal} {Phys. Rev.}\ }\textbf {\bibinfo {volume} {147}},\
  \bibinfo {pages} {392} (\bibinfo {year} {1966})}\BibitemShut {NoStop}%
\bibitem [{\citenamefont {Thouless}(1965)}]{Thouless1}%
  \BibitemOpen
  \bibfield  {author} {\bibinfo {author} {\bibfnamefont {D.~J.}\ \bibnamefont
  {Thouless}},\ }\href {<Go to ISI>://WOS:A19657040100001} {\bibfield
  {journal} {\bibinfo  {journal} {Proc. Phys. Soc. London}\ }\textbf {\bibinfo
  {volume} {86}},\ \bibinfo {pages} {893} (\bibinfo {year} {1965})}\BibitemShut
  {NoStop}%
\bibitem [{\citenamefont {Tasaki}(1989)}]{Tasaki2}%
  \BibitemOpen
  \bibfield  {author} {\bibinfo {author} {\bibfnamefont {H.}~\bibnamefont
  {Tasaki}},\ }\href {<Go to ISI>://WOS:A1989AX80800073} {\bibfield  {journal}
  {\bibinfo  {journal} {Phys. Rev. B}\ }\textbf {\bibinfo {volume} {40}},\
  \bibinfo {pages} {9192} (\bibinfo {year} {1989})}\BibitemShut {NoStop}%
\bibitem [{\citenamefont {Wang}\ and\ \citenamefont
  {Capogrosso-Sansone}()}]{inprogress1}%
  \BibitemOpen
  \bibfield  {author} {\bibinfo {author} {\bibfnamefont {W.}~\bibnamefont
  {Wang}}\ and\ \bibinfo {author} {\bibfnamefont {B.}~\bibnamefont
  {Capogrosso-Sansone}},\ }\href@noop {} {\ }\bibinfo {note} {In
  progress}\BibitemShut {NoStop}%
\bibitem [{Note1()}]{Note1}%
  \BibitemOpen
  \bibinfo {note} {Particle number operators commute with both $H_0$ and
  $H$.}\BibitemShut {Stop}%
\bibitem [{Note2()}]{Note2}%
  \BibitemOpen
  \bibinfo {note} {A linear operator $X$ is symmetric if, for arbitrary states
  $\mathinner {|{\phi }\delimiter "526930B }$ and $\mathinner {|{\psi
  }\delimiter "526930B }$, $\mathinner {\delimiter "426830A {\phi |X|\psi
  }\delimiter "526930B }=\mathinner {\delimiter "426830A {\psi |X|\phi
  }\delimiter "526930B }$.}\BibitemShut {Stop}%
\bibitem [{Note3()}]{Note3}%
  \BibitemOpen
  \bibinfo {note} {A relation $\protect \mathfrak {R}_X$ on a set $O$ is a
  collection of ordered pairs $(\mathinner {|{\xi ,\gamma }\delimiter "526930B
  },\mathinner {|{\xi ',\gamma '}\delimiter "526930B })$ in $O$. If
  $(\mathinner {|{\xi ,\gamma }\delimiter "526930B },\mathinner {|{\xi ',\gamma
  '}\delimiter "526930B })\in {O}$, we say $\mathinner {|{\xi ,\gamma
  }\delimiter "526930B }\protect \mathfrak {R}_X\mathinner {|{\xi ',\gamma
  '}\delimiter "526930B }$. $\protect \mathfrak {R}_X$ is an equivalence
  relation on a set $O$ if it satisfies \par (i) reflexivity, i.e. $\mathinner
  {|{\xi ,\gamma }\delimiter "526930B }\protect \mathfrak {R}_X\mathinner
  {|{\xi ,\gamma }\delimiter "526930B }$, (ii) symmetry, i.e. $\mathinner
  {|{\xi ,\gamma }\delimiter "526930B }\protect \mathfrak {R}_X\mathinner
  {|{\xi ',\gamma '}\delimiter "526930B }$ $\rightarrow $ $\mathinner {|{\xi
  ',\gamma '}\delimiter "526930B }\protect \mathfrak {R}_X\mathinner {|{\xi
  ,\gamma }\delimiter "526930B }$, (iii) transitivity, i.e. $\mathinner {|{\xi
  ,\gamma }\delimiter "526930B }\protect \mathfrak {R}_X\mathinner {|{\xi
  ',\gamma '}\delimiter "526930B }$, $\mathinner {|{\xi ',\gamma '}\delimiter
  "526930B }\protect \mathfrak {R}_X\mathinner {|{\xi '',\gamma ''}\delimiter
  "526930B }$ $\rightarrow $ $\mathinner {|{\xi ,\gamma }\delimiter "526930B
  }\protect \mathfrak {R}_X\mathinner {|{\xi '',\gamma ''}\delimiter "526930B
  }$. \par For a given relation $\protect \mathfrak {R}_X$, $\mathinner {|{\xi
  ,\gamma }\delimiter "526930B }/\protect \mathfrak {R}_{X}$ denotes the set of
  all $\mathinner {|{\xi ',\gamma '}\delimiter "526930B }$ related to
  $\mathinner {|{\xi ,\gamma }\delimiter "526930B }$, and $O/\protect \mathfrak
  {R}_{X}$ denotes the collection of all $\mathinner {|{\xi ,\gamma }\delimiter
  "526930B }/\protect \mathfrak {R}_{X}$'s. An important property of
  equivalence relations is that $O/\protect \mathfrak {R}_{X}$ is a partition
  of $O$~\cite {settheory}. It's easy to check that $\protect \mathfrak {R}_X$
  is well-defined here.}\BibitemShut {Stop}%
\bibitem [{Note4()}]{Note4}%
  \BibitemOpen
  \bibinfo {note} {A symmetric matrix is irreducible if and only if it cannot
  be block-diagonalized by permuting the indices.}\BibitemShut {Stop}%
\bibitem [{Note5()}]{Note5}%
  \BibitemOpen
  \bibinfo {note} {$\protect \mathbf {G}$ is connected if any two sites can be
  linked by a path. Two sites $i$ and $j$ are linked if there exists a path
  $\protect \{i,\protect \cdots ,k_l,\protect \cdots ,j\protect \}$ in which
  every neighboring pair in the sequence forms a bond.}\BibitemShut {Stop}%
\bibitem [{Note6()}]{Note6}%
  \BibitemOpen
  \bibinfo {note} {Every site in $K$ is not linked to any site in its
  complement.}\BibitemShut {Stop}%
\bibitem [{Note7()}]{Note7}%
  \BibitemOpen
  \bibinfo {note} {A vector is positive (in terms of the basis) if its
  expansion coefficients are all positive.}\BibitemShut {Stop}%
\bibitem [{Note8()}]{Note8}%
  \BibitemOpen
  \bibinfo {note} {These rules can also be stated formally: $\mathinner
  {\delimiter "426830A {\sigma ,\lambda |W_g|\sigma ',\lambda '}\delimiter
  "526930B }$ is nonzero if and only if either set $\lambda =\lambda '$ while
  sets $\sigma $, $\sigma '$ only differ by sites $i$ and $j$ belonging to the
  bond $\protect \{i,j\protect \}$, or set $\sigma =\sigma '$ while sets
  $\lambda $, $\lambda '$ only differ by sites $k$ and $l$ belonging to the
  bond $\protect \{{k,l}\protect \}$.}\BibitemShut {Stop}%
\bibitem [{Note9()}]{Note9}%
  \BibitemOpen
  \bibinfo {note} {Formally, it's a sequence of pairs of sets $\protect
  \{(\sigma ,\lambda ),\protect \cdots ,(\chi _l,\theta _l),\protect \cdots
  ,(\sigma ',\lambda ')\protect \}$.}\BibitemShut {Stop}%
\bibitem [{\citenamefont {Diestel}(2010)}]{Graph_Theory}%
  \BibitemOpen
  \bibfield  {author} {\bibinfo {author} {\bibfnamefont {R.}~\bibnamefont
  {Diestel}},\ }\href {http://books.google.com/books?id=KGYTSwAACAAJ} {\emph
  {\bibinfo {title} {Graph Theory}}},\ Graduate Texts in Mathematics\ (\bibinfo
   {publisher} {Springer},\ \bibinfo {year} {2010})\BibitemShut {NoStop}%
\bibitem [{Note10()}]{Note10}%
  \BibitemOpen
  \bibinfo {note} {A circle is a path where the two ends are the
  same.}\BibitemShut {Stop}%
\bibitem [{Note11()}]{Note11}%
  \BibitemOpen
  \bibinfo {note} {Note that in the remainder of this Subsection all added
  paths are non trivial in the sense that they add new sites to the already
  constructed lattice. Adding trivial bonds does not change our conclusions as
  it keeps the 2-connectivity properties of the lattice.}\BibitemShut {Stop}%
\bibitem [{Note12()}]{Note12}%
  \BibitemOpen
  \bibinfo {note} {Two states on a circle with $4$ or less sites are always
  connected, as shown in Lemma 4.6 of Tasaki's work~\cite {Tasaki1}. In
  Ref.~\cite {Tasaki1} the authors discuss the degeneracy problem of the
  ground-state energy of Fermi-Hubbard model with infinite $U$ at fixed number
  of spin-up(down) fermions and in the presence of a single hole. It is
  interesting noting that the basis $\mathinner {|{\sigma ,\lambda }\delimiter
  "526930B }$ that we define here is equivalent to the corresponding basis
  defined in Ref.~\cite {Tasaki1}.}\BibitemShut {Stop}%
\bibitem [{Note13()}]{Note13}%
  \BibitemOpen
  \bibinfo {note} {Nothing would change if more than two grey sites are present
  since the grey color can be freely moved on the lattice.}\BibitemShut {Stop}%
\bibitem [{Note14()}]{Note14}%
  \BibitemOpen
  \bibinfo {note} {If the two ends of any added path form a bond on the already
  constructed lattice as shown in Fig.~\ref {Fig39}, where pink indicates the
  already constructed lattice and green indicates the added path, then, since
  the path adds at least one new site, the ``new'' lattice we consider includes
  all sites as shown in Fig.~\ref {Fig40} by pink bonds with an added green
  trivial path. Now, in order to apply Proposition~\ref {Prps5}, we regard the
  green bond in Fig.~\ref {Fig40} as the added path.}\BibitemShut {Stop}%
\bibitem [{\citenamefont {Katsura}\ and\ \citenamefont
  {Tanaka}(2013)}]{Katsura2}%
  \BibitemOpen
  \bibfield  {author} {\bibinfo {author} {\bibfnamefont {H.}~\bibnamefont
  {Katsura}}\ and\ \bibinfo {author} {\bibfnamefont {A.}~\bibnamefont
  {Tanaka}},\ }\href {http://link.aps.org/doi/10.1103/PhysRevA.87.013617}
  {\bibfield  {journal} {\bibinfo  {journal} {Phys. Rev. A}\ }\textbf {\bibinfo
  {volume} {87}},\ \bibinfo {pages} {013617} (\bibinfo {year}
  {2013})}\BibitemShut {NoStop}%
\bibitem [{Note15()}]{Note15}%
  \BibitemOpen
  \bibinfo {note} {In~\cite {Katsura2} the authors study the degeneracy of the
  ground-state energy of the $SU(n)$ Fermi-Hubbard model with $U=\infty $ and
  with exactly one hole. Another sufficient condition for the $SU(2)$
  Fermi-Hubbard model requiring the lattice to be constructed by ``exchange
  bond'' was given in~\cite {Tasaki2}.}\BibitemShut {Stop}%
\bibitem [{Note16()}]{Note16}%
  \BibitemOpen
  \bibinfo {note} {A unitary operator is always bounded. Then, if an operator
  $S_r$ is bounded, $\left (\mathinner {|{\psi ^0}\delimiter "526930B }+\DOTSB
  \sum@ \slimits@ _{n=1}^\infty {\epsilon }^n\mathinner {|{\psi ^n}\delimiter
  "526930B }\right ) \rightarrow \mathinner {|{\Psi }\delimiter "526930B }$
  implies $\left (S_r\mathinner {|{\psi ^0}\delimiter "526930B }+\DOTSB \sum@
  \slimits@ _{n=1}^\infty {\epsilon }^n{S_r}\mathinner {|{\psi ^n}\delimiter
  "526930B }\right ) \rightarrow S_r\mathinner {|{\Psi }\delimiter "526930B
  }$.}\BibitemShut {Stop}%
\bibitem [{\citenamefont {Godsil}\ and\ \citenamefont
  {Royle}(2001)}]{Irreducible_Graph}%
  \BibitemOpen
  \bibfield  {author} {\bibinfo {author} {\bibfnamefont {C.}~\bibnamefont
  {Godsil}}\ and\ \bibinfo {author} {\bibfnamefont {G.~F.}\ \bibnamefont
  {Royle}},\ }\href {http://books.google.com/books?id=pYfJe-ZVUyAC} {\emph
  {\bibinfo {title} {Algebraic Graph Theory}}},\ Graduate Texts in Mathematics\
  (\bibinfo  {publisher} {Springer New York},\ \bibinfo {year}
  {2001})\BibitemShut {NoStop}%
\bibitem [{\citenamefont {Halmos}(1960)}]{settheory}%
  \BibitemOpen
  \bibfield  {author} {\bibinfo {author} {\bibfnamefont {P.~R.}\ \bibnamefont
  {Halmos}},\ }\href {http://books.google.com/books?id=x6cZBQ9qtgoC} {\emph
  {\bibinfo {title} {Naive Set Theory}}},\ Undergraduate Texts in Mathematics\
  (\bibinfo  {publisher} {Springer},\ \bibinfo {year} {1960})\BibitemShut
  {NoStop}%
\end{thebibliography}%

\end{document}